%% file: main.tex
\documentclass[usenatbib]{mnras}

\usepackage[T1]{fontenc}

\DeclareRobustCommand{\VAN}[3]{#2}
\let\VANthebibliography\thebibliography
\def\thebibliography{\DeclareRobustCommand{\VAN}[3]{##3}\VANthebibliography}

\usepackage{graphicx}  
\usepackage{amsmath}  
\usepackage{amssymb}  
\usepackage{float}
\usepackage{xcolor}
\usepackage{hyperref}
\usepackage{array}
\usepackage[most]{tcolorbox}
\usepackage{subcaption}
\usepackage{xspace}
\usepackage{orcidlink}
\numberwithin{equation}{section}

\newcommand{\M}{{\rm{M_\odot}}} 
\newcommand{\hp}{\rm{H_2^+}} 
\newcommand{\hmn}{\rm{H^-}} 
\newcommand{\pt}{{Pop~III~}}

\newtcolorbox{caveat}{colback=red!5!bg,colframe=orange!75!black,boxrule=0.45mm,coltext=fg}
\newtcolorbox{note}{colback=blue!10!white,colframe=blue!75!black,boxrule=0.45mm,coltext=black, sharp corners=downhill}
\newtcolorbox{highlight}{colback=yellow!10!white,colframe=yellow!75!black,boxrule=0.45mm,coltext=black, sharp corners=downhill,top=0mm,bottom=0mm,left=1mm,right=1mm}
\newtcolorbox{warning}[1][]{
  colback=yellow!5!white,
  colframe=yellow!75!black,
  top=4mm,
  fonttitle=\bfseries,
  colbacktitle=yellow!75!black,enhanced,
  attach boxed title to top left={yshift=-4mm, xshift=2mm},
  title={\faExclamationTriangle\quad}, 
  #1
}
\newtcolorbox{swarning}{ 
  colback=yellow!20!white,
  colframe=yellow!80!black,
  coltext=black,
  sharp corners=west,
  fonttitle=\bfseries,
  boxrule=0.5mm,
  leftrule=6mm,
  toprule=0pt,
  bottomrule=0pt,
  enhanced,
  overlay={
    \node[anchor=center, text=white, inner sep=1pt, circle] at ([xshift=3mm]frame.west) {\bfseries\LARGE !};
  }
}

\newcommand{\pcc}{\rm{cm}^{-3}}  

\title[Modelling Pop III stars]{A framework for modelling Population III stars in cosmological simulations}
\author[B. Saha et al.]{
  Bipradeep Saha \orcidlink{0000-0002-2329-7340}, $^{1,2}$\thanks{E-mail: bisaha@mpia.de, bipradeepsaha04@gmail.com}
  Rahul Kannan \orcidlink{0000-0001-6092-2187}, $^{3}$\thanks{E-mail: kannanr@yorku.ca}
  and Giovanni M. Mirouh \orcidlink{0000-0003-0238-8435}$^4$
  \\
  $^{1}$Max-Planck-Institut f{\"u}r Astronomie, K{\"o}nigstuhl 17, D-69117 Heidelberg, Germany\\
  $^{2}$Department of Physical Sciences, Indian Institute of Science Education and Research - Kolkata, India\\
  $^{3}$Department of Physics and Astronomy, York University, 4700 Keele Street, Toronto, ON M3J 1P3, Canada\\
  $^4$Departamento de fisica teórica y del cosmos, Universidad de Granada, Spain
}

\date{Accepted XXX. Received YYY; in original form ZZZ}

\pubyear{\the\year{}}

\begin{document}
\label{firstpage}
\pagerange{\pageref{firstpage}--\pageref{lastpage}}
\maketitle

\begin{abstract}
Population~III (Pop~III) stars are the first generation of stars to form in the universe, emerging from primordial gas composed mainly of hydrogen and helium. They play a crucial role in ending the cosmic dark ages and initiating reionization. In this work, we present a comprehensive framework for modelling Pop~III stars in cosmological simulations. This includes three key components: (1) an enhanced thermochemical network that tracks the equilibrium abundances of key catalytic species such as $\rm{H_2^+}$ and $\rm{H^-}$, which are crucial for forming molecular hydrogen in primordial gas; (2) detailed stellar spectra of Pop~III stars computed from {\tt MESA} evolutionary tracks and {\tt TLUSTY} atmosphere models; and (3) comprehensive supernova feedback, including both Core-Collapse and Pair-Instability supernovae, with detailed elemental yields. We implement these improvements in {\tt AREPO-RT} and test them using cosmological zoom-in simulations of a $1.95 \times 10^9 \M$ halo at $z=3$. Our results show that Pop~III stars form at $z > 13$ and continue forming until $z \sim 5$, significantly affecting early galaxy evolution through radiation and energetic supernova feedback. The enhanced thermochemistry enables more efficient gas cooling, while Pop~III feedback creates photo-heated diffuse gas and drives distinct metal enrichment patterns at $10 < z < 6$. The choice of IMF for Pop~III stars critically determines the balance between radiative and mechanical feedback, with top-heavy choices producing stronger feedback and more metals but retaining less metal-enriched gas within the halo. Finally, we show that high-energy radiation from Pop~III stars is necessary to explain the recent high-equivalent-width observations of the \ion{He}{ii} line from a galaxy at $z\sim11$.
\end{abstract}




\section{Introduction}
The Cosmic Microwave Background (CMB) decoupled from baryonic matter at a redshift of about 1100, allowing structures to grow unimpeded. Structure formation was then driven by gravitational instabilities in a hierarchical, bottom-up fashion. This led to the formation of the first collapsed structures in the cosmic dark ages, "minihaloes" with masses $ \lesssim 10^7 \M $ \citep{Hartwig_2022}. These minihaloes served as sites for the formation of the first stars in the universe, at $ z \approx 30 $ \citep{Klessen_2023}. This first generation of stars is known as Population III (Pop III) stars, which are characterized by forming in primordial gas that contains only elements produced in Big Bang nucleosynthesis, or, more precisely, is free of metals except for trace amounts of lithium. The birth of these stars marks the end of the cosmic dark ages and the start of Cosmic Dawn (CD) and the Epoch of Reionization (EoR).

As galaxy formation progresses, haloes with virial temperatures above $ 8000 $ K cool efficiently through de-excitation lines of atomic hydrogen \citep[{\it atomic cooling haloes}; ][]{Sutherland_1993}. However, the pristine gas can cool further to $ 200 $ K through excitation of the vibrational states of molecular hydrogen \citep[$ \rm{H_2} $;][]{Greif_2015}. Further cooling can occur only under special conditions in very massive haloes that allow efficient formation of deuterated hydrogen (HD) \citep{Glover_2008, Nagakura_2005}. Nevertheless, these temperatures remain higher than the typical temperature of the present-day interstellar medium (ISM) ($ \sim 10 $ K), achieved through metal line cooling \citep{Klessen_2016}. This increases the Jeans mass for the formation of \pt stars and reduces fragmentation, leading to a comparatively top-heavy Initial Mass Function (IMF) \citep{Klessen_2023}. While low-mass stars can still form by fragmentation of the gaseous disk \citep[see, e.g.,][]{Greif_2012,Stacey_2013b,Prole_2022a} or due to turbulence in star-forming clouds \citep[see, e.g.,][]{Turk_2009,Clark_2011}, the overall consensus is that \pt stars are, on average, more massive than their present-day counterparts.

Like all stars, the fate and evolution of the \pt stars strongly depend on the initial mass. Massive stars ($\gtrsim 8~\mathrm{M}_\odot$) either end their lives as a Core Collapse Supernova (CCSN) or, in more extreme cases, as the more energetic Pair-instability Supernova (PISN) \citep{Fowler_1964, Heger_2002, Woosley_2017}. These supernovae inject energy and metals into their host haloes, which can, in some cases, disrupt the halo and enrich the surrounding intergalactic medium (IGM) with metals. The enriched ISM/IGM cools more efficiently, initiating the formation of Population II (PopII) stars \citep{Greif_2010,Jaacks_2018,Chiaki_2019}. While the lack of metals in \pt stars prevents mass loss through stellar winds \citep{Kudritzki_2002, Krticka_2006a, Krticka_2006b, Krticka_2009}, they produce a large amount of Lyman-Werner (LW) radiation ($11.2-13.6~\rm{eV}$) and hydrogen- and helium-ionizing radiation \citep{Heger_2010, Schauer_2017, Hartwig_2022}. LW photons can penetrate the IGM and dissociate $ \rm{H_2} $, preventing gas cooling and delaying star formation in distant minihaloes \citep{Haiman_2000,Ahn_2009}. In addition to regulating the star formation rate (SFR) of the \pt stars, this process can also lead to the formation of extremely massive stars up to $ 10^5 ~\M $ and can seed supermassive black holes (SMBH) at high redshifts \citep{Wise_2008, Smith_2019, Inayoshi_2020}.

Despite extensive searches, \pt stars have yet to be directly observed in the nearby local Universe \citep{Christlieb_2002, Cayrel_2004, Keller_2014, Howes_2015}. This is because their formation peaks at $ z \approx 15-30 $ \citep{Klessen_2023}, making it unlikely that massive \pt stars would survive to the present day. Moreover, even a small amount of metal enrichment leads to more efficient cooling and the formation of Pop~II stars, which will dominate the observed flux \citep{Johnson_2013, Sarmento_2019, Liu_2020b}. At high redshifts, even a massive $ 1000~\M $ star will be too faint to be detected by the {\it James Webb Space Telescope} ({\it JWST}) \citep{Schauer_2020}. However, \pt stars can be detected by observing the energetic SN explosions that occur at the end of their lifetime \citep{Scannapieco_2005, Hummel_2012, Venditti_2024a}, by observing highly lensed caustic transits of individual star clusters \citep{Welch_2022}, or through observations of galaxies with significant \pt populations \citep{Scannapieco_2003, Maiolino_2026}. Unfortunately, their short lifetimes limit the number of expected observations. \pt stars can also be studied by quantifying their impact on the $ 21 $ cm signal from neutral hydrogen \citep{Madau_2014, Jones_2022, Jones_2025, Magg_2022a}, or by studying the gravitational wave signatures from the mergers of \pt star remnants \citep{Dayal_2019, Tang_2020, Tanikawa_2022}. Until now, we have only general observational clues about the history of \pt star formation from deep Hubble Space Telescope (HST) observations up to $ z \sim 8 $ \citep{McLure_2013, Finkelstein_2016, Bouwens_2021, Mason_2015, McLeod_2015}, which are now being complemented by {\it JWST} observations \citep{Donnan_2023, Davis_2024, Adams_2024, Harikane_2023, Maiolino_2026} pushing out to $ z \sim 15 $. Therefore, increasing the probability of observing \pt stars requires better modelling and understanding of how galaxies hosting them are distributed, both spatially and as a function of redshift.

Modelling the formation and evolution of \pt stars presents considerable challenges, as these stars originate in minihaloes with masses $ \lesssim ~ 10^7~\M $, at redshifts ranging from $10\leq z \leq 30$ \citep{Klessen_2023}. Consequently, simulating their formation necessitates a combination of high spatial resolution and extensive volume coverage to accurately capture the earliest structures formed in the universe.  Most existing simulation efforts have focused on the high-resolution `zoom-in' technique to capture the collapse of primordial gas clouds into \pt stars. These simulations include a variety of physical processes like magnetic fields which are theorized to reduce gas fragmentation \citep{Peters_2014, Sharda_2021}, a detailed thermochemical network that includes $ \hmn $, HD \citep{Glover_2005,Nishijima_2024}, streaming velocity between the dark and baryonic matter \citep{Tseliakhovich_2010,Schauer_2021, Lake_2024} and radiation transport of LW photons \citep{Haiman_1997, Greif_2011, Schauer_2021, Jaura_2022, Sugimura_2023}. Most of these simulations predict that the lack of efficient cooling gives rise to a top-heavy IMF  \citep{Hosokawa_2011, Chon_2024, Sugimura_2020}. The exact metallicity ($ Z_{\rm{crit}}  $) at which the IMF switches from being top-heavy to a regular one is still unclear. If gas cooling is dominates then $ Z_{\rm{crit}}\sim 10^{-3} - 10^{-4}~Z_\odot$ \citep{Bromm_2001, Maio_2010}, or if the dust cooling is dominant, then $ Z_{\rm{crit}} \sim  10^{-6} - 10^{-4} ~Z_\odot$ is needed \citep{Bromm_2003,Schneider_2003,Schneider_2006}.

Large-scale cosmological simulations build upon the insights gained from small-scale studies of the \pt initial mass function (IMF) to model the complete Epoch of Reionization (EoR) of the intergalactic medium (IGM). These models are typically executed through either semi-analytical frameworks or full hydrodynamical simulations \break \citep{Somerville_2012, Mason_2015, Becker_2015, Bosman_2022, OShea_2015}. The vast volumes of these simulations facilitate robust statistical analyses and investigations into the redshift dependence of the \pt star formation rate (SFR). However, these large cosmic volumes inherently restrict mass resolution, preventing the explicit resolution of the minihaloes where the first stars form. Consequently, this introduces significant uncertainty regarding the \pt SFR and the overall contribution of these stars to cosmic reionization. Furthermore, the treatment of stellar feedback is often inconsistent; while some models implement explicit sub-grid prescriptions for \pt stars based on the metallicity of newly formed star particles, others simply apply standard Population II feedback parameters to \pt populations. Moreover, the omission of Lyman-Werner (LW) radiative transfer or detailed molecular hydrogen chemistry in many studies limits the physical fidelity of their results. Even when simulations feature exceptionally high resolutions, detailed sub-grid modelling, and comprehensive thermo-chemical networks, computational expense typically dictates that they only run down to $ z > 10 $. This early termination restricts their predictive power and precludes a holistic understanding of early galaxy formation and the transition from the \pt $\to$ PopII/I regimes. Examples of studies that analyse the formation and evolution of low-metallicity stars in these contexts include {\tt THESAN-HR} \citep{Borrow_2023}, simulations with dustyGadget \citep{Venditti_2023, DiCesare_2023}, {\tt FLARES} \citep{Lovell_2021, Vijayan_2021}, the Renaissance Simulations \citep{OShea_2015}, LYRA simulations \citep{Gutcke_2021}, GIZMO-based simulations \citep{Jaacks_2019, Liu_2020b}, RAMSES based simulations \citep{Pallottini_2014, Sarmento_2018, Sarmento_2022, Sarmento_2025}, {\tt AEOS} simulations \citep{Brauer_2025a, Brauer_2025b}, and \textsc{Thesan-Zoom} simulations \citep{Zier_2025}.

In this paper we aim to develop a comprehensive framework for modelling \pt stars in a cosmological context. We implement three key elements relevant to \pt star formation and feedback:
\begin{enumerate}
  \item an enhanced thermochemistry network based on \citet{gnedin2011environmental} that enhances molecular hydrogen chemistry by including the equilibrium abundances of $\rm{H_2^+}$ and $\rm{H^-}$ species, which act as catalysts for the formation of molecular hydrogen in primordial environments.
  \item detailed spectra for \pt stars, computed using the stellar evolution code {\tt MESA} \citep[Modules for Experiments in Stellar Astrophysics; ][]{Mirouh_2023} coupled with stellar atmospheric modelling using {\tt TLUSTY} \citep{Hubeny_1988}, to accurately model the radiative feedback from the first stars.
  \item comprehensive supernova feedback from \pt stars, including both Core-Collapse and Pair-Instability Supernovae based on \citep{Heger_2002, Heger_2010} with detailed elemental yields.
\end{enumerate}
We implement these improvements in the {\tt AREPO-RT} \citep{Kannan_2019a} radiation-hydrodynamics code and test the model using cosmological zoom-in simulations within the {\tt Thesan-Zoom} framework \cite[introduced in ][]{Kannan_2025}.

The paper is structured as follows: In Section \ref{sec:methods}, we describe our methods, including the new thermochemistry network (\S\ref{sec:thermochem}), modelling of \pt stellar spectra (\S\ref{sec:\pt_spectra}), and feedback and yields from \pt stars (\S\ref{sec:pt_end-state}). We test the new \pt model using cosmological simulations in \S\ref{sec:comso-sim}, and analyse the gas and stellar mass evolution and the \pt star formation history. We also describe how our new physics modules affect the evolution of gas phases and metallicity patterns. Section \ref{sec:discussion} discusses the effects of our enhanced thermochemistry network and \pt stellar feedback, compares our findings with existing literature (\S\ref{subsec:comparison_existing_works}), and addresses limitations of our approach (\S\ref{subsec:caveats}). Finally, Section \ref{sec:conclusion} summarizes our conclusions and their implications for understanding early-universe galaxy formation and future {\it JWST} observations.

\section{Methods}\label{sec:methods}
We use the   {\tt Thesan-Zoom} framework \citep{Kannan_2025} to model the impact of \pt stars. Briefly, the simulations are performed using {\tt AREPO-RT} \citep{Kannan_2019a, Zier2024}, a radiation-hydrodynamic (RHD) extension to the moving mesh code {\tt AREPO} \citep{Springel_2010}. {\tt AREPO} solves the equations of (radiation-) hydrodynamics on an unstructured mesh that adapts to gas flows. This mesh is constructed by dividing space into Voronoi cells, which are constantly adjusted as the gas moves. Solutions to the hydrodynamic equations are obtained in a quasi-Lagrangian manner, by  solving them at interfaces between moving mesh cells in the rest frame of the interface. The use of second-order Runge-Kutta time integration in conjunction with a least square fit (LSF) gradient estimate, which performs well even on highly distorted meshes \citep{Pakmor_2015}, ensures higher-order accuracy. Additionally, the mesh is regularly regularised using a method described in \citet{Vogelsberger_2012}.

The gravitational forces are calculated using the Hybrid Tree-PM approach. This method estimates short-range forces through a hierarchical oct-tree algorithm \citep{Barnes_1986}, while long-range forces are computed using the particle mesh (PM) method. The PM method bins particles onto a grid, and the gravitational potential is derived by solving the Poisson equation using the Fourier method. Furthermore, if the count of active particles falls below a specific threshold, the gravitational force is computed through direct summation. This approach is particularly beneficial in large-scale simulations, where the computational cost of direct summation is less than the tree algorithm for the lowest time-bins, typically populated by only a few active particles. A hierarchical time integration approach is employed to construct the tree only for the currently active particle set, accelerating gravity calculations. This is especially useful as the time-bin hierarchy can become very deep in our simulations \citep{Springel_2020}.

Star formation is treated stochastically, with self-gravitating, Jeans-unstable gas cells with densities above $10\,\text{cm}^{-3}$ considered eligible to form stars.
Each star particle (non-Pop~III stars) represents a stellar population, with an assumed Chabrier initial mass function \citep[]{Chabrier_2003} with a minimum and maximum stellar mass of $0.1~\rm{M}_\odot$ and $100~\rm{M}_\odot$ respectively. Supernovae (SN) feedback and stellar winds are implemented using the Stars and MUltiphase Gas in GaLaxiEs (SMUGGLE) model \citep[]{Marinacci_2019}, with further improvements discussed in \citet{Kannan_2025}. Stars with mass $M_\ast > 8\,\text{M}_\odot$ explode as SN type II, and energy, momentum, and mass are injected into the closest gas cells, along with metals and dust \citep{Vogelsberger_2013, McKinnon_2016}.
The simulations also include feedback from SN type Ia and stellar winds from O-B and asymptotic giant branch stars.
Moreover, early stellar feedback, which disrupts molecular clouds, is added to regulate feedback before the first SN explosions, injecting momentum for the first $5$\,Myr after star particles are formed.

Radiation fields are evolved \citep{Kannan_2019a} by solving a set of coupled hyperbolic conservation equations for photon number density and photon flux, closed using the M1 scheme \citep{Levermore_1984,Dubroca_1999}. Radiation fields are coupled to the gas through a six-species non-equilibrium thermochemical network that models the ionization state and cooling/heating from molecular and atomic hydrogen and helium. This framework allows us to self-consistently follow the photoionization, photoheating, and radiation pressure from local sources on the surrounding medium. Further improvements to this framework for modelling the formation and evolution of \pt stars are detailed below.

\begin{table}
  \centering
  \renewcommand{\arraystretch}{1.2}
  \begin{tabular}{llc}
    \hline
    & Reactions                                             & Energy band/Threshold $ [\rm{eV}] $ \\
    \hline\hline
    $\Gamma_{\rm A}$    & $\rm{H^-} + \gamma \rightarrow \ion{H}{i} + e$          & $ > 0.755  $           \\
    $\Gamma_{\rm B}$    & $\rm{H_2^+} + \gamma \rightarrow \ion{H}{I} + \ion{H}{II}$ & $ [2.65-21] $          \\
    $\Gamma_{\rm C}$    & $\rm{H_2^+} + \gamma \rightarrow 2\ion{H}{II} + e$      & $ [30-90] $            \\
    $\Gamma_{\rm D}$    & $\rm{H_2} + \gamma \rightarrow \rm{H_2^+} + e$       & $ > 15.42  $           \\
    $\Gamma_{\rm E}$    & $\rm{H_2} + \gamma \rightarrow 2\ion{H}{I}$             & $ [14.159-17.6]$       \\
    $\Gamma_{\rm LW}$   & $\rm{H_2} + \gamma \rightarrow 2\ion{H}{I}$             & $ [11.18-13.6] $       \\
    $\Gamma_{\rm HI}$   & $\ion{H}{I} + \gamma \rightarrow \ion{H}{II} + e$               & $ > 13.6  $         \\
    $\Gamma_{\rm HeI}$  & $\ion{He}{I} + \gamma \rightarrow \ion{He}{II} + e$              & $ > 24.59 $          \\
    $\Gamma_{\rm HeII}$ & $\ion{He}{II} + \gamma \rightarrow \ion{He}{III} + e$           & $ > 54.42 $          \\
    \hline
  \end{tabular}
  \caption{Summary of photon-mediated reactions and their corresponding energy bands/thresholds.}
  \label{eq:photo-ion-coeffs}
\end{table}

\subsection{A new thermochemical network for early universe}\label{sec:thermochem}

As outlined in \citet{Kannan_2020}, {\tt Arepo-RT} includes a thermochemical network that models the abundance of molecular hydrogen $(\rm H_2)$ using a simplified network that performs well in high-metallicity dusty environments \citep{Nickerson_2018}. However, in the primordial gas which is devoid of metals and dust,
$\rm{H_2}$ formation relies on additional gas phase reactions modulated by species like $ \rm{H_{2}^+}$ and $\rm{H^-} $. $ \rm{H^-} $ facilitates a 2 step process for $ \rm{H_2} $ formation \citep{Peebles_1968}:
\begin{equation}
  \rm{H + e}   \rightarrow \rm{H^- + \gamma} \hspace{2ex};\hspace{2ex} \rm{H + H^-} \rightarrow \rm{H_2 + e^-}~.
\end{equation}
Similarly, $ \rm{H_2^+} $ facilitates $ \rm{H_2} $ formation \citep{Saslaw_1967} by:
\begin{equation}
  \rm{H + H^+} \rightarrow \rm{H_2^+ + \gamma } \hspace{2ex};\hspace{2ex} \rm{H + H_2^+} \rightarrow \rm{H_2 + H^+}~.
\end{equation}
We therefore, update the thermochemical network to include these species following the model outlined in \citet{gnedin2011environmental}.
The new reaction network is  described in full in Appendix~\ref{app:therm} (Eqs. \ref{eq:HI}-\ref{eq:temperature-evolution}), where $\dot{\mathcal{M}}_j$ is the rate of change of the total number of ions/molecules for the $j$-th species in $ [{\ion{H}{i}, \ion{H}{ii}, \rm{H}_2, \rm{H}_2^+, \rm{H}^-, \ion{He}{I}, \ion{He}{II}, \ion{He}{III}}] $  and internal energy `U', $\Gamma_{\rm [A-E, LW, \ion{H}{i}, \ion{He}{i}, \ion{He}{ii}]}$ are photo-ionization rates for the different reactions described in Table \ref{eq:photo-ion-coeffs}, $\alpha^D_{\rm H_2}$ is the formation rate of molecular hydrogen on dust grains, \textit{D} is the dust-to-gas ratio (modelled self-consistently using the dust model described in \citealt{McKinnon_2016}) and $D_{\rm MW}$ is the dust-to-gas ratio in the Milky-Way (MW; 0.01).

The rate coefficients $ k_1 - k_{31}  $ are obtained from \citet{Glover_2008}, the frequency dependence of photoionization cross-sections $(\sigma)$ for (listed in Table \ref{eq:photo-ion-coeffs}) $\Gamma_{\rm{A}}$ to $\Gamma_{\rm{D}}$ are taken from \citet{Shapiro_1987}, for  $\Gamma_{\rm{E}}$ we use the rate from \citet{Abel_1997}, and we use a fixed value of $ 2.47\times10^{-18} ~\rm{s}^{-1} $ for the  $\rm H_2 $ dissociation by LW photons \citep{Nickerson_2018}. The photo-ionization rates are then given by:
\begin{equation}
  \Gamma _j = -\tilde{c} n_j \sum_i \bar{\sigma}_{i j} N_{\gamma }^i~,
\end{equation}
where $ \bar{\sigma}_{ij} $ is the mean photo-ionization cross-section of the $j^{th} $ species in the $i^{th} $ bin \citep[see][]{Kannan_2019a}, and $ N_{\gamma }^i $ is the number density of photons in the $i^{th} $ bin. The $ ``-" $ sign indicates that the $ j^{th}  $ species is being destroyed by the photo-ionization process. The values of $\Gamma_{\rm{HI} }, \Gamma_{\rm{HeI}}, \Gamma_{\rm{HeII}}, \alpha_{\rm{HII}}, \sigma_{\rm{eHI}}$ and $ \alpha^D_{\rm{H_2}} $ are taken from \cite{Kannan_2019a, Kannan_2020}.

This set of thermochemical equations can be simplified by noting that, in the regimes we are interested in, the abundances of $ \hp $ and $ \hmn $ are always extremely small, so they can be assumed to be in kinetic equilibrium, i.e., $\dot{\mathcal{M}}_{\rm{H}_2^{+}}, \dot{\mathcal{M}}_{\rm{H}^{-}} = 0$. We also assume that there are no cross-species reactions (i.e., no reactions involving H and He together) and neglect quadratic terms proportional to $ n_{\rm H^-}n_{\rm H_2^+} $ and three-body terms, which are relevant only at very high densities not achieved in our simulations and have very low interaction cross-sections. With these assumptions, we arrive at a simple eight-species model, which is closed using the following relations:
\begin{align}
  n_{\rm{H}} &= n_{\rm{HI}}+n_{\rm{HII}}+2 n_{\rm{H}_2} ~,\label{eq:species-closure}\\
  n_{\rm{He}} &= n_{\rm{HeI}}+n_{\rm{HeII}}+n_{\rm{HeIII}}~, \\
  n_e &= n_{\rm{HII}}+n_{\rm{HeII}}+2 n_{\rm{HeIII}}~.\label{eq:electron-closure}
\end{align}

Then, the total gas cooling is given by:
\begin{align}
  \Lambda _\text{tot} = &~\Lambda _p\left(n_j, N_\gamma ^i, T\right) + \frac{Z}{Z_\odot } \Lambda _M(T, \rho , z) \nonumber\\
  &+ \Lambda _\text{PE}\left(D, T, N_\gamma ^\text{FUV}\right) + \Lambda _D\left(\rho , T, D, N_\gamma ^\text{IR}\right)~, \label{eq:cooling-function}
\end{align}
where $ \Lambda_p $ is the primordial cooling from Hydrogen and Helium, $ \Lambda_M $ is the metal line cooling, $ \Lambda_{\rm{PE} } $ is the photo-electric heating, and $ \Lambda_D $ is the dust cooling, $ \rho  $ is the density of the gas cell, $n_j$ is the number density of the $ j^{th} $ ionic species tracked in our thermochemistry network, and $ N_\gamma^{\rm{IR}} $  and $ N_\gamma^{\rm{UV} } $  are the photon number density in the infra-red (IR) and far ultraviolet (FUV) bands, respectively. In the simulations, instead of the temperature, we evolve the internal energy ($\rm{U}$) of the gas, which is related to the temperature through the specific heat at constant volume, $C_{\rm v}$. The evolution of the internal energy ($\dot{\mathcal{M}}_{\rm{U}}$) is described in Eq. \ref{eq:temperature-evolution}, where $ \Lambda\left( \rm{H_2}  \right) = \Lambda(n\to 0)_{\rm{H_2HI} }n_{\rm{H_2} }n_{\rm{HI} } + \Lambda(n \to 0)_{\rm{H_2H_2}}n_{\rm{H_2} }^2  $ is the cooling from molecular hydrogen due to collisions in the low density limits. We add two additional cooling channels, with $\Lambda_{\rm{H_2^+}e}$ denoting the cooling due to the collision of $ \rm{H_2^+} $ with electrons, and $ \Lambda_{\rm{H_2^+ HI}} $ the cooling due to the collision of $ \rm{H_2^+} $ with $ \rm{HI} $. All these cooling rates are calculated in the low density limit as outlined in \citet{Glover_2008}. $\Lambda_{\text{C}} $ is the cooling due to inverse Compton scattering of the electrons with the background CMB photons. Finally we note that the photoheating and momentum injection rates through photon absorption  remains unchanged from the prescriptions outlined in \cite{Kannan_2019a}.

\begin{table*}
  \centering
  \input{Tab-thermochem}
  \caption{Values of different reactions rates used in the thermochemistry calculations. The first segment lists the mean ionization cross-section for the various reactions in units of $10^{-18}~\rm{cm}^2$. The second segment lists the mean photoheating rate for different reactions in units of eV and the third section lists the mean momentum transfer per photon absorption for different reactions in units of eV.}
  \label{tab:Initialized_RT_values}
\end{table*}

Therefore the final thermochemical network tracks the non-equilibrium abundances of six species $ \left( \rm{H_2, HI, HII, HeI, HeII, HeIII},   \right)$  and internal energy, in addition to equilibrium abundances of \ion{H}{$_2^+$} and \ion{H}{$^-$}.  This involves solving five (instead of seven, because we can use the closure relations to reduce the dimensionality of the problem) highly-coupled non-linear equations for:
\begin{align}
  \left[ \dot{\mathcal{M}}_{\rm{H_2}},
  \dot{\mathcal{M}}_{\rm{HII}},  \dot{\mathcal{M}}_{\rm{HeII}}, \dot{\mathcal{M}}_{\rm{HeIII}}, \dot{\mathcal{M}}_{\rm{U}}  \right] ~.
\end{align}
The numerical integration of this thermochemistry network is quite challenging as small changes in the photon density can lead to rapid changes in the ionization state and temperature of the gas. Therefore, an explicit time integration of these equations would require very small time-steps making the thermochemistry step computationally expensive. Therefore we resort to a semi-implicit scheme (an extension of the one outlined in Appendix B of \citealt{Kannan_2019a}), which first solves for the number density evolution of the ionic species implicitly using the values of the $ n_e $  and $ N_{\gamma } ^i $  from the previous time-step. $ n_e $  is then updated with the revised values of the number density of the ionic species using Eq \ref{eq:electron-closure}. This semi-implicit solution is only used if the changes in abundances and internal energy is small, less than $10\%$ of the previous value. If the change is larger, then  we solve the coupled differential equations using the publicly available {\tt SUNDIALS CVODE} \citep{Hindmarsh_2005} solver, which employs a variable order, variable step, and multi-step backward difference scheme to compute the new temperature and chemical abundances.

Table \ref{tab:Initialized_RT_values} lists the ionization cross sections, photoionization rates, and momentum injection rates used in this work. Although we evolve the abundances of $\rm{H_2^+, H^-}$, the corresponding photoheating and momentum injection from these reactions are not included, as their number densities are always very small. In the optical and FUV bands, the photo-detachment of $ \rm{H^-} $ (reaction A) and the photo-dissociation of $ \rm{H_2^+} $ (reaction B) have the two highest reaction rates because they are the main catalysts for $ \rm{H_2} $ formation in primordial gas. In the LW band, even though $ \overline{\sigma }_{ij} $ is higher for reactions A and B than for LW dissociation (i.e., photo-dissociation of $ \rm{H_2} $), most photons dissociate $ \rm{H_2} $, as $ n_{\rm{H_2}} \gg n_{\rm{H_2^+}}, n_{\rm{H^-}} $. In the EUV1 band, both the dissociation of $ \rm{H_2} $ and the ionization of $ \rm{HI} $ occur, as the cross-sections are comparable. In the EUV2 band, most photons are absorbed by $ \rm{HeI} $, and in the EUV3 band, photons are absorbed by $ \rm{HeII} $.

\subsection{Radiation output from \pt stars}\label{sec:\pt_spectra}

A substantial portion of the energy output from massive stars is emitted as radiation during their long-lived main-sequence (MS) phase \citep[see, e.g.,][]{Heger_2010,Agertz_2013}. In general, the spectra of \pt stars are thought to differ significantly from those of present-day metal-enriched stars, as we expect them to be more massive on average. In this work, we combine stellar evolutionary tracks from {\tt MESA} \citep{Paxton2011,Paxton2013,Paxton2015,Paxton2019,Jermyn2023} with {\tt TLUSTY} \citep{Hubeny_1988} stellar atmosphere models to accurately model \pt spectra. A brief outline of this process is given below:
\begin{enumerate}
  \item Model the evolution of metal-free stars at $ 120 $ different masses covering $[0.1, 1000]~\M $, initialized at BBN proportions, using {\tt MESA}.
  \item Using the evolutionary tracks from {\tt MESA}, and calculate the spectra of the stars at each point during the MS evolution using {\tt TLUSTY}.
  \item Combine the evolutionary track history to compute stellar spectra for each star over a given time period, $L_{\nu }(M, t)$.
  \item Integrate over $ L_\nu (M,t) $, weighted by the initial mass function (IMF), to compute the \pt spectrum for a single stellar population.
\end{enumerate}

\subsubsection{Initial Mass Function of \pt stars}
The IMF describes the distribution of initial masses for a population of stars formed from the same molecular cloud. The most widely used form of the IMF is a power-law function, originally proposed by \cite{Salpeter_1955}:
\begin{equation}\label{eq:IMF}
  \Phi (m) := \frac{\rm{d} N}{\rm{d}m} \propto m^{-\alpha} ~ ~ ; ~ ~ M_{\rm{\min} } \leq m \leq M_{\rm{\max} }
\end{equation}
where $ N $ is the number of stars with given mass $ m $, $ \alpha  $ is the power-law coefficient, and as the names suggest $ M_{\rm{\min} }, M_{\rm{\max} } $ bracket the range of possible stellar masses. Recent advancements have refined the functional form of the IMF, specifically at the low mass end. One such example is the Chabrier IMF  \citep{Chabrier_2004} which predicts a log-normal distribution below $1~\rm{M}_\odot$.

\begin{table}
  \centering
  \renewcommand{\arraystretch}{1.3}
  \begin{tabular}{ccccccc}
    \hline
    IMF      & $\alpha$ & $M_{\rm min}$ & $M_{\rm max}$ & $\overline{M}$ & $\bar{t}_{\rm life}$ & $\bar{z}_{\rm death}$  \\
    \hline\hline
    Salpeter & $-2.35$  & $2.0$ & $150$ & $6.03$  &  $615.2$  & $6.79$    \\
    Log-flat & $-1$     & $2.0$ & $150$ & $34.28$ &  $351.7$  & $9.37$   \\
    Log-flat & $-1$     & $2.0$ & $250$ & $51.36$ &  $336.9$  & $9.58$\\
    \hline
  \end{tabular}
  \caption{
    The various IMF and their parameters used in this study. For each IMF, we show the mean stellar mass in units of $ \M $, mass-weighted mean stellar lifetime $ \overline{t} _{\rm{life} } $ in units of Myr, and the redshift at which a star born at $ z = 20 $ would die ($\bar{z}_{\rm death}$) after living for $ \overline{t}_{\rm{life} }$,   $ z_{\rm{death} } $. $\bar{z}_{\rm death}$ is calculated assuming best fit parameters for $\Lambda CDM$ model from Planck 2018 \citep{Planck_2018}.
  }
  \label{tab:IMF_params}
\end{table}

Table \ref{tab:IMF_params} lists the model parameters for the different IMFs used for \pt stars in this study. For each IMF, we list the mean mass ($\overline{M}$) of the stellar population
\begin{equation}
  \overline{M} = \frac{\displaystyle \int_{M_{\rm{ \min} }}^{M_{\rm{\max} }} m \cdot \Phi (m) ~ \text{d}m}{\displaystyle\int_{M_{\rm{ \min} }}^{M_{\rm{\max} }}\Phi (m) ~ \text{d}m}~,
\end{equation}
the mass-weighted mean stellar lifetime
\begin{equation}
  \overline{t}_{\rm{life} } = \frac{\displaystyle\int_{M_{\rm{ \min} }}^{M_{\rm{\max} }} m \cdot t_{\rm{ life}}(m) \cdot\Phi(m) ~ \text{d}m}{\overline{M}}~,
\end{equation}
and the redshift at which a star born at $ z = 20 $ would die after living for $ \overline{t}_{\rm{life} }$ ($\bar{z}_{\rm death}$), assuming the best-fit parameters for the $\Lambda\rm{CDM}$ model from Planck 2018 \citep{Planck_2018}.

We note that the three IMFs have very different properties and are intended to span large uncertainties in the properties of the \pt stars and their resultant spectra. For a Salpeter IMF, the distribution of stellar masses is skewed towards the low-mass end, and as a result the population has a long $\overline{t}_{\rm{life} }$. The top-heavy Log-flat IMFs have a higher probability of forming higher-mass stars and hence a short $\overline{t}_{\rm{life} }$.
In contrast to a Pop~II/I IMF, the minimum mass ($M_2{\rm{min}}$) of the \pt stars is higher ( $\sim 2~\rm{M}_\odot$) because cooling in primordial environments is less efficient, leading to less fragmentation and higher stellar masses. This is why we also expect the \pt IMF to be more top-heavy, with a larger $M_{\rm{max}}$ \citep[see, e.g.,][]{Klessen_2023}. Therefore, we use the Log-flat IMF (with $M_{\rm{max}} = 150~\rm{M}_\odot$) as the fiducial IMF for the \pt stars in our simulations, but also perform simulations assuming different IMFs and test their impact on galaxy properties, as shown in \S\ref{sec:comso-sim}.

\begin{figure}
  \centering
  \includegraphics[width=0.99\linewidth]{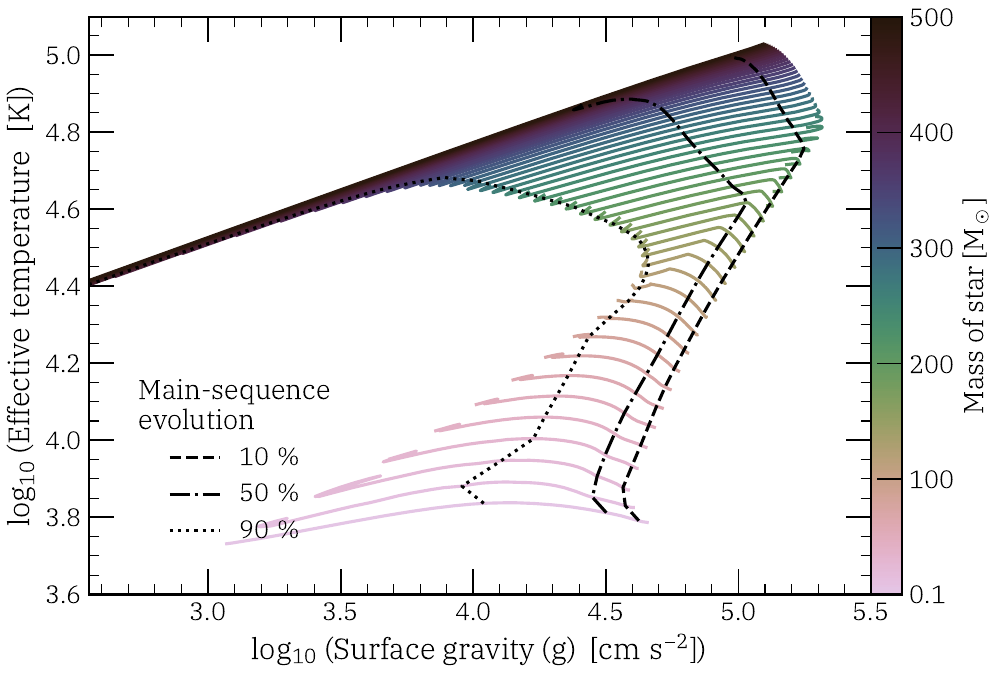}
  \caption{MESA stellar evolutionary tracks plotted in the effective temperature ($T_{\rm{eff}}$) - surface gravity (log(g)) phase space for \pt stars as a function of $\rm M_\star \in [0.8, 500]~\M$, with $10, 50, 90\%$ lifetimes of the stars indicated by dashed, dot-dashed and dotted black lines.}
  \label{fig:MESA_tracks}
\end{figure}

\subsubsection{Stellar Tracks of \pt stars}
The IMF-averaged radiation output from a population of \pt stars is calculated by first estimating the radiation output from individual stars of different masses. As the stellar spectrum evolves over a star's lifetime \citep{Schaerer_2002}, it is necessary to model the evolutionary histories of individual stars of different masses and then sum the stellar spectra, weighted by the IMF, to obtain the expected radiation at any given time. We follow the method outlined in \citet{Mirouh_2023} and use the MESA stellar evolution code to model 120 evolutionary tracks for non-rotating stars with masses between $0.1$ and $1000\,\M$. These models are metal-free with BBN proportions of hydrogen and helium, namely $X=0.76, Y=0.24, Z=0$ (we point the reader to \citealt{Skillman_2026} for a thorough review). As the parametrization of convection through the mixing-length parameter is unconstrained for \pt stars, we use a fiducial value of $\alpha_{\rm MLT} =2$ \citep[see][for more details]{Joyce_2023}. We also include semi-convection following the prescription of \citet{Langer1985} with $\alpha_{\rm sc} = 0.1$ and core step overshooting following the solar calibration of \citet{Christensen-Dalsgaard_2011}, which extends from $0.05 H_P$ within the convection zone to a thickness of $0.33 H_P$ (where, $H_P$ is the pressure scale height), using the same diffusion coefficient. Mass loss is not included in the models to ease the interpolation of the resulting grid \citep[see \S2.2 of][for more details]{Mirouh_2023}.

The pre-main-sequence (PMS) evolution is modelled as a radial contraction of an isothermal sphere with a mass equal to the star's zero-age main-sequence (ZAMS) mass. This widely used approach yields an early PMS evolution that is somewhat unrealistic but provides robust ZAMS parameters. As the PMS phase is quite short-lived we do not include it in our spectrum calculations. Once on the MS, low- and high-mass stars exhibit different behaviours. Stars with masses $m \lesssim 0.8 {\rm M}_\odot$ are either fully convective or have both a convective core and envelope, until radiation dominates in the core, leaving a deep convective envelope. Conversely, stars with masses $m \gtrsim 0.8 {\rm M}_\odot$ have a radiative envelope and a convective core for most of their MS evolution, with stars in the range $0.8\,{\rm M}_\odot < m < 7\,{\rm M}_\odot$ even experiencing a fully radiative phase in the second half of their MS evolution.

\begin{figure}
  \centering
  \includegraphics[width=0.99\linewidth]{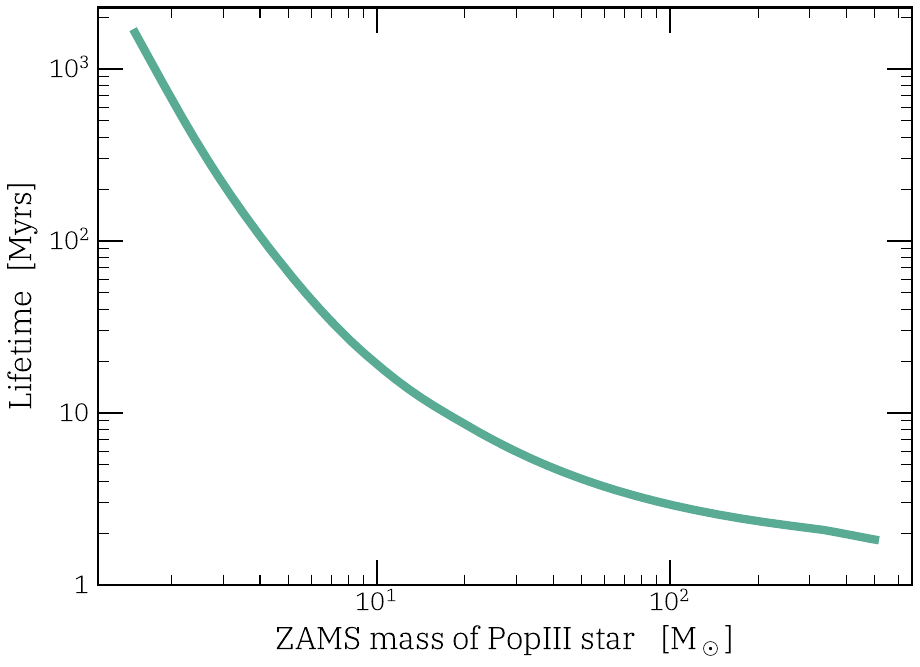}
  \caption{Lifetime of \pt stars as a function of stellar mass. The lifetimes are computed from the MESA evolutionary tracks and is defined as the time spent by the star on the main-sequence.}
  \label{fig:Pop3Lifetimes}
\end{figure}

Stars with $m < 310 {\rm M}_\odot$ are evolved until the end of their MS evolution, but more massive stars are simulated only until they reach the Eddington luminosity limit and begin to photo-evaporate. During the photo-evaporation phase, emission from these stars will be negligible, as the radiation output decreases rapidly before photo-evaporation. This is attributed to a sharp drop in effective temperature as the outer layers expand \citep[e.g.][]{Sanyal_2015}. These stars are also expected to lose a significant portion of their mass before photo-evaporation due to strong stellar winds \citep[e.g.][]{Vink_2018,Higgins_2022}, which is not included in our modelling. We note that similar models have been used in previous studies examining the impact of \pt stars and X-ray binaries on the cosmological 21-cm signal \citep{Jones_2022,Jones_2025,Sartorio_2023}.

Fig.~\ref{fig:MESA_tracks} shows the resultant evolutionary tracks plotted in the effective temperature ($T_{\rm{eff}}$) - surface gravity (log(g)) phase space for \pt stars with different masses. The dashed, dashed-dotted and dotted lines representing $ 10, 50$ and $90\%$ of the MS evolution respectively. The $x$-axis is clipped for visual clarity. Fig.~\ref{fig:Pop3Lifetimes} plots the corresponding MS lifetime of \pt stars as a function of their mass.

\subsubsection{Flux from \pt stars}
The properties of the stars predicted by MESA stellar evolutionary tracks are used as input for the stellar atmosphere code {\tt TLUSTY} \citep[cersion $205$][]{Hubeny_1988,Hubeny_2017a, Hubeny_2017b, Hubeny_2017c}. Specifically, three important parameters, the effective surface temperature $ T_{\rm{eff}} $, surface gravity $ \log(g) $, and the composition of the star at every point in its evolution, are used to estimate the stellar fluxes. An iterative approach outlined in \citet{Hubeny_2017a} is used, where the code first computes the structure of the stellar atmosphere assuming Local Thermodynamic Equilibrium (LTE), then estimates the non-local thermodynamic equilibrium (NLTE) solution without lines, and finally calculates the NLTE atmosphere with the lines. Each step inherits the result of the previous step as its initial atmosphere during this iterative process.

Using the effective temperature ($t_{\rm{eff}}$), surface gravity ($\log(g)$), and chemical composition of the star, {\tt TLUSTY} computes the \textit{Stellar flux}\footnote{{\tt TLUSTY} computes the Eddington flux $ H_\nu $ for a given configuration, i.e., the first moment of the specific intensity. We convert to flux by assuming spherical symmetry, i.e., $ F_\nu = 4\pi H_\nu $.} ($ F_\nu $). The radiation field of the stars is predominantly determined by the composition of the outermost radiating layer. Since the stars are mostly radiative in the mass range covered by our IMFs, changes in surface composition are minimal. Hence, throughout their evolutionary history, we fix the composition of the stars to BBN proportions of hydrogen and helium. To calculate the spectra, we initialize a grid of size $ 400 \times 240 $ that spans $ T_{\rm{eff}} $ from $ 4000~\rm{K}$ to $110000~\rm{K}$ and $ \log(g) $ from $ -0.5 $ to $ 5.5 $, and calculate the spectra at each point on the grid. On this grid, we compute $ F_\nu $ between $ 0.004 - 165~\rm{eV} $ with $ 10000 $ linearly spaced points. This grid is sufficient to cover the atmospheres of the stellar tracks shown in Fig \ref{fig:MESA_tracks}. We then use linear interpolation in $ T_{\rm{eff}} $ and $ \log(g) $ to calculate the stellar flux.

\begin{figure}
  \centering
  \includegraphics[width=0.99\linewidth]{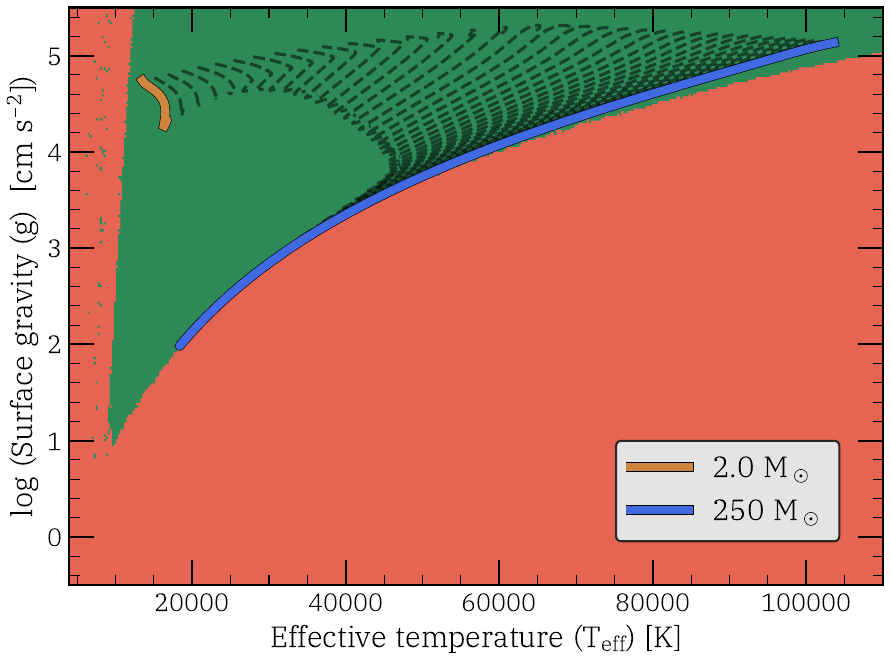}
  \caption{Convergence of the stellar atmosphere across a grid of $T_{\rm eff}, \log(g)$ values. The green points indicate converged regions, while the red ones indicate unconverged regions. The orange and blue curves show the evolutionary tracks of the least- and most-massive stars whose stellar atmosphere models are stable and converged, with dashed lines showing the tracks for intermediate masses. For the rest of this work, we consider only stars in the converged regime.}
  \label{fig:Tlusty_grid}
\end{figure}

We note that not all stars within our parameter space have stable atmospheres because they are either super-luminous or not Eddington stable. The stars mainly affected by this are less massive than $ 2~\M $ or more massive than $250~\M$. In the rest of our calculations, we use only stars with converged atmospheric properties, as shown in  Fig.~\ref{fig:Tlusty_grid}. The green region marks the parameter space with converged stellar atmospheres, whereas the red region shows the parameter space where the stellar atmospheres do not converge. The thick solid lines indicate MS tracks of the least and most massive stars with stable atmospheres, with dashed lines for everything in between. We found that ignoring stars without converged atmospheres (i.e., $M_\star < 2~\M$ and $M_\star > 250~\M$) does not significantly change the resultant spectra because of the low luminosity at the low-mass end and the very short MS lifetime at the high-mass end.

The stellar flux grid is then used to evaluate the spectra of individual stars at any point during their MS evolution. Specifically, we calculate the spectral emission rate ($\dot{\varepsilon}$) in units of $ [\rm{photons}~s^{-1}~Hz^{-1}] $
\begin{equation}\label{eq:spectral-emission-rate}
  \dot{\varepsilon} \left(\nu;M , t\right) = A(t)\times F_\nu\left( \nu ; T_{\rm{eff}}(t), \log[g(t)] \right)
\end{equation}
where $ A(t) $ is the star's surface area at time $ t $, and $ F_\nu\left( \nu ; T_{\rm{eff}}(t), \log[g(t)] \right) $ is the star's flux at frequency $ \nu $ at time $ t $, obtained by interpolating the pre-computed grid.  \footnote{Often, we will switch to another measure of stellar radiation, namely luminosity $ L_\nu $, in units of $ [\rm{ergs~s^{-1}~Hz^{-1} }] $}

\begin{figure}
  \centering
  \includegraphics[width=0.99\linewidth]{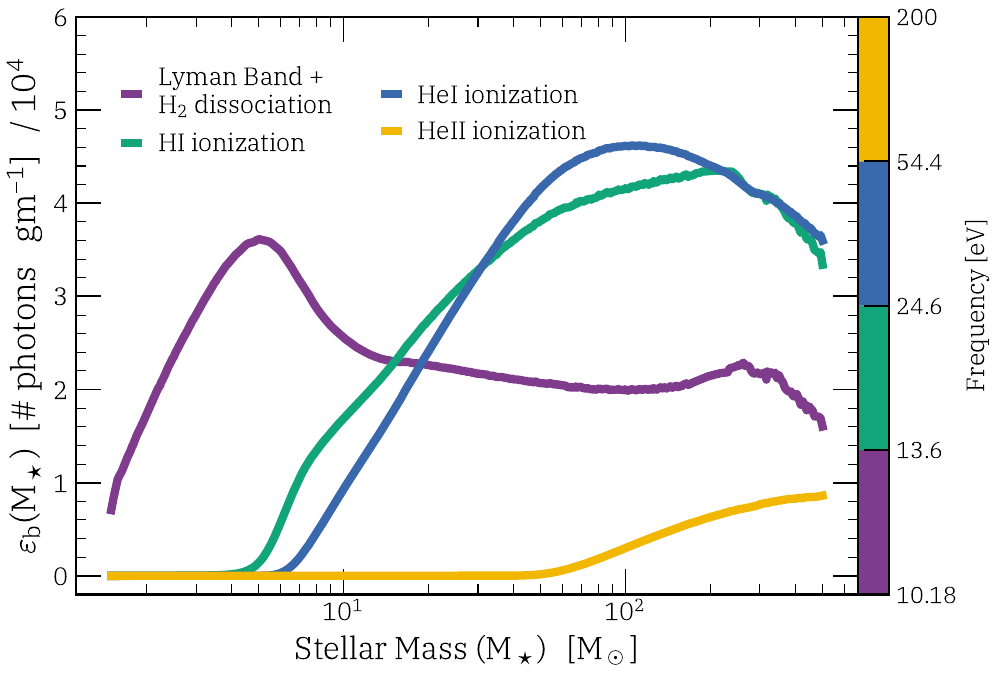}
  \caption{Lifetime photon emissivity per stellar baryon for \pt stars across different masses and frequency bands. The Lyman band emission peaks at $5~\rm{\M}$, decreases to $10~\rm{\M}$, plateaus until $\sim 200~\rm{\M}$, then declines at higher masses due to photo-evaporation of very massive stars. The HI and HeI ionization bands emerge at $\gtrsim 5~\rm{\M}$ and peak at $100~\rm{\M}$ and $250~\rm{\M}$, respectively, while HeII band emission is significant only above $50~\rm{\M}$.}
  \label{fig:ionization-contribution}
\end{figure}

\begin{figure*}
  \centering
  \includegraphics[width=\textwidth]{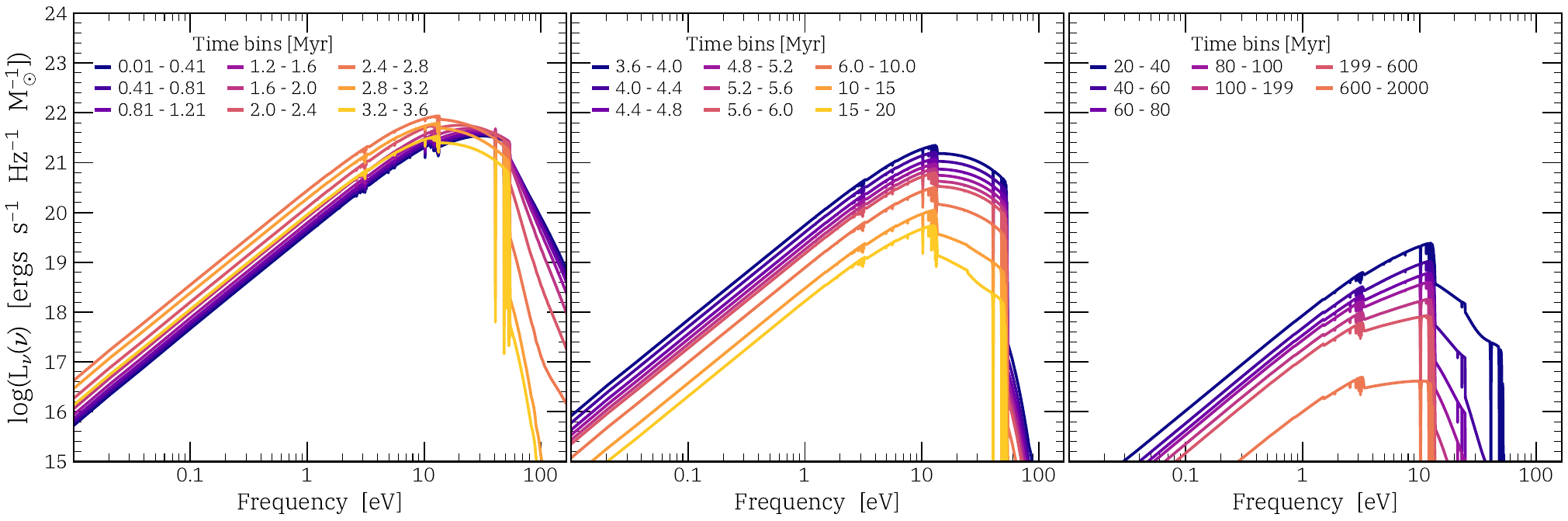}
  \caption{IMF-averaged luminosity per unit stellar mass, $ L_{\rm \nu,IMF} $, for \pt stars with a log-flat IMF, $m_\star \in [2.0, 150]~\M$. Each curve shows the luminosity at different ages of the stellar population, as indicated. Massive young stars emit a large amount of high-energy radiation during the first few million years, but the radiation output decreases as these stars die out quickly.}
  \label{fig:IMF_spec}
\end{figure*}

Fig.~ \ref{fig:ionization-contribution} shows the total emission per stellar baryon ($ \varepsilon_b $), defined as
\begin{equation}
  \varepsilon_b(m) = \frac{\displaystyle\int_{t_{\rm{life}}} \displaystyle\int_{\nu,\rm{start}}^{\nu,\rm{end}} \dot{\varepsilon} \left(\nu;m , t\right)~\text{d}\nu ~ \text{d}t}{m} ~ ,
\end{equation}
as a function of the \pt star's mass ($m$) in the LW (purple curve), HI ionizing (green curve), HeI ionizing (blue curve), and HeII ionizing (yellow curve) bands. Emission in the LW band peaks at $ 5~\rm{\M} $ and decreases rapidly for lower-mass stars. Emission also decreases gradually up to $ 10~\rm{\M} $, then remains nearly constant to about $\thickapprox 200~\rm{\M} $ before showing another peak. This smaller peak arises from the photo-evaporation of massive stars ($ > 310~\rm{\M} $), which reduces $ T_{\rm{eff}} $ below the Ly$-\alpha $ temperature, thereby exponentially suppressing the LW flux. Because this suppression is exponential, it dominates the effect of increasing surface area, leading to an overall decline in the LLW band flux of high-mass stars \citep[for a detailed discussion, see][Appendix A]{Jones_2022}. Only stars more massive than $ \thickapprox 5~\rm{\M} $ are hot enough to produce $ \rm{HI,HeI} $ ionizing band radiation. Emission gradually increases and then peaks at $ 100 (250)~\rm{\M} $ in the EUV1 (EUV2) band. The decline in $ \varepsilon_b $ in the HI and HeI ionization bands for high-mass stars is again attributed to photo-evaporation of the most massive stars. High-energy photons in EUV3 are emitted only by extremely massive stars with $\rm{M}_\star > 50~\rm{\M} $, with a gradual increase up to $ 500~\rm{\M} $. This also suggests that the IMF-averaged stellar flux will be determined by a complex balance between the age of the stellar population and the shape of the mass function.

\subsubsection{IMF Averaged Stellar Spectra}
The IMF averaged spectra for a single stellar population (SPP) of \pt stars at any point in time is obtained by summing over the contributions from all stars that are on the main-sequence (MS) at that time. As we do not have continuous data, instead of determining the stellar radiation at some time $ t_i $ , we calculate the mean stellar radiation(luminosity) in a small window, $ t_i \pm \Delta t $:
\begin{equation}
  \overline{L}_\nu(t_i, m) = \frac{\displaystyle \int_{\mathrm{t_i - \Delta t} }^{\mathrm{t_i + \Delta t} } L_\nu(t, m, T_{\mathrm{eff} }, \log(g)) \text{d}t}{2\Delta t}~.
\end{equation}
Then the IMF averaged luminosity per stellar mass at time $ t_i $ ($ L_{\rm{\nu, IMF}} (\nu , t_i) [\rm{ergs~s^{-1}~Hz^{-1}}]$) is :
\begin{equation}\label{eq:IMF_averaged_spectra}
  L_{\mathrm{  \nu, IMF}}(\nu, t_i) = \dfrac{\displaystyle\int_{M_{\mathrm{min}}}^{M_{\mathrm{max}}} \overline{L}_{\nu}(t_i, m) ~ \Phi(m) \text{d}m}{\displaystyle\int_{M_{\mathrm{min}}}^{M_{\mathrm{max}}} m ~ \Phi(m) ~ \text{d}m} ~.
\end{equation}
For simplicity, we refer to this IMF average luminosity as $ L_\nu $, as we do not work with the individual stellar luminosity in the rest of this work).

\begin{figure}
\centering
\includegraphics[width=\linewidth]{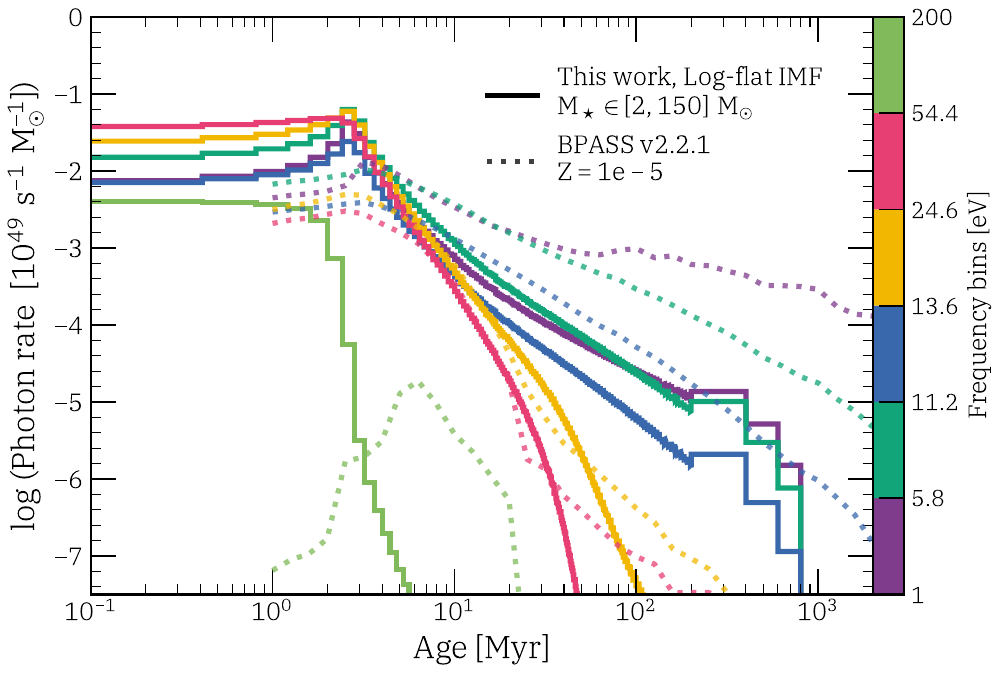}
\caption{Photon injection rate in different frequency bins as a function of time for a \pt stellar population with the fiducial Log-Flat IMF (solid lines) compared with BPASS (v2.2.1) spectral models at their lowest metallicity ($Z=10^{-5}$, dotted lines). The Log-Flat IMF population injects more photons at $t < 4~\rm{Myr}$ than the BPASS models, as the initial radiation budget is dominated by very massive \pt stars. At $t > 4~\rm{Myr}$, the photon injection rate of the Log-Flat IMF drops rapidly below that of the BPASS model as massive stars die and radiation becomes dominated by long-lived low-mass stars.}
\label{fig:Photon_injection_rate}
\end{figure}

Fig.~\ref{fig:IMF_spec} shows the IMF-averaged luminosity per unit solar mass of \pt stars formed as a function of SSP age, assuming the fiducial log-flat IMF with minimum and maximum stellar masses of $2.0$ and $150~\M$, respectively.\footnote{A more detailed comparison of the emergent \pt spectra for different IMF assumptions is outlined in Appendix \ref{sec:Spectra_Comparison}.} The majority of radiation from \pt stars is released during the first few Myr ($ t \lessapprox 4~{\rm{Myr}} $) and becomes negligible after approximately $2$ Gyrs of evolution. Fig.~\ref{fig:Photon_injection_rate} plots the photon injection rate ($Q$, $[{\rm{s^{-1}~\M^{-1}}}]$) in the radiation bands listed in Tab.~\ref{tab:Initialized_RT_values}. The solid curves show the results for the fiducial Log-Flat IMF, whereas the dotted lines show the energy injection if one assumes a BPASS (v2.2.1) SPS model for low-metallicity ($Z=10^{-5}$) Pop~II/I stars with a Chabrier IMF and $M_{\rm{min}}$ and $M_{\rm{max}}$ set to $0.1$ and $100$ M$_\odot$ respectively. The Log-Flat IMF population injects more photons at $ t < 4~\rm{Myr} $ than the BPASS models, as the radiation budget is initially dominated by very massive \pt stars. After $ t > 4~\rm{Myr} $, the photon injection rate of the stellar population with a Log-Flat IMF drops significantly below that of the BPASS model, as the most massive stars die and the radiation budget is dominated by long-lived low-mass stars that provide much less radiation than the more massive stars. This highlights the importance of properly modelling the radiation feedback from \pt stars in simulations of the early universe, as they can produce a significant number of high-energy photons in a short time, which can impact the ionization and thermal history of the gas surrounding them.

\subsection{The end stages of \pt stars}\label{sec:pt_end-state}
Devoid of heavy elements in their outer envelopes, \pt stars remain stable against opacity-induced pulsations that affect metal-enriched stars \citep[see, e.g.,][]{Heger_2010}. Their hydrogen-burning shells are comparably weak, contributing to their tendency towards compact, blue-star morphology throughout their evolution. The end state of a \pt star depends on the mass of the He core at the end of the star's life. It can die as a white dwarf, neutron star, or black hole after exploding as a supernova, or it can directly collapse into a black hole without undergoing an explosion. The mass ranges for different end states remain uncertain. For non-rotating \pt stars with negligible mass loss, the helium core mass ($M_{\rm{He}}$) scales linearly with the initial mass as $ M_{\rm{He}} = (13 / 24) \times (M_{\star}  - 20) $ \citep{Heger_2002}. Thus, it is common to discuss end states in terms of the star's initial mass.

For stars with masses $ 10 - 25~\M $, the core collapse of the central degenerate core triggers a type II supernova explosion (hereafter referred to as a core-collapse supernova or CCSN) that leaves behind a neutron star remnant. For stars with $ 25 - 40~\M $, the core collapse is unable to completely unbind the stellar envelope, and fallback of the material is expected to collapse it into a black hole \citep{Klessen_2023, Heger_2010}. Between $ 40 - 70~\M $, the stars also directly collapse into a black hole, but in the process they eject a significant amount of energy and metals into the surrounding medium \citep{Heger_2010}. For stars with masses of about $ 70-100~\M $ to $ 260~\M $, pair instabilities occur, and they die as supernovae that may or may not leave behind a compact remnant \citep{Heger_2002, Chen_2014, Woosley_2017}. These pair instabilities occur during the post-carbon-burning stage. The core of the star reaches a high-temperature, low-density regime that allows the production of electron-positron pairs. Pair production reduces pressure support and effectively lowers the adiabatic index $ (\gamma) $ in the radiation-dominated plasma. This results in rapid contraction of the core, with free-fall acceleration in the inner part of the C/O core. The increase in temperature following the collapse triggers rapid thermonuclear burning that can fuse heavier elements, and the energy released in this process is comparable to or higher than the binding energy of the star. The contraction to higher temperature occurs with little resistance compared with what the star would otherwise encounter, making the star unstable. This instability results in an implosion of the core, which is reversed by nuclear burning. If enough burning occurs, the star may be completely disrupted in a single pulse -- this only happens for initial mass $ > 140~\M $ and is called a pair-instability supernova (PISN), which is one of the most energetic events in the universe. Otherwise, the core expands for a while, kicks off the outer layers, including any residual hydrogen layers, and then collapses until it encounters the instability again -- this is called a pulsational pair-instability supernova (PPISN). For stars more massive than $ 260~\M $, the star directly collapses into a black hole without any supernova explosion.

\subsubsection{Supernova feedback from \pt stars}

\begin{figure}
\centering
\includegraphics[width=0.99\linewidth]{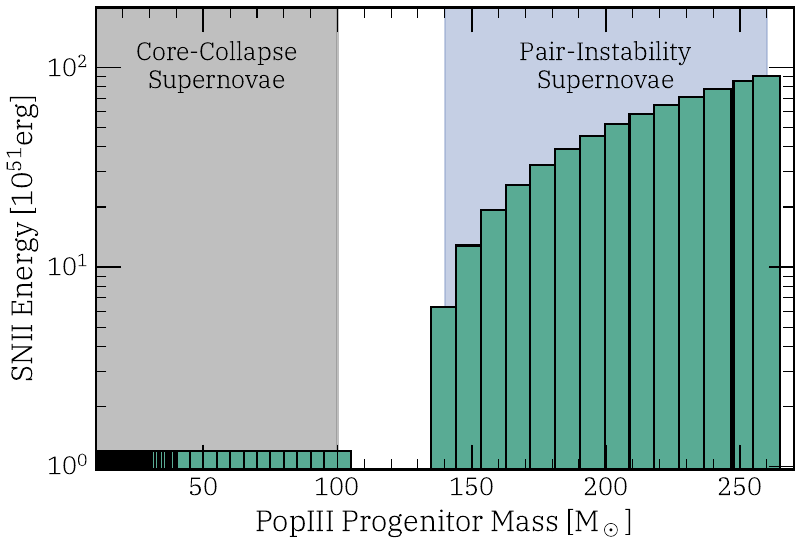}
\caption{SN explosion energy as a function of stellar mass for \pt stars in the mass range $ 10 - 260 ~\M $. Highlighted are the area where the stars explode as core-collapse SN and Pair-Instability SN.}
\label{fig:PopIII_SN_energy_vs_mass}
\end{figure}

\pt supernovae inject large amounts of energy, momentum, and newly synthesized metals into their surroundings. Several theoretical studies have modelled the evolution of \pt stars and provided detailed estimates of elemental enrichment during supernova explosions \citep[for e.g.][]{Nomoto_2006,Heger_2002,Heger_2010}. In this work, we adopt the elemental yields from \citet{Heger_2002} for SN with initial mass between $ 10-100~\M $ and from \citet{Heger_2010} for SN with initial mass between $ 140-260~\M $. We refer the reader to Appendix~\ref{sec:yields} for detailed element-by-element yield estimates used in this work from \pt supernovae.
\citet{Heger_2010} explored a wide parameter space for explosion energy, mixing, and fallback. By comparing their models with the chemical abundances of extremely metal-poor stars, they determined that an explosion energy of $1.2 \times 10^{51}$\,erg and a mixing length parameter equivalent to $10\%$ of the helium core mass best reproduced the observations for CCSN ($10-100~\rm{M}_\odot$). We adopt these fiducial values for the elemental yields and energy injection in our model. For PISNs, we adopt the mass-dependent energy injection from \citet{Heger_2002}:
\begin{equation}\label{eq:PISN-energy}
E_{\rm{PISN}} =  10^{51} \times \left[ 5.0 + 1.304\left( \frac{M_{\rm{He}}}{\M} - 64 \right)  \right]~\rm{erg} ~.
\end{equation}
We apply this for stars in the $ 140 - 260~\M $ range. For the missing range between $ 100-140~\M $, we don't inject any energy or metals, as the star is expected to directly collapse into a BH. Fig.~\ref{fig:PopIII_SN_energy_vs_mass} shows the SN explosion energy as a function of stellar mass for \pt stars in the mass range $ 10 - 260 ~\M $. The regions where stars die as CCSN and PISN are highlighted.

Modelling the impact of injected SN energy is non-trivial, particularly in cosmological simulations with limited resolution \citep[for e.g. see discussions in][]{Scannapieco_2012}. The primary challenge is to properly capture the {\it Sedov-Taylor phase} of the SN remnant -- an early expansion phase that conserves energy. During this phase, a highly pressurized central gas bubble expands outward, engulfing and accelerating surrounding ISM gas. This expansion significantly amplifies the radial momentum of the SN shell, potentially reaching 10 times the initial value \citep{Martizzi_2015}. This substantial momentum boost plays a crucial role in regulating star formation \citep{Ostriker_2011} and propelling galactic winds \citep{Kim_2018}. Hence, an effective SN implementation that accurately captures both the total momentum and energy injection is essential for realistic galaxy formation simulations, particularly when the Sedov-Taylor phase is not resolved. For the \pt stars, we follow the same prescription as \citet{Marinacci_2019}. The only caveat is that for \pt stars, we assume no mass loss during the MS evolution (as \pt stars do not have metal-line driven winds), and hence chemical enrichment occurs only after a star dies.

\begin{table}
\centering
\renewcommand{\arraystretch}{1.5}
\begin{tabular}{|m{1.5cm}| m{1.1cm}m{1.1cm}m{1.1cm}m{1.1cm}|}
  \hline
  & & \pt Physics & & \\
  Name &
  Tracked & Spectrum & Feedback &
  Thermo-chemistry \\
  \hline\hline
  {\tt Thesan-Zoom} 
  & Yes & No & No & \cite{Kannan_2025} \\
  {\tt TZ + Th. Chem} 
  & Yes & No & No & $ + $ \S\ref{sec:thermochem} \\
  {\tt Pop3 (fiducial)} 
  & log-flat IMF & Yes & Yes & $ + $ \S\ref{sec:thermochem} \\
  {\tt Pop3 M250} 
  & log-flat IMF & Yes & Yes & $ + $ \S\ref{sec:thermochem} \\
  {\tt Pop3 Salpeter} 
  & Salpeter IMF & Yes & Yes & $ + $ \S\ref{sec:thermochem} \\
  \hline
\end{tabular}
\caption{Simulation suite used to test the \pt physics modules. Left to right the columns indicate: Name of the simulation, whether \pt stars are tracked, whether \pt spectrum is included, if feedback from \pt stars are enabled or not, and the thermochemistry network used.}
\label{tab:sims}
\end{table}

\section{Cosmological Simulations including \pt stars}\label{sec:comso-sim}

\begin{figure*}
\centering
\begin{subfigure}[t]{0.49\textwidth}
  \includegraphics[width=\linewidth]{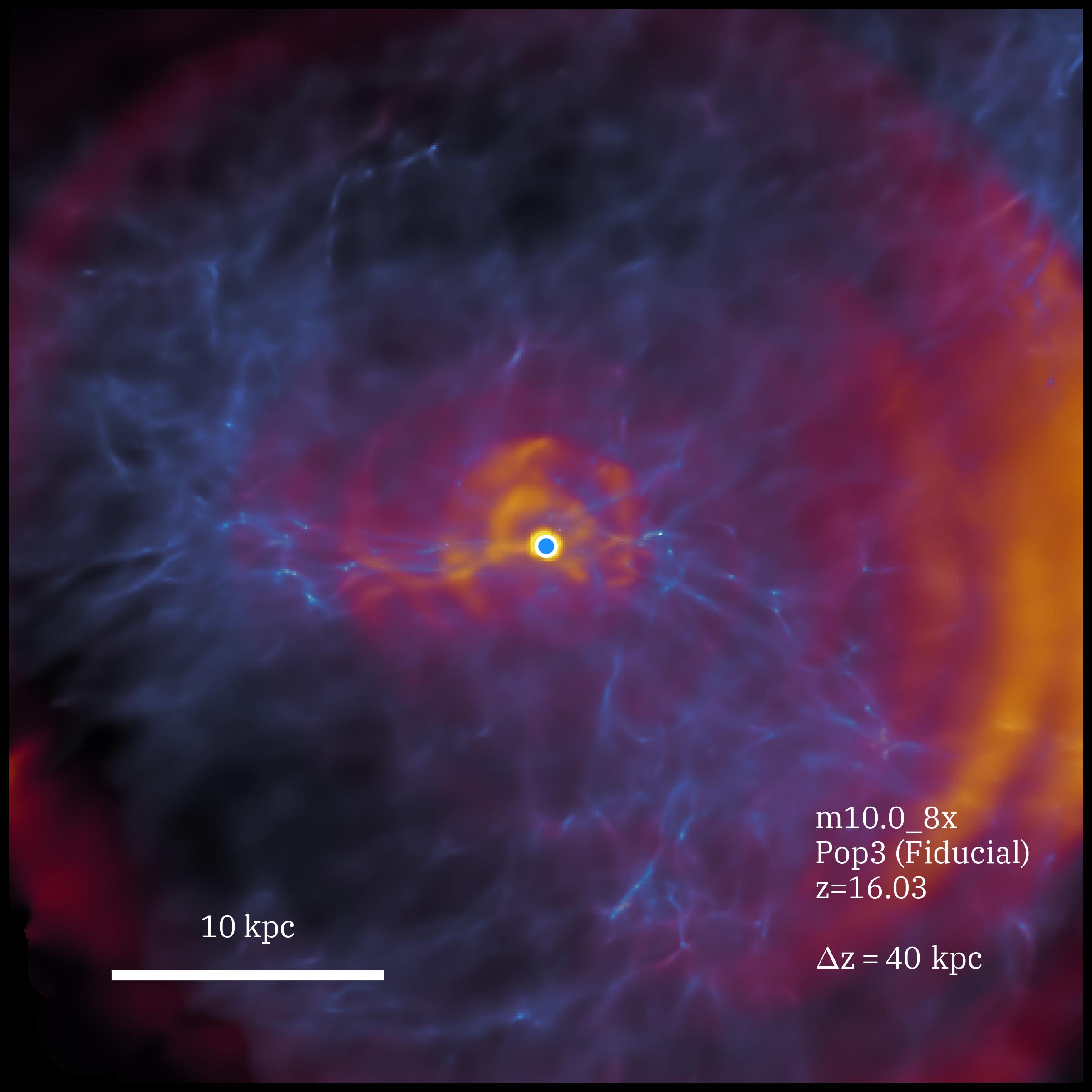}
\end{subfigure}%
\hspace{0pt}%
\begin{subfigure}[t]{0.49\textwidth}
  \includegraphics[width=\linewidth]{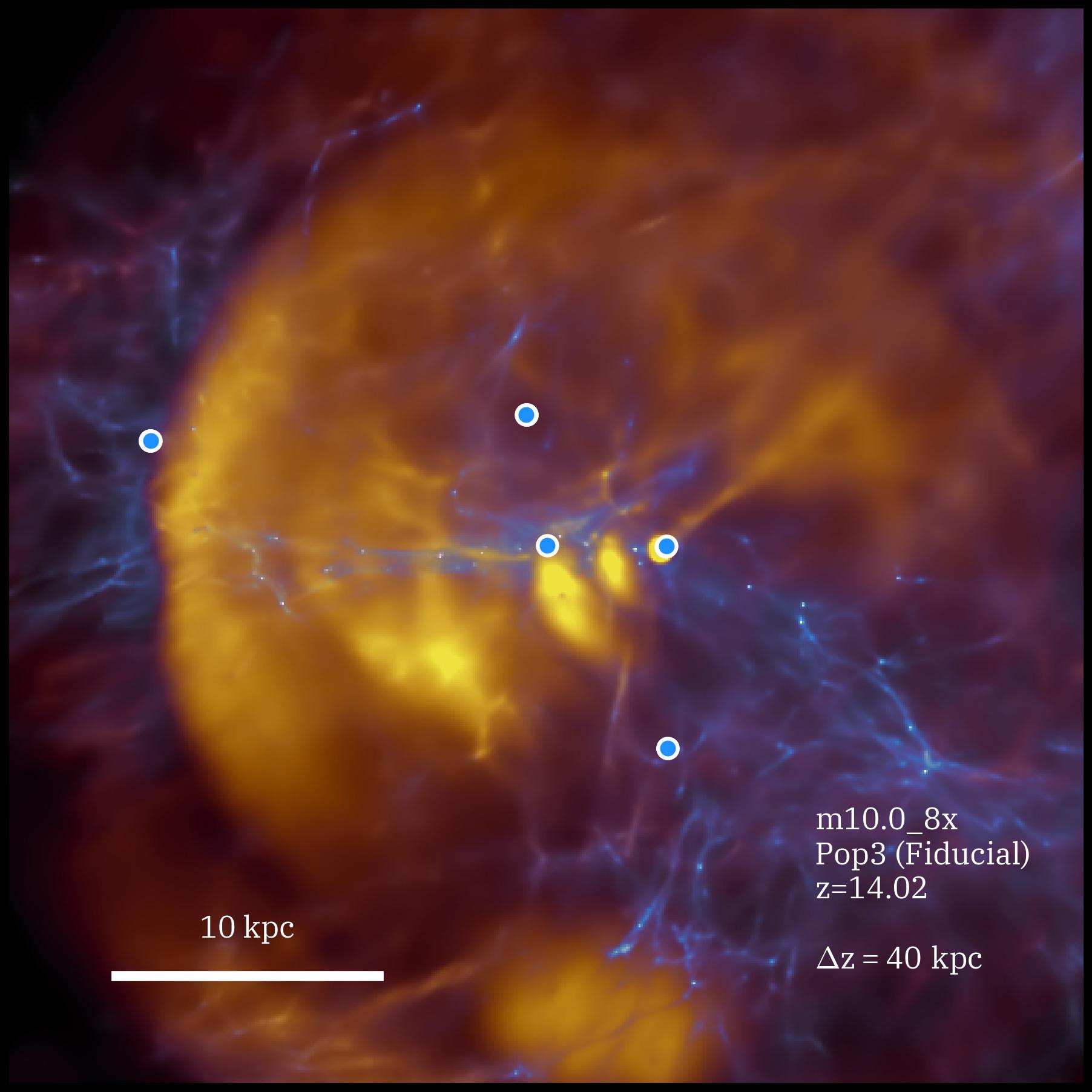}
\end{subfigure}
\begin{subfigure}[t]{0.49\textwidth}
  \includegraphics[width=\linewidth]{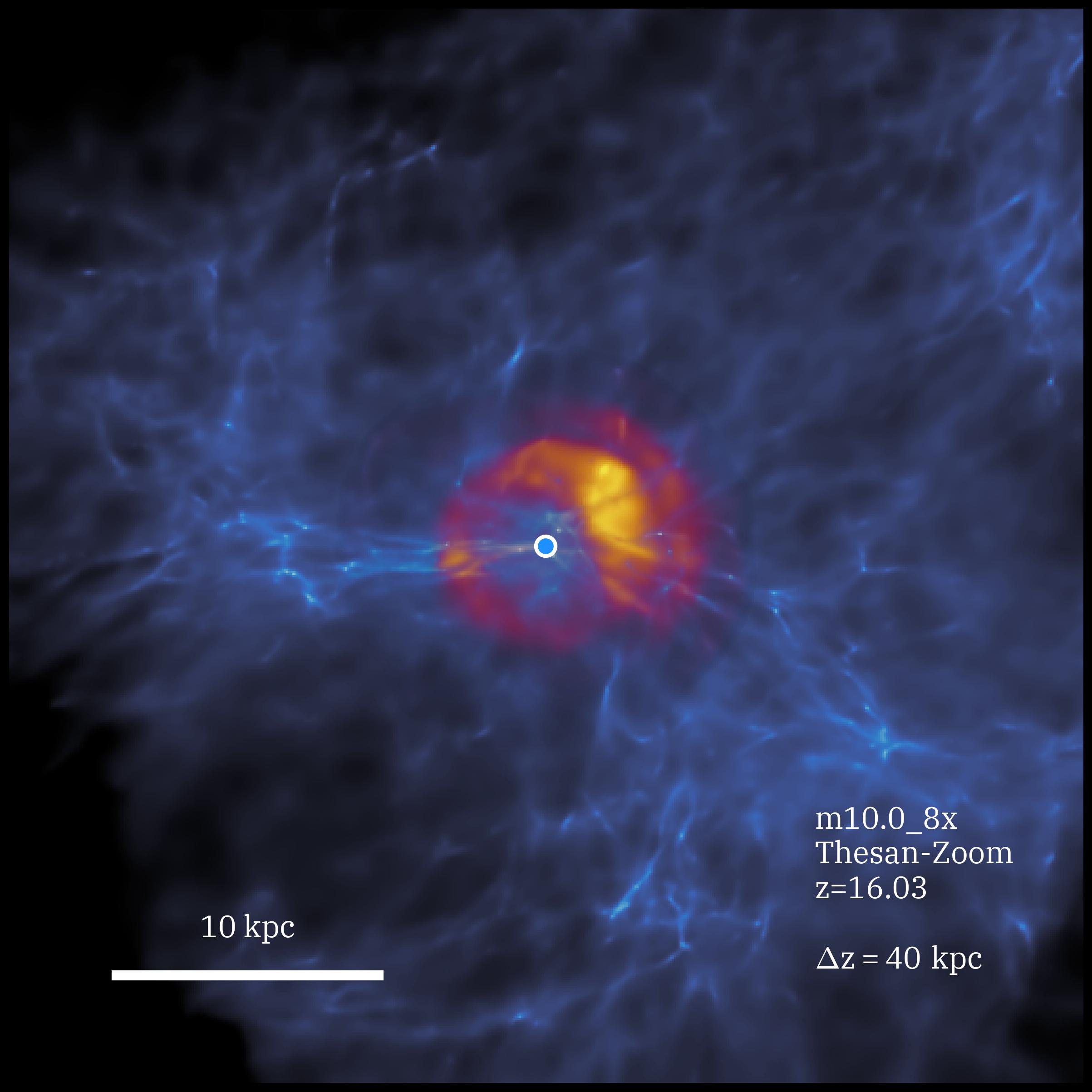}
\end{subfigure}%
\hspace{0pt}%
\begin{subfigure}[t]{0.49\textwidth}
  \includegraphics[width=\linewidth]{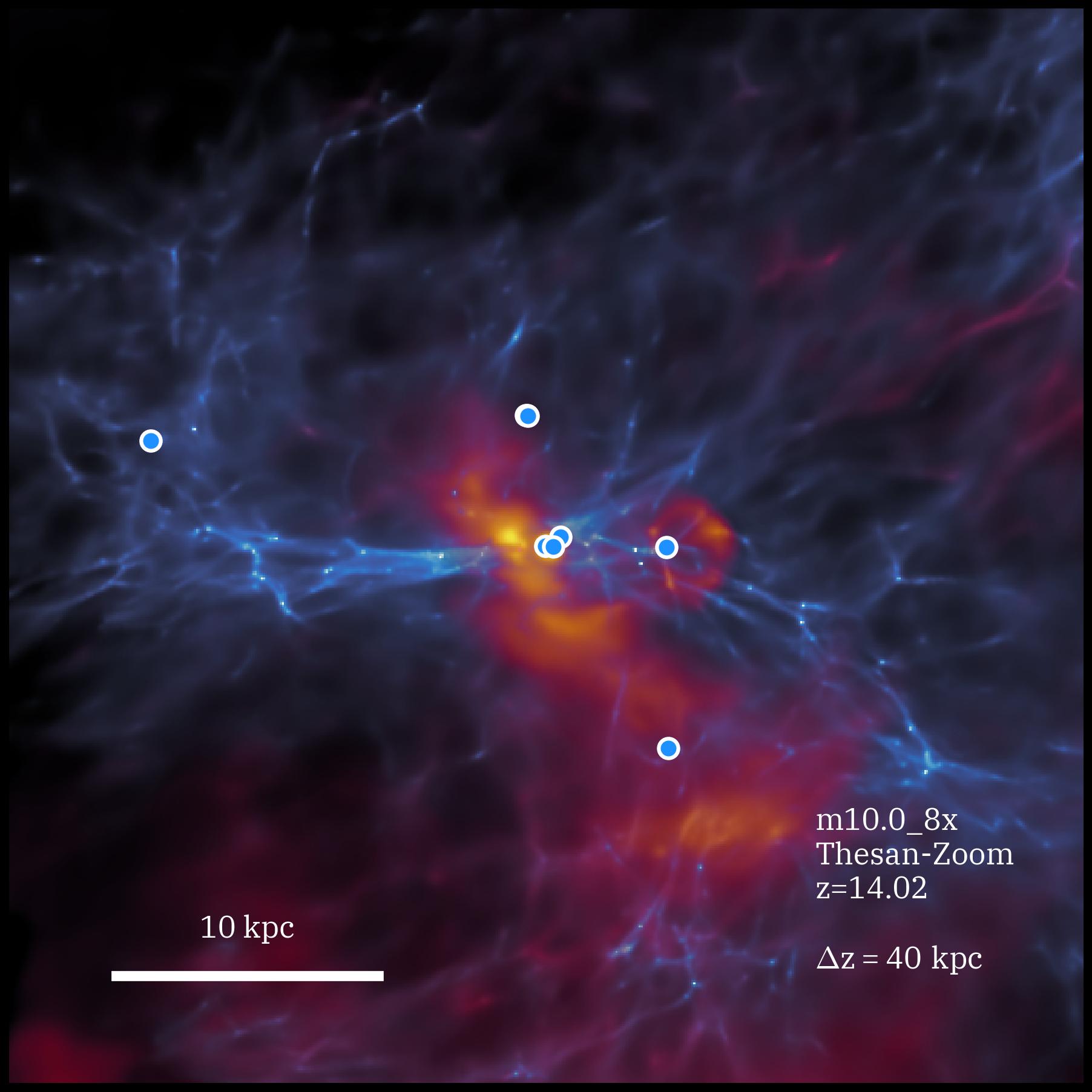}
\end{subfigure}
\caption{
    Mass weighted mean Lyman-Werner flux $(J_{21})$ blended with the underlying gas surface density across two different model variations {\tt Pop3 (fiducial)} (top panels) and {\tt Thesan-Zoom} (bottom panels). Left and right columns show two different snapshots at $z \approx 16$ (the time of first star formation) and  $z\approx14$. Blue markers indicate the location of stars. The images are centred on the minimum of the gravitational potential of the subhalo identified as the target galaxy of the zoom-in region. All images are normalized to a fixed range of $\log(J_{21}~/~[\rm erg~s^{-1}~cm^{-2}~Hz^{-1}~sr^{-1}]) \in [-1, 1.7]$ and  $\log(\Sigma_{\rm gas}~/{\rm \M~kpc^{-2}}) \in [5.7, 7.4]$. We clearly see that at $z \approx 16$ the {\tt Pop3 (fiducial)} run has 2 orders of magnitude stronger LW flux, compared to the {\tt Thesan-Zoom} run.
}
\label{fig:ResultsDemo}
\end{figure*}

Finally we test the new \pt star formation, cooling, metal enrichment, and feedback modules in a cosmological setup. We select the target halo {\tt m10.0\_8x}, from the parent {\tt Thesan-Zoom} suite, which is then re-simulated with the new prescriptions for \pt stars - \S\ref{sec:thermochem}, \ref{sec:\pt_spectra}, \ref{sec:pt_end-state}. The selected halo has a group mass of $ 1.07 \times 10^{10} $ at $ z=3.0 $. The halo is simulated at the $ 8\times $ zoom level, corresponding to a baryonic mass resolution of $ 1.14 \times 10^3~\M $ and a dark matter mass resolution of $ 6.09 \times 10^3~\M $, respectively. The gravitational softening length for DM and stars is set to $276.79~\rm{cpc}$ with the minimum comoving gas softening set to $34.60~\rm{cpc}$. We run five model variations of the same halo at the given resolution level -- fiducial {\tt Thesan-Zoom}, one with only the new thermochemistry network (i.e. \S\ref{sec:thermochem}), and three with \pt radiation and feedback but with different IMFs for the \pt stars (i.e., two use a log-flat IMF and the other uses a Salpeter IMF). Table \ref{tab:sims} lists the properties of the physics variations.

We differentiate between stellar populations solely based on the metallicity of the star particles relative to the solar metallicity $ Z_\odot =0.0127 $ \citep{Wiersma_2009}. \pt stars have metallicity $ Z_{\star} < 10^{-4} Z_\odot $, and non-\pt stars have metallicity $ Z_{\star} \geq 10^{-4} Z_\odot $. This choice is motivated by previous works such as \citet{Jaacks_2019}, \citet{Liu_2020}, and \citet{Venditti_2023}.

For each simulation, we produce $ 189 $ Snapshot files that contain all the information about the gas, dark matter and star particles in the simulations, with a time cadence of $ \Delta t \sim  10~\rm{Myr} $, from $z=16$ to $ z = 3 $. The halo catalogues are generated using the friends-of-friends (FoF) algorithm \citep{Davis_1985}, with the {\tt SUBFIND-HBT} algorithm \citep{Springel_2001, Springel_2021} used to identify the subhaloes.  We define the most massive subhalo as the central galaxy, while are other subhaloes are defined as the satellite galaxies.

\subsection{Results}\label{sec:results}
Before diving into the quantitative assembly and star formation history of our central galaxy, we first provide a visual demonstration of the impact of \pt stars. Fig.~\ref{fig:ResultsDemo} presents a visual overview of the early Lyman-Werner (LW) radiation field. Because our simulation self-consistently tracks the spatial and temporal propagation of this flux, we capture the highly inhomogeneous nature of early stellar feedback, rather than relying on a uniform background. The figure contrasts the mass-weighted mean $J_{21}$ flux in our {\tt Pop3 (fiducial)} model with the baseline {\tt Thesan-Zoom} run at two epochs: the onset of the first star formation ($z \approx 16$) and a slightly later stage of early assembly ($z \approx 14$). The LW flux is overlaid on the gas surface density maps\footnote{We would like to remind the readers that because the two maps are blended together a usual notion of colour bar does not apply here. One should only compare the final intensity of the image. For the blending step, all the images are normalized to a fixed range: $\log(J_{21}~/~[\rm erg~s^{-1}~cm^{-2}~Hz^{-1}~sr^{-1}]) \in [-1, 1.7]$, $\log(\Sigma_{\rm gas}~/{\rm \M~kpc^{-2}}) \in [5.7, 7.4]$}.

\begin{figure}
\centering
\includegraphics[width=1\linewidth]{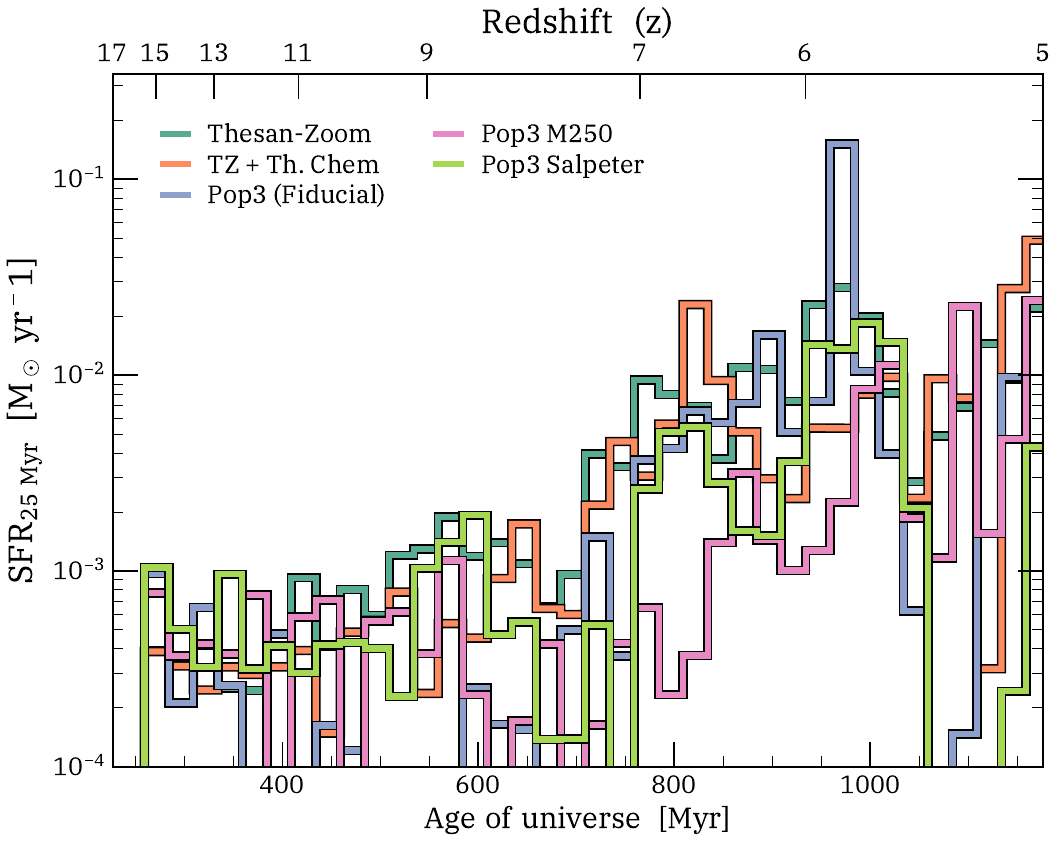}
\caption{Star formation rate (SFR) as a function of redshift for different simulation runs, computed using star particles that end up in the primary zoom-in halo at $z \sim 5$. While runs show qualitatively similar SFRs with a scatter of $\sim 0.5$~dex at low redshift, \pt models show consistently lower SFRs at high-z..
}
\label{fig:SFH}
\end{figure}

The consequences of explicitly modelling \pt stars are evident at $z \approx 16$, where the birth of the first stars produces intense, highly localized LW radiation bubbles roughly two orders of magnitude stronger than in the non-\pt baseline model. By $z \approx 14$, these early radiation bubbles have expanded and overlapped, creating a pervasive, highly structured LW field throughout the region. As quantitatively corroborated by the radial flux profiles in Appendix Fig.~\ref{fig:J21_shells}, this enhanced local LW background stifles early cooling in nearby minihaloes by efficiently photodissociating molecular hydrogen. In the following sections, we will quantify these differences and examine the broader effects of early \pt feedback on the formation and evolution of the central galaxy in more detail.

\subsection{Star Formation History and \pt stars}
Fig.~\ref{fig:SFH} shows the star formation rate (SFR) as a function of redshift for the different simulations. The SFR is computed by selecting all star particles that end up in the primary zoom-in halo at $ z\sim 5$ and constructing a mass-weighted histogram of their formation times, with $\delta t = 25$ Myr. The stochastic nature of the underlying sub-grid model for star formation and feedback precludes exact one-to-one comparisons; however, qualitatively, all the runs exhibit a similar SFR, with a scatter of about $ \sim 0.5~\rm{dex} $ between the runs. The {\tt Thesan-Zoom} and {\tt TZ~+~Th.~Chem} runs are strikingly similar, suggesting no major effect of the additional $ \hp $ cooling on star formation. However, runs with \pt included predict a lower average SFR than the other runs at $ z > 5.5 $, with the {\tt Pop3 M250} run predicting the lowest SFR. The stronger feedback in the {\tt Pop3*} runs better regulates star formation. Correspondingly, the {\tt Pop3 M250} run includes more massive \pt stars $(M_{\max}=250~\M)$ and hence suppresses star formation more due to PISN at the high-mass end.

\begin{figure}
\centering
\includegraphics[width=1\linewidth]{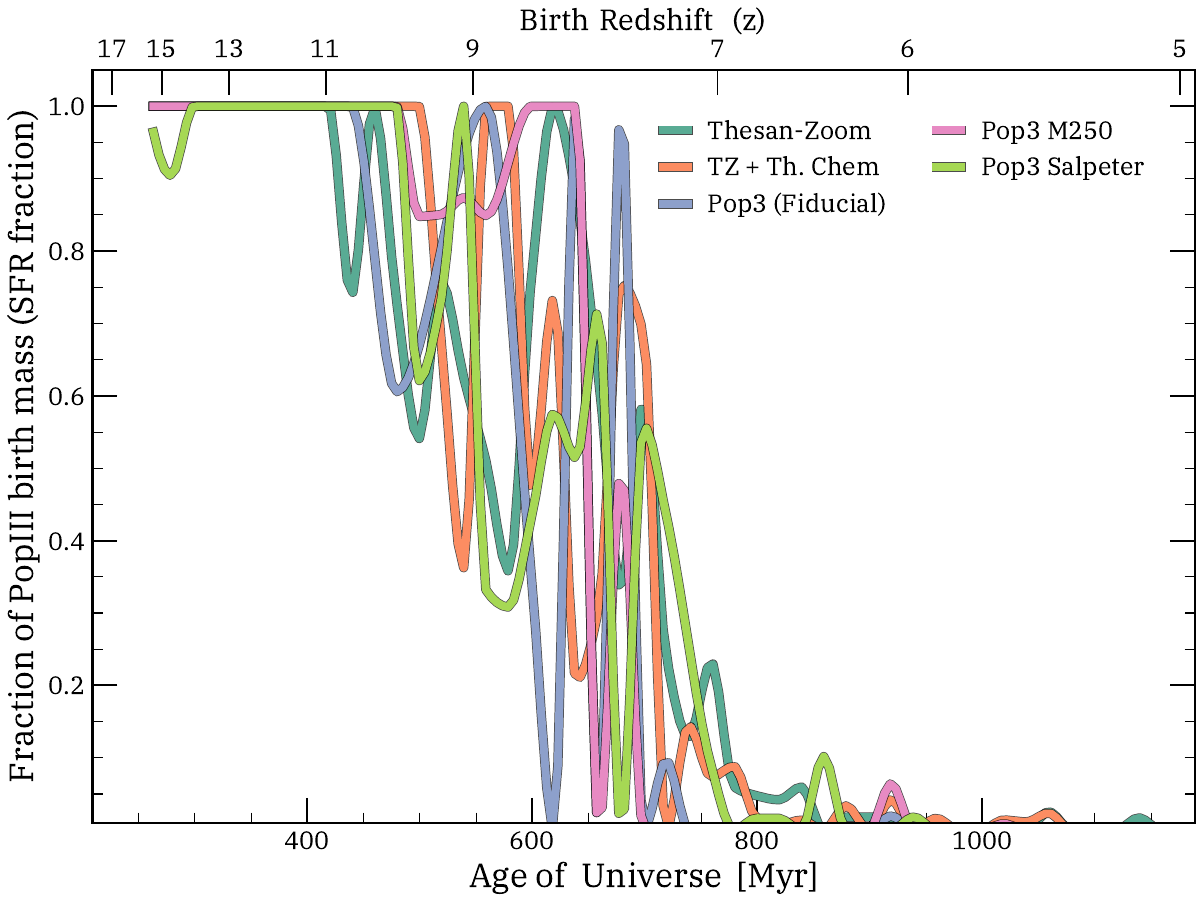}
\caption{
    Mass fraction of stars forming as \pt at different times, which end up in the central group for different variations.  By $z\sim6$, all the stars forming are non-\pt.
}
\label{fig:Pop3_fraction}
\end{figure}

Fig.~\ref{fig:Pop3_fraction} shows the mass fraction of \pt stars forming at any given time (i.e. SFR fraction of \pt stars) in the various simulations, for stars that end up in our target halos at $z=5$. The simulations predict that the first stars form around $ z \sim 16 $ in all the runs, and by $ z \sim 6 $, less than $ 1 \% $ of all stars forming are \pt stars, regardless of the model variations. While the different simulations predict different times for the drop in the fraction of \pt stars, the stochastic nature of the sub-grid SF model does not allow for a direct comparison of the point at which the fraction of \pt stars drops considerably.
Across the variations, we observe episodes of rejuvenation of \pt stars -- i.e., the fraction of \pt stars increases again (sometimes to even SFR fraction $\thickapprox 1$) after dropping. This is because, within the high-resolution region, there are pockets of pristine gas, and if they cool down enough, they can form \pt stars even at later times (see also Fig.~\ref{fig:metallicity_shells})

\begin{figure*}
\centering
\includegraphics[width=\textwidth]{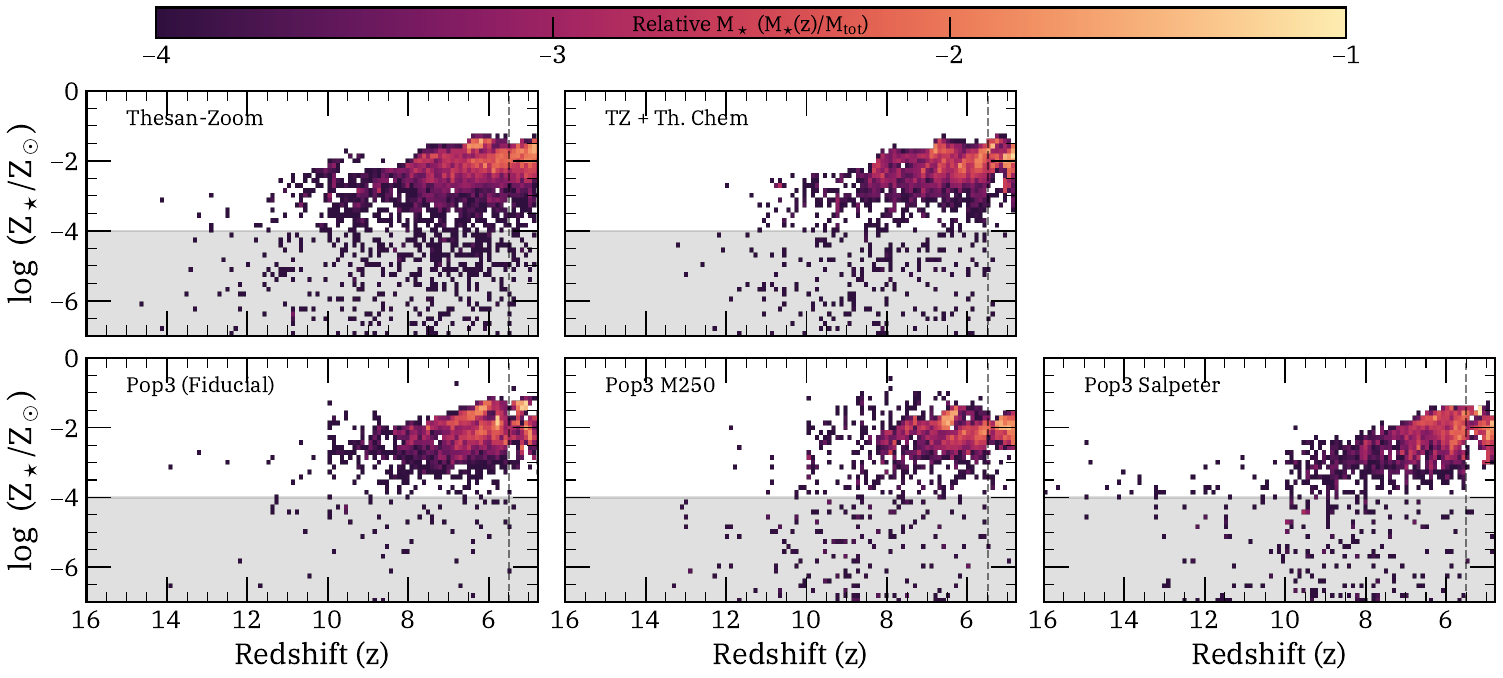}
\caption{
  Birth redshift and metallicity distributions of stars ending up in the primary zoom-in region at $z\sim 5$ for different simulation. The shaded gray region marks the \pt metallicity threshold ($Z_{\star} < 10^{-4}Z_\odot$); vertical dashed lines indicate the end of reionization. In {\tt Thesan-Zoom} and {\tt TZ + Th. Chem} runs, there are more stars forming below the metallicity threshold compared to the {\tt Pop3*} runs. {\tt Pop3*} runs drive bursty, episodic metal enrichment, sharply curtailing continuous pristine star formation compared to the {\it smoother} evolution in the baseline models.
}
\label{fig:Stars_Z_vs_redshift}
\end{figure*}

Fig.~\ref{fig:Stars_Z_vs_redshift} shows the 2D histogram of birth redshift and stellar metallicity for all stars that end up in the primary halo at $z=5.0$. The shaded gray region is the very low-metallicity \pt regime, while the dashed vertical lines mark the end of reionization. All runs reach similar stellar metallicities by $z=5$; however, differences in the SFRs lead to different metallicity evolution histories. The main difference between the simulations is that star formation in the {\tt Thesan-Zoom} and {\tt TZ + Th. Chem} runs is smoother than in the {\tt Pop3*} runs. In the {\tt Pop3*} runs, the number of stars forming below the Pop3 metallicity threshold $(Z = 10^{-4}Z\odot)$ is much lower than in the {\tt Thesan-Zoom, TZ + Th. Chem} runs, a direct consequence of stronger metal enrichment from the \pt stars (see Sec.~\ref{subsec:effects_Pop3_feedback}, Appendix.~\ref{sec:chemical_feedback} for a detailed discussion of \pt feedback and metal enrichment due to \pt stars).

\subsection{Evolution of gas phase and metallicity trends}
\begin{figure*}
\centering
\includegraphics[width=1\textwidth]{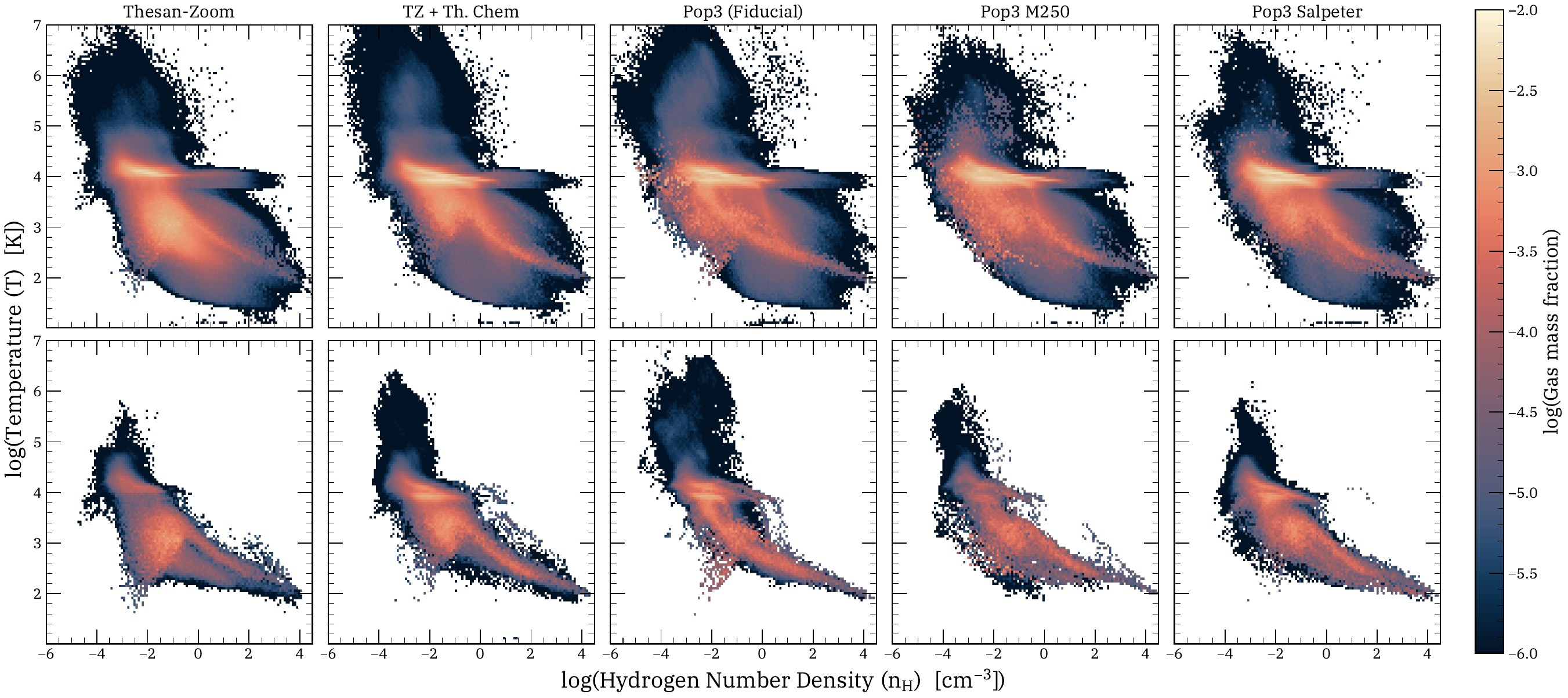}
\caption{
  Mass-weighted phase-space diagrams for all high-resolution gas (top row) and pristine gas ($Z_{\rm{gas}} < 10^{-4}Z_\odot$; bottom row) in the central halo, averaged over the simulation runtime. Different columns correspond to different model variations: {\tt Thesan-Zoom}, {\tt TZ + Th. Chem}, {\tt Pop3 (fiducial)}, and {\tt Pop3 Salpeter} (from left to right). Simulations with the updated thermochemistry network ({\tt TZ + Th. Chem} and {\tt Pop3*} runs) show a higher fraction of cooler, denser gas ($T < 100~\rm{K}, n_{\rm{H}} > 10^{3}~\pcc$) relative to the fiducial run, demonstrating the effect of additional $\hp$ cooling. The {\tt Pop3*} runs also exhibit a slightly higher fraction of photo-heated diffuse gas ($T \sim 10^4~\rm{K}, n_{\rm{H}} < 10^{-1}~\pcc$) due to  stronger ionizing radiation from \pt stars.
}
\label{fig:PhaseSpace}
\end{figure*}

Fig.~\ref{fig:PhaseSpace} shows temperature-density phase-space histograms for all {\tt HighRes} gas across the different runs as indicated. The top row shows the phase-space diagram for all gas in the high-resolution region, whereas the bottom row corresponds to pristine gas (i.e., $ Z_{\rm{gas}} < 10^{-4}Z_\odot $). The diagrams are weighted by gas-cell mass and averaged over the duration of the simulations. The simulations predict a slightly higher fraction of cooler, denser gas (i.e., $ T < 100~\rm{K}, n_{\rm{H}} > 10^{3} \pcc $) in runs with the updated thermochemistry network (i.e., {\tt TZ + Th. Chem, Pop3*} runs) compared to the fiducial {\tt Thesan-Zoom} run. This is due to additional cooling from $ \hp $, which allows gas to cool more efficiently, and is more evident in the pristine gas phase diagrams (i.e., the bottom rows). This also leads to more $\rm H_2$-rich gas in the galaxy, especially at high ($z\gtrsim14$) redshifts (See Appendix~\ref{app:h2} and the figures within for a more detailed discussion of the impact of the new thermochemistry network).
{\tt Pop3*} runs also show a higher fraction of gas in the photo-heated regime (i.e., $ T \sim 10^4~\rm{K}, n_{\rm{H}} < 10^{-1} \pcc $), which can be attributed to stronger radiative feedback from the \pt stars. Across all runs, a large fraction of pristine gas ends up in the low-density, high-temperature regime (i.e., $ T \sim 10^4~\rm{K}, n_{\rm{H}} < 10^{-1} \pcc $) due to photo-heating from the stars and SN feedback.
The violent disruption of pristine gas is inextricably linked to the chemical enrichment of the ISM. The same \pt supernovae that drive the gas into the diffuse, photo-heated phase simultaneously seed the surrounding medium with the first heavy elements. Consequently, the distinct thermodynamic evolution seen in the Pop3 variations directly dictates the chemical abundance patterns locked into the next generation of stars, which will be explored in future works.
For now, we refer the reader to \citet{Zier_2025} for additional discussion of the origin and evolution of metal-free gas in {\tt Thesan-Zoom} galaxies.

\begin{figure*}
\centering
\begin{subfigure}[t]{0.47\textwidth}
  \includegraphics[width=\linewidth]{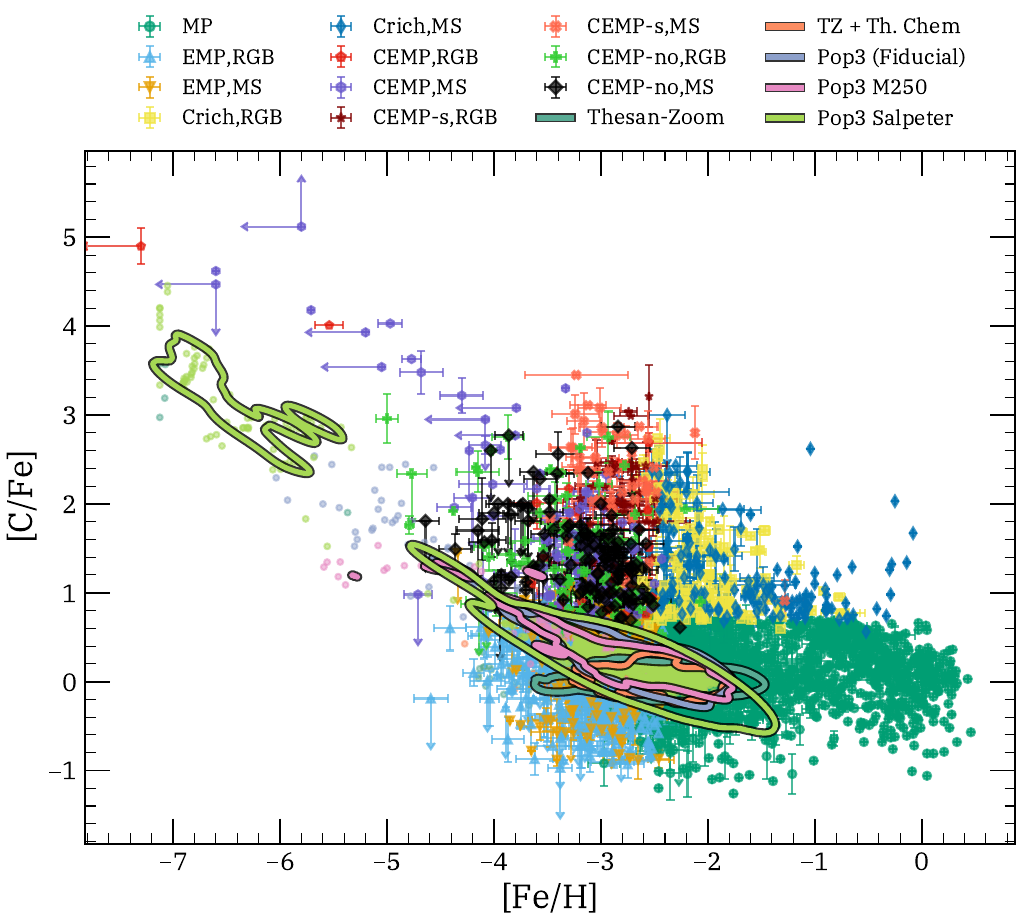}
\end{subfigure}%
\hspace{20pt}%
\begin{subfigure}[t]{0.47\textwidth}
  \includegraphics[width=\linewidth]{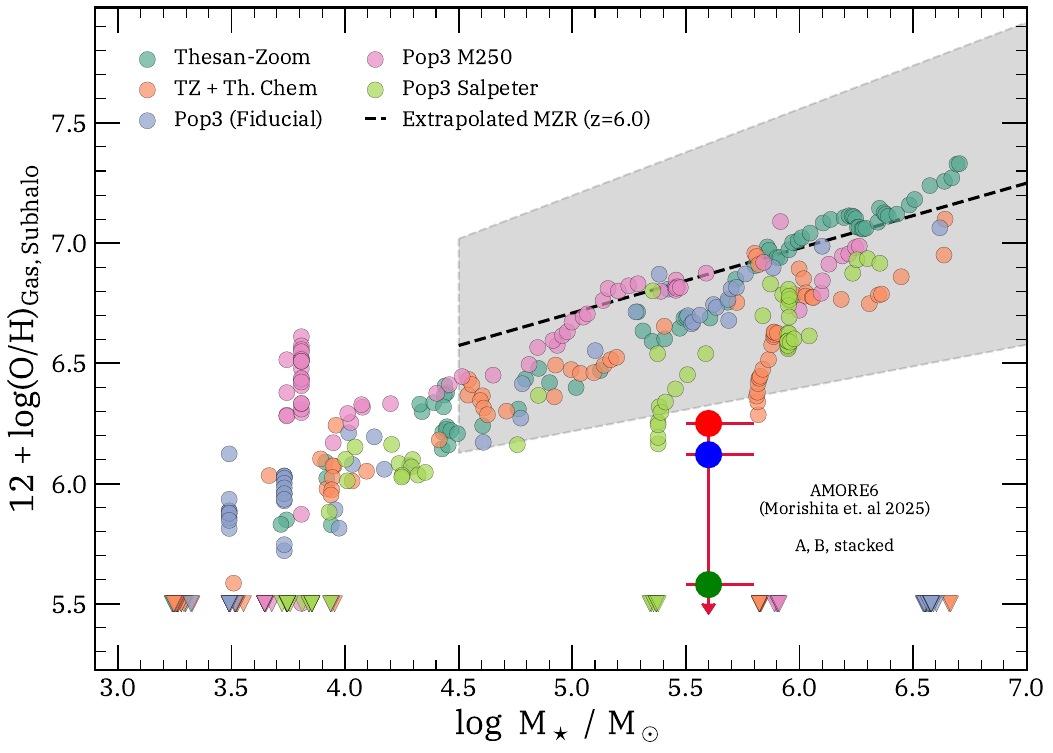}
\end{subfigure}

\caption{
  {\bf Left:} Carbon and Iron abundances of second-generation stars (i.e., non-\pt stars) across the model variations. Contours show the $2.5-97.5^{th}$ confidence interval. The scatter points with error bars are observations of various populations of metal-poor stars in our own Milky Way galaxy, taken from the SAGA database (indicated in the legends). Only the {\tt Pop3*} runs predict an extended distribution that is qualitatively similar to CEMP stars, whereas the {\tt Thesan-Zoom, TZ + Th. Chem} runs predict a very confined distribution of Carbon and Iron.
  \break
  {\bf Right:} Oxygen abundances measured from gas within $ 2r_{\star} $ of the central galaxy as a function of stellar mass for different model variations. The black dashed line with the shaded region shows the extrapolated mass-metallicity-relation from observations of galaxies at $z=6$ \citep{Morishita_2024}. The red, blue and green scatter points with the error-bars plots the observed oxygen abundance from the {\tt AMORE 6} system $(z\approx5.72)$, which is believed to host pristine star formation. All the model variations predict a similar MZR with a scatter $<0.5$ dex between the variations.
}
\label{fig:MetallicityTrends}
\end{figure*}

Fig.~\ref{fig:MetallicityTrends} plots metallicity trends in the primary zoom-in halo across different model variations. We start with the left panel, which shows the carbon-to-iron abundance ([C/Fe]) as a function of iron abundance (relative to solar; [Fe/H]) for non-\pt stars. Coloured markers indicate the various model variations, as shown, and the corresponding contours mark the $2.5-97.5^{th}$ confidence interval. The scatter points with error bars are the observed abundances of metal-poor stars (across different sub-populations) within the Milky Way from the SAGA survey \citep{Suda_2011}. The simulations predict that metal enrichment in the non {\tt Pop3*} runs (i.e., {\tt Thesan-Zoom} and {\tt TZ + Th. Chem}) can produce only a population of iron (metal)-poor stars that are also carbon-poor within the first billion years. By contrast, the {\tt Pop3*} enrichment models produce a population of carbon-enhanced-metal-poor (CEMP) stars alongside carbon-poor-metal-poor stars, as clearly seen by the extended contours in the {\tt Pop3*} runs. The {\tt Pop3 Salpeter} run predicts the most extreme values for the CEMP stars. This can be attributed to more numerous but relatively low-mass \pt stars that enrich the ISM with higher carbon-to-iron ratios (as iron is primarily produced by \pt stars with $M_\star > 100 \M$). While the predictions do not directly overlap with the observations, they are relatively close (within $1$ dex). Additionally, we note that metal enrichment in the local universe has had much longer time to evolve than in our simulation (meaning the stars we observe today evolved from the remnants of multiple stellar generations because of the large cosmic time, whereas in the simulations we only simulate the first billion years and only a few stellar generations), which can also drive the observed differences. A more accurate comparison, obtained by forward modelling the emission-line diagnostics, will be presented in future work.

The right panel, on the other hand, shows the gas-phase oxygen abundance (within twice the stellar half-mass radius) as a function of stellar mass for the simulated galaxies, compared with the extrapolated mass-metallicity relation (MZR) from \cite{Morishita_2024} at $z=6$ and a recent measurement of a metal-poor galaxy at $z=5.725$ \citep{Morishita_2025}. Triangular markers indicate points that have been offset for visual clarity; the direction of the triangle points to the true location of the scatter point, which is the value from a single snapshot. The MZR converges across all runs and agrees with the extrapolated MZR from the literature, with a maximum scatter of $\sim 0.5~\rm{dex}$. Most of the scatter in $12 + \log (\rm{O / H})$ (and correspondingly in the metallicity) occurs at low stellar masses. While we cannot make definitive statements based on a single halo, we qualitatively find that the {\tt Pop3 M250} run predicts higher metallicities at low stellar masses (or high redshifts) than the other {\tt Pop3*} runs. This is governed by the interplay between the yields from massive \pt stars and the number of \pt stars formed -- for instance, although the {\tt Pop3 Salpeter} run forms a larger number of \pt stars, they are predominantly low-mass and contribute fewer metals to the ISM. We also note that, while a direct comparison of our predicted oxygen abundance to individual observations is challenging, it serves as a good sanity check. We do not exactly match the recent measurement from \cite{Morishita_2025} of an extremely metal-poor galaxy with $12+\log(\rm{O/H}) \approx 6.25$, which is $\approx 0.5~\rm{dex}$ lower than our simulation predictions.

\subsection{Predictions for the \ion{He}{II} $1640$\AA\,line}
\begin{figure*}
\centering
\includegraphics[width=1\linewidth]{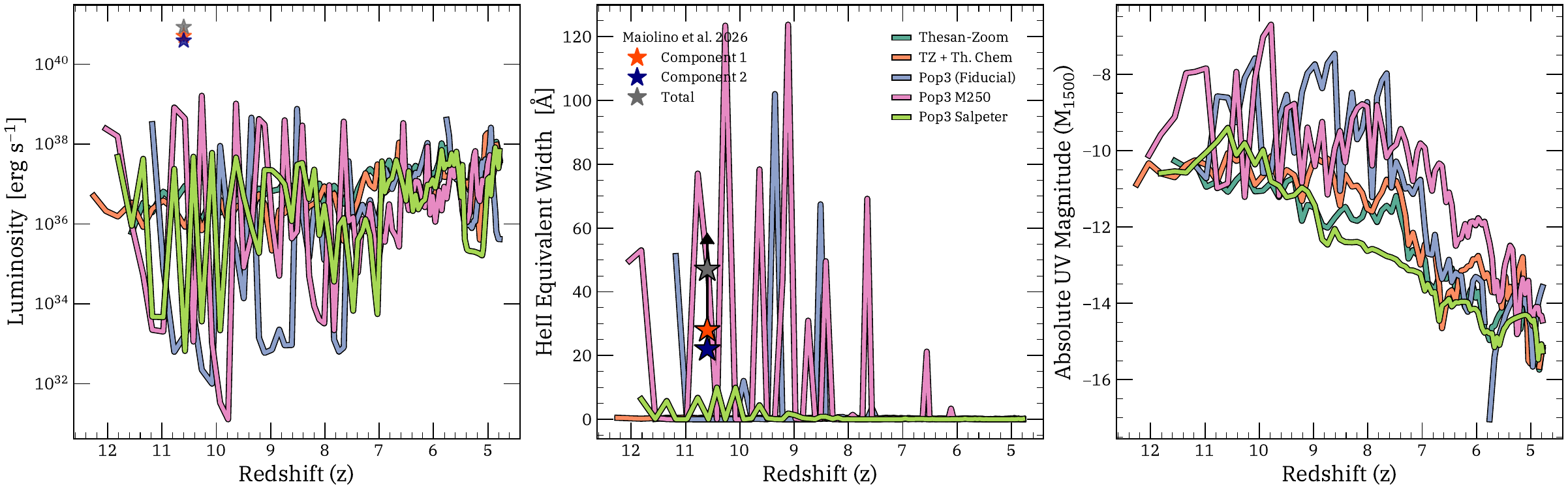}
\caption{
  {\bf Left : } Predicted luminosity of the \ion{He}{II} $1640$\AA~line across different model variations for the central galaxy.
  {\bf Middle : } Equivalent width of the \ion{He}{II} $1640$ \AA~line from the central galaxy.
  {\bf Right : } The absolute UV magnitude (in the AB system) of the central galaxy.
  The $\star$ markers indicate the first spectroscopically confirmed detection of a galaxy hosting \pt stars (i.e., no metal lines in the observed spectra). The simulations predict that only the {\tt Pop3*} runs produce observable signatures similar to the {\it JWST} observations outlined in \citet{Maiolino_2026}.
}
\label{fig:EW_HeII}
\end{figure*}

Next, we make predictions for a major emission-line diagnostic used to detect \pt stars, namely the presence of a strong (high-equivalent-width) $1640$\AA~\ion{He}{ii} line. Fig.~\ref{fig:EW_HeII} plots, from left to right, the expected luminosity of the $1640$\AA~\ion{He}{ii} line, its equivalent width, and the rest-frame UV continuum at 1500 \AA~(i.e., $M_{1500}$) for the target simulated galaxies. We refer the reader to Appendix~\ref{sec:methods_heII} for more details on the approach used to calculate these quantities. For comparison, the plots also show the results from the first spectroscopic detection of a pristine, metal-free emitter at $z\sim10.6$ around GN-z11 \citep{Maiolino_2026} using {\it JWST}. The simulations predict a very bursty line luminosity for the {\tt Pop3*} runs at $z > 7$; in contrast, the line luminosity changes very mildly with time for {\tt Thesan-Zoom} and the {\tt TZ + Th. Chem} runs. At even higher redshifts $(z>10)$, the {\tt Pop3 M250} run predicts the highest luminosity, arising from the more massive \pt stars. For this particular simulated low-mass galaxy, the predicted line luminosity is at least two orders of magnitude lower (the predicted line luminosity fluctuates between $10^{32} - 10^{38}~{\rm erg~s^{-1}}$) than that of the observed galaxy.

The middle panel shows the \ion{He}{ii} EW. The simulations predict that the non-{\tt Pop3} runs have negligible \ion{He}{ii} EWs $({\rm EW} \sim 0)$. The EW is highest for the {\tt Pop3 M250} run, followed by the {\tt Pop3 (fiducial)} run. The {\tt Pop3 Salpeter} run predicts weak \ion{He}{ii} emission $({\rm EW}\sim 10 {\text \AA})$. This is primarily because \ion{He}{ii} emission from a star is proportional to its mass and is present only for \pt stars with $M_\star \gtrsim 60 \M$ (see Fig.~\ref{fig:ionization-contribution}). In the {\tt Pop3 Salpeter} run, the SSP is dominated by low-mass stars, whereas in the {\tt Pop3 (fiducial), Pop3 M250} runs, the log-flat IMF allows the SSP to sample the high-mass end of the stellar population, resulting in a stronger emission-line EW. The predicted EW from the {\tt Pop3*} runs is in good agreement with the observed lines and is often much stronger. The {\tt Pop3 (fiducial), Pop3 M250} runs predict strong EWs (EW $>20$~\AA) down to $z \sim 8$. We note that the predicted signal is quite bursty because of the short lifetimes of the massive \pt stars responsible for high EWs (see also Fig.~\ref{fig:Expected_HeII_EW}). Finally, the right panel plots $M_{\rm{UV}}$ as a function of redshift for the different model variations. The current luminosity cut-off from {\it JWST} is $M_{\rm UV} \lesssim -17$ \citep{Whitler_2025}, which implies that the simulated galaxies are too faint to be detected by {\it JWST}, at least at high-z. In future work, we will simulate more massive halos to enable better comparison with recent observations.

\section{Discussion}\label{sec:discussion}
\subsection{Feedback from \pt stars}\label{subsec:effects_Pop3_feedback}
\begin{figure}
  \centering
  \includegraphics[width=\linewidth]{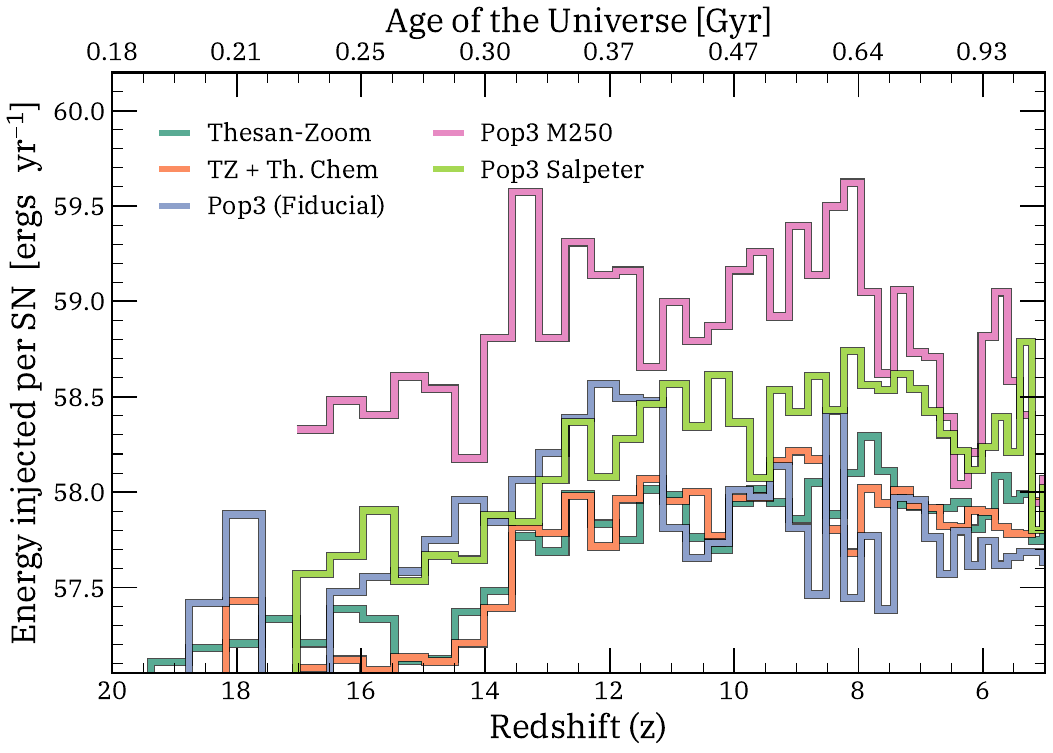}
  \caption{Total supernova feedback energy injected into the ISM per SN across the full simulation domain as a function of time for different simulation variants. On average the {\tt Pop3*} runs have higher energy injection rate compared to {\tt Thesan-Zoom} and {\tt TZ + Th. Chem} runs, with the {\tt Pop3 M250} run with log-flat IMF exhibiting the highest energy injection rate.}
  \label{fig:Total_Feedback_Energy}
\end{figure}

Differences across model variations, such as gas and stellar mass, metal enrichment patterns, and the evolution of phase-space properties, can generally be attributed to variations in SN feedback. Fig.~\ref{fig:Total_Feedback_Energy} plots the energy injected per SN (both SNII and SNIa) as a function of redshift for different model variations. We note that direct one-to-one comparisons between runs are unfair due to the stochastic nature of the subgrid model, so we focus on qualitative differences. The simulations predict that for $ z > 7 $, the {\tt Pop3 M250} run, on average, has the highest energy injection per SN, followed by the {\tt Pop3 (fiducial)} and {\tt Pop3 Salpeter} runs. A top-heavy IMF allows more massive stars to form, which die quickly and inject more energy (see \S\ref{sec:pt_end-state}). For $ z < 7 $, the fraction of \pt stars drops significantly, metal-rich stars begin to form, and differences between stars diminish. Finally, we note that the energy injection rate between the {\tt Thesan-Zoom} and {\tt TZ + Th. Chem} runs is very similar.

The impact of varying SN feedback is also evident in the metal enrichment patterns. This is shown clearly in Fig.~\ref{fig:Elemental_Mass_Abundance}, which plots the evolution of specific elemental yields (defined as the ratio of metals in the gas phase to total stellar mass) in the central galaxy (i.e. all the bound gas of the subhalo) as a function of redshift. Significant metal build-up occurs only for $z \lesssim 11$, even though the first stars form around $z \sim 14-16$, because SN feedback from the first stars efficiently expels gas and metals, delaying the enrichment of the central galaxy. The simulations also predict that metal build-up is slower in the {\tt Pop3*} runs, as stronger feedback in these runs expels gas much farther, and it takes longer for the enriched gas to accrete back into the galaxy (see also the associated discussion for Fig.~\ref{fig:metallicity_shells}). Between $ 10 < z < 6$, the simulations predict that the {\tt Pop3*} runs produce more {\tt C, O, Ne, Mg} and {\tt Si}. The simulations also predict that, for $z > 8$, the central galaxy lacks {\tt N, Fe} in the {\tt Pop3} runs, as these are produced by AGB stars and SNIa, which occur only for Population II/I populations in our model.

The {\tt Pop3 (fiducial)} and {\tt Pop3 M250} runs show higher enrichment of all metals except Iron compared with other runs. Notably, the {\tt Pop3 (fiducial)} run achieves the highest levels of Carbon and Nitrogen enrichment, while the {\tt Pop3 M250} run shows the highest abundances of Oxygen, Neon, Magnesium, Silicon, and Iron. This distinction arises because the IMF in {\tt Pop3 M250} extends to $250~\M$, enabling the formation of more massive stars that undergo PISN, thereby synthesizing heavier elements (see Fig.~\ref{fig:PopIII_SN_yields_comparison} for more details). Conversely, Carbon and Nitrogen are also produced in CCSN, and their higher abundance in the {\tt Pop3 (fiducial)} run reflects the greater total number of low-mass stars, given its lower maximum mass compared with the {\tt Pop3 M250} simulation. In contrast, the {\tt Pop3 Salpeter} run shows minimal differences from the non-\pt runs ({\tt Thesan-Zoom}, {\tt TZ + Th. Chem}) for all elements except Carbon. This is because the Salpeter IMF is dominated by lower-mass \pt stars, which produce fewer metals overall.

Below $ z \lesssim 6 $, metal abundances converge across all simulations, with differences $\lesssim 0.1~\rm{dex}$. This convergence occurs as star formation becomes dominated by metal-rich populations, diminishing the relative impact of \pt feedback and yields. Furthermore, efficient metal mixing wipes out remaining discrepancies \citep[see][for detailed discussions on metal transport and mixing]{Garcia_2025,Sarrato-Alos_2023, Ritter_2015}. Nevertheless, our results indicate that including \pt physics yields a different ISM enrichment pattern than the baseline {\tt Thesan-Zoom} model, particularly during the transition from \pt to PopII star formation ($ 6 < z < 10 $). This scenario provides favourable conditions for the formation of Carbon-Enhanced Metal-Poor (CEMP) stars \cite[e.g.][]{Lucey_2026}. As discussed earlier, searches for \pt signatures often rely on emission-line ratios dominated by nebular emission lines \citep[e.g.][]{Scholtz_2025,Morishita_2025,Cullen_2025}. However, if the chemical signatures in the gas phase are subtle, these diagnostic signals may fall below current detection thresholds. Future work will explore how this early metal enrichment manifests in the observable properties of high-redshift galaxies.

\subsection{Comparison with existing works}\label{subsec:comparison_existing_works}
Theoretical work on \pt star formation broadly falls into two categories. One focuses on the detailed microphysics of collapse within minihaloes during the era when \pt star formation dominates the cosmic star formation rate density(typically $z\gtrsim 15$), examining internal halo conditions and, in some cases, zoom-ins that attempt to resolve an emergent IMF \citep{Abel_2002,Yoshida_2003, Smith_2015, Smith_2024, Lenoble_2024, Stacy_2016}. The second emphasizes statistical properties derived from cosmological-volume simulations. Our study belongs to the latter but uses zoom-ins to achieve higher resolution, which, however,  precludes a direct measurement of the cosmic-averaged star formation rate density (SFRD).

\cite{Wise_2012} studied the transition from \pt (with a top-heavy IMF, like ours) to Pop~II star formation using high resolution RHD simulations. They found that a single PISN from \pt stars can enrich the entire minihalo to a value of about $ 10^{-3}Z_\odot  $, and transition it to Pop~II regimes. The {\tt AEOS} project \citep{Brauer_2025a, Brauer_2025b} aims to model the early chemical enrichment and galaxy formation, which includes star-by-star treatment and detailed prescription for \pt stars (formation and SN feedback) in a cosmological setting (until $ z=14.5 $  ). They found that the choice of IMF for \pt stars has a significant impact on the chemical enrichment and ionization history of the ISM, but bulk properties like total star formation history remains robust. Additionally they found that compared to using  subgrid models that make use of SSP approximation for star particles, star-by-star treatment allows for more metal accumulation in the ISM.

Consistent with our results, Fire2 simulations (without radiative transfer) with distinct \pt/II feedback models found continued \pt activity \citep{Jaacks_2018,Jaacks_2019} to $z=7.5$ (dominated by low-mass $M_{\rm halo}\!<\!10^9\,\M$ haloes). On the other hand, \citet{Liu_2020,Liu_2020b} show continued \pt star formation down to $z=4$ (shifting to $M_{\rm halo}\!\sim\!10^{10}\,\M$ at later times). Both works predict that late-time \pt star formation occurs preferentially in halo outskirts. Zoom-ins by \citet{Yajima_2022,Yajima_2023} found that galaxies with lower stellar mass have higher \pt fractions {($ \gtrsim 10\%, M_\star < 10^5~\M  $)} at $z\sim 10$ and a spatial preference for recently formed \pt stars to be present in minihaloes around larger systems. The Renaissance project \citep{OShea_2015} (zoom-in simulations with {\tt ENZO}) reported ongoing \pt formation in void regions to $z=7.6$, concentrated in $5\times 10^7$--$10^8\,\M$ haloes, with LW radiation suppressing smaller systems while inefficient metal transport left some dense clumps of pristine gas amenable to star formation \citep{Xu_2016}.

\cite{Zier_2025} analysed galaxies in the {\tt Thesan-Zoom} simulations \citep{Kannan_2025}, where \pt stars are identified using a more aggressive metallicity threshold of $ 10^{-6}~Z_\odot$. However, the model does not include stellar radiation and SN feedback (including metal returns) from the \pt stars. They find a low \pt SFR beyond reionization, dense pristine gas surrounding the target haloes, and residual \pt stars that can persist in satellites surrounding the target halo at lower redshifts. {\tt THESAN-HR} \citep{Borrow_2023} likewise found \pt to the end of reionization using an effective ISM model and no LW transport, and \cite{Pakmor_2022} even reported \pt down to $z=0$ in IllustrisTNG (using an effective equation-of-state ISM prescription).

Our findings regarding the spatial and temporal distribution of \pt stars are also consistent with recent high-resolution Ramses based simulation MEGATRON which also includes a model for \pt stars \citep{Storck_2026}. Similar to our methodology, MEGATRON self-consistently couples radiation and non-equilibrium chemistry at near-parsec resolution. They observe that while the initial Pop III stars emerge in molecular hydrogen-cooling minihaloes, the rapid establishment of a Lyman-Werner (LW) background shifts the majority of subsequent Pop III formation into more massive atomic cooling haloes. This aligns perfectly with our observations in Section 3.1 and Appendix \ref{sec:chemical_feedback}, where early, intense LW radiation from the first stars severely suppresses cooling in the surrounding IGM and neighbouring substructures. Furthermore, MEGATRON identifies the LW background as the most dominant quenching mechanism for minihaloes, outweighing gas starvation and external chemical enrichment—a conclusion vividly supported by the $\rm H_2$-poor voids generated by our {\tt Pop3*} runs. Finally, MEGATRON predicts that Pop III stars form at a wide range of distances from UV-bright central galaxies, with only a fraction forming within the virial radius; this spatial preference strongly corroborates our findings that early Pop III formation is heavily offset from the central galaxy.

Overall, the literature converges on low \pt SFRs at low redshift, typically concentrated in small haloes or in the outskirts or satellites of larger haloes, especially in recently accreted minihaloes, in broad agreement with our findings. The precise endpoint remains uncertain and sensitive to reionization, feedback, and metal mixing. In our simulations, the fraction of dense primordial gas drops sharply during reionization, suggesting that Pop~III formation largely ceases near the end of the EoR, though the exact timing can also be strongly environment-dependent.

Several observational \pt galaxy candidates have recently been detected using {\it JWST} \citep{Maiolino_2026}. They have low stellar masses and are spatially offset from massive hosts. For example, \citet{Vanzella_2023} detected a potential source at $z=6.639$ with $M_\star<10^4\,\M$. A couple of galaxies at $z=10.6$ with $M_\star\simeq 2$--$2.5\times 10^5\,\M$ about $\sim$2\,kpc from GN-z11 have recently been spectroscopically confirmed as strong \ion{He}{II} emitters \citep{Jiang_2021,Tacchella_2023, Maiolino_2026}. Similarly, a cluster at $z\approx 8.2$ with $M_\star=(7.8\pm 1.4)\times 10^5\,\M$ at a distance of $\sim$1\,kpc from its host \citep{Wang_2024} and a candidate \pt galaxy at $z\approx 6.5$ with $M_\star\sim 10^5\,\M$ \citep{Fujimoto_2025} have also been proposed as potential \pt candidates. These systems likely host mixtures of \pt and enriched stars and align with the expectation that late-time \pt formation occurs in low-mass haloes or satellites offset from central galaxies. A better comparison of stellar masses and demographics will require an explicit \pt stellar evolution model simulated over large cosmological volumes to chart the \pt$\rightarrow$PopII transition and the epochs over which \pt can dominate within individual haloes.

\subsection{Caveats}\label{subsec:caveats}
Our simulations include molecular hydrogen chemistry, radiative transfer (including LW), and sufficient resolution to resolve the smallest star-forming haloes ($\sim 10^6\,\M$). Modelling the escape of LW photons self-consistently removes the need for an imposed, spatially uniform LW background. Nonetheless, several relevant processes are absent or uncertain:

\begin{figure*}
  \centering
  \includegraphics[width=\linewidth]{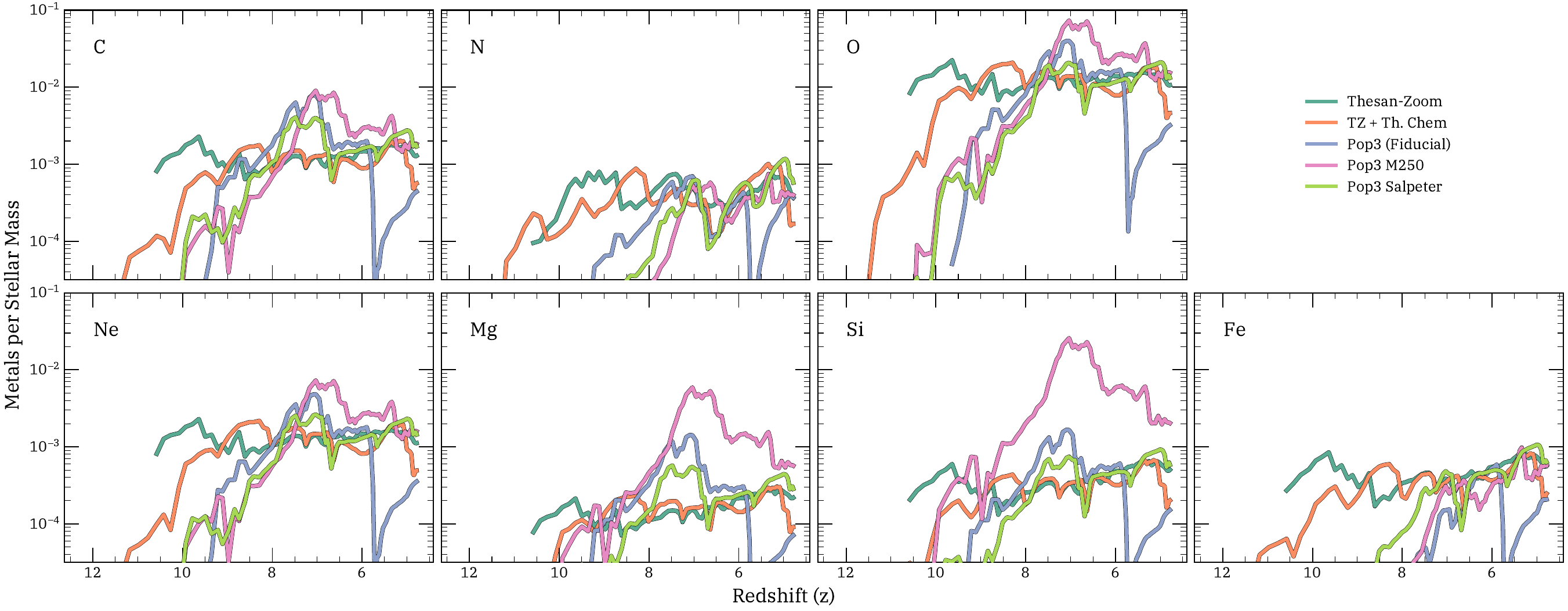}
  \caption{
  Evolution of elemental abundances (mass fractions of metal species relative to total gaseous or stellar mass) in the central galaxy    across the different simulation runs. The panels show the evolution of Hydrogen, Helium, Carbon, Nitrogen, Oxygen, Neon, Magnesium, Silicon, and Iron. The {\tt Pop3 M250} run (pink) shows the highest enrichment in O, Ne, Mg, Si, and Fe due to Pair-Instability Supernovae of massive stars ($>140\,\M$). The {\tt Pop3 (fiducial)} run (blue) leads in C and N production, reflecting higher CCSN rates from a top-heavy IMF. By $z \lesssim 6$, abundances converge across all models as metal-rich star formation dominates, though \pt physics predicts earlier ISM enrichment at $z > 6$.}
  \label{fig:Elemental_Mass_Abundance}
\end{figure*}

\begin{itemize}
  \item Baryon--DM streaming velocities: We do not include the coherent relative streaming that decays as $(1+z)$ and preferentially suppresses collapse in low-mass haloes, shifting the \pt threshold to higher masses \citep{Schauer_2021}. This effect is most relevant for $M_{\rm halo}\!\lesssim\!10^7\,\M$, which contribute substantially to late-time \pt formation.

  \item Deuterium/HD cooling: Our network does not model the formation and impact of HD. HD can form efficiently below $\sim$200\,K and cool the gas to the CMB floor \citep{Nagakura_2005}. This can be achieved via external LW fields \citep{Nishijima_2024}, higher-mass haloes, or slower collapse \citep{Greif_2011}. Cooler gas would reduce Jeans scales and could trigger earlier star formation.

  \item LW background: We model local LW sources via radiative transfer, and most \pt formation occurs in $M_{\rm halo}\!>\!10^6\,\M$ haloes that are less sensitive to a uniform LW field. However, we do not impose an extragalactic LW background \citep[such as][]{Incatasciato_2023} at the domain boundaries. This extragalactic LW background can suppress \pt star formation in low-mass minihaloes by dissociating $ \rm{H_2} $.
  \item Metal mixing: Within resolution elements, we assume perfect mixing, without a subgrid turbulent metal-diffusion model. Such models can boost the \pt SFR by factors of $\sim$4--10 by allowing partially pristine fractions in otherwise enriched gas \citep{Sarmento_2017,Sarmento_2022}. Given resolution differences (and AMR vs.\ moving mesh), direct comparison is non-trivial \citep[see][for more comparison of {\tt THESAN-ZOOM} galaxies with other methods]{Kannan_2025,Zier_2025,McClymont_2025}. Exploring subgrid mixing in our setup will be important for quantitative forecasts.
  \item We do not include magnetic fields in the {\tt Thesan-Zoom} framework. It has been shown that magnetic fields slow down gravitational collapse, which can delay the onset of \pt star formation \citep{Saad_2022,Sharda_2025}. Including magnetic fields in our model with non-ideal MHD effects would be necessary to fully capture their impact on \pt formation and evolution.
  \item Stellar Rotation: Our MESA tracks and assumed supernova yields are based on non-rotating stellar models. \pt stars are theorized to be fast rotators, which induces internal mixing that can significantly extend main-sequence lifetimes and alter the mass of the final helium core. Furthermore, rotational flattening alters the surface temperature ($T_{\rm eff}$) and gravity ($\log g$), which would modify the emergent spectra, and could also potentially drive metal free winds (e.g. \citealt{Liu_2021}).
  \item X-ray Binaries: We currently treat all \pt stars as single stars. However, massive stars overwhelmingly form in binary or multiple systems. Binary interactions dictate mass transfer and the eventual formation of High-Mass X-ray Binaries (HMXBs). As emphasized by recent studies (e.g., \citealt{Sartorio_2023}), X-rays from these binaries provide a hard radiation field that can deeply penetrate and ionize the diffuse IGM, profoundly altering the global 21-cm signal. Incorporating a mass-dependent binary fraction to capture both the modified UV/LW spectra and this hard X-ray component remains a critical avenue for future work.
  \item Impact of resolution: It is unclear whether a subgrid model that assumes each \pt star particle represents a single stellar population with a specified IMF remains valid at resolutions lower than that used in this work. For instance, if a giant molecular cloud (GMC) collapses to produce approximately $10^5$ to $10^6$~M$_\odot$ of stars, it is ambiguous whether all the stars will be metal-poor or whether the initial stars formed from the GMC will promptly enrich the subsequent stellar populations generated during gravitational collapse. The critical GMC mass scale above which the entire cloud is unlikely to form solely pristine Population III stars is not well established.  This issue has been partially addressed in this work by adopting a subgrid SSP model coupled with high resolution ($\sim 1000~\rm{M}_\odot$). However, for lower-resolution simulations, we plan to address this issue either by assuming that the star particles are a mixture of Pop III and Pop II stars, or by forming lower-mass star particles for \pt stars. We plan to investigate this in future work.
\end{itemize}

\section{Conclusion and Summary}\label{sec:conclusion}
This study presents a comprehensive framework for modelling Population III (\pt) stars in cosmological simulations to elucidate their impact on early galaxy formation and evolution.
\subsection{Key Methodological Advances}
We have developed three major improvements to the {\tt Thesan-Zoom} simulation framework:

\subsubsection{Enhanced Thermochemistry Network}
We implemented an improved thermochemistry network for primordial gas (\S\ref{sec:thermochem}) that includes equilibrium $\hp$ and $\hmn$ species. These species are critical for $\rm{H_2}$ formation and gas cooling in primordial, metal-free environments. The network tracks the non-equilibrium evolution of 6 species ($\rm{H_2}$, $\rm{HI}$, $\rm{HII}$, $\rm{HeI}$, $\rm{HeII}$, $\rm{HeIII}$) and the equilibrium evolution of 2 species ($\hp, \hmn$), along with the evolution of temperature.

\subsubsection{Detailed \pt Stellar Spectra}
We used {\tt MESA} to model the stellar evolutionary tracks of 120 metal-free stars in the range $(0.1-1000\,\M)$ (Fig.~\ref{fig:MESA_tracks}) and {\tt TLUSTY} to compute detailed spectra using 1D, NLTE radiative transfer calculations of the stellar atmospheres. The IMF-averaged spectra for three Initial Mass Functions: a log-flat IMF ($\alpha = -1$) with $M_{\rm{min}} = 2~\rm{M}_\odot$ and $M_{\rm{max}} = 150~\rm{M}_\odot$ (fiducial model), a log-flat IMF with $M_{\rm{min}} = 2~\rm{M}_\odot$ and $M_{\rm{max}} = 250~\rm{M}_\odot$ (Fig.\ref{fig:IMF_spec}), and a Salpeter IMF ($\alpha = -2.35$) with $M_{\rm{min}} = 2~\rm{M}_\odot$ and $M_{\rm{max}} = 150~\rm{M}_\odot$ were calculated and integrated into the simulation framework.

\subsubsection{Supernova Feedback Implementation}
Elemental yields from type II, \pt supernovae (\S\ref{sec:pt_end-state}) were incorporated for the mass ranges $10-100\,\M$ and $140-260\,\M$, modelling both Core-Collapse Supernovae (CCSN) and Pair-Instability Supernovae (PISN). PISN inject significantly more energy (up to $\sim 10^{52}$\,erg) than typical CCSN, and we track the production and evolution of 9 chemical species: H, He, C, N, O, Ne, Mg, Si, and Fe.

\vspace{2ex}
We then use the {\tt AREPO-RT} radiation-hydrodynamics code to simulate a $1.95 \times 10^9~\M$ halo at $z = 3$. Using the zoom-in technique, we achieve a baryonic mass resolution of $1.14 \times 10^3~\M$. We conduct five simulation variants: {\tt Thesan-Zoom} (baseline), {\tt TZ + Th. Chem} (enhanced thermochemistry only), {\tt Pop3} (full \pt physics with a log-flat IMF), {\tt Pop3 M250} (full \pt physics with a log-flat IMF but with higher $M_{\max}$), and {\tt Pop3\_Salpeter} (full \pt physics with a Salpeter IMF) (see Tab.~\ref{tab:IMF_params} for the different choices of IMF).

\subsection{Major Findings}
\subsubsection{\pt Star Formation and Evolution}
The first stars appear at $z \sim 16-14$ across different models, with the \pt fraction dropping significantly after $z \sim 10$ as metals from SN explosions enrich the gas in galaxies (Fig.~\ref{fig:Pop3_fraction}). While the formation of \pt stars becomes subdominant by $z \sim 8$, pockets of pristine gas allow them to form down to $z \sim 6$ (Fig.~\ref{fig:Stars_Z_vs_redshift}). The choice of IMF critically affects the star formation history; top-heavy IMFs produce stronger feedback that suppresses star formation more effectively, leading to lower overall stellar masses than in models with Salpeter-like IMFs or without \pt feedback (Fig.~\ref{fig:SFH}).

\subsubsection{Gas Phase Evolution and Thermochemistry}
The updated thermochemistry network enables more efficient cooling, yielding a higher fraction of dense gas ($n_{\rm H} > 10^{3} \pcc$) and lower overall temperatures (Fig.~\ref{fig:PhaseSpace}) than the baseline model. The stronger \pt radiative feedback also introduces a distinctive photo-heated diffuse gas phase ($T \sim 10^4$\,K, $n_{\rm H} < 10^{-4}~\pcc$).

\subsubsection{\pt Feedback and Chemical Enrichment}
Metal enrichment from \pt stars proceeds differently than from later generations of stars. The {\tt Pop3*} runs predict the formation of Carbon-Enhanced Metal-Poor (CEMP) stars, which are absent in the non-\pt runs (Fig.~\ref{fig:MetallicityTrends}). Higher energy injection from \pt supernovae expels gas and metals further from the galaxy, leading to a slower build-up of metals in the ISM (Fig.~\ref{fig:Elemental_Mass_Abundance}) at high redshifts $ (z > 8) $. The specific elemental abundance patterns also differ, with top-heavy IMFs producing higher abundances of Oxygen, Neon, Magnesium, Silicon, and Iron due to contributions from PISN.

\subsubsection{Predictions for {\it JWST}}
Our model predicts strong, bursty \ion{He}{ii} $1640${\AA} equivalent widths (EW $> 20$\,{\AA}) for models with top-heavy IMFs ({\tt Pop3 fiducial} and {\tt Pop3 M250}) down to $z \sim 8$, consistent with recent early {\it JWST} detections of potential metal-free galaxies. In contrast, models without \pt stars or with a Salpeter IMF produce negligible or weak \ion{He}{ii} emission. Although the specific simulated galaxy is currently too faint for direct {\it JWST} detection, our results support the use of \ion{He}{ii} emission as a critical diagnostic of high-mass \pt stellar populations in more massive systems.

\subsection{Implications}
We present a comprehensive framework for modelling the formation and impact of \pt stars in cosmological simulations. This will enable more accurate predictions for {\it James Webb Space Telescope} observations of high-redshift galaxies. Our technical innovations include a self-consistent treatment of \pt radiation transport, including Lyman-Werner photons; proper accounting for momentum injection during the Sedov-Taylor phase of supernova feedback; and detailed tracking of ionization states and molecular chemistry in primordial gas. This work advances the modelling of the transition from the first stars to subsequent generations, providing a more complete picture of how \pt stars influenced early galaxy formation and left imprints on the intergalactic medium.

\section*{Acknowledgements}
Computational resources for this project were enabled by a grant to RK from Compute Canada/Digital Research Alliance of Canada (\url{alliancecan.ca}) and carried out on the Trillium supercomputer at the SciNet HPC Consortium \citep{2010Loken-SciNet}. We acknowledge constructive discussions with Thomas Gessy-Jones on spectral modelling of \pt stars and with Aaron Smith and William McClymont on estimating the luminosity of the \ion{He}{II} line. RK acknowledges support from the Natural Sciences and Engineering Research Council of Canada (NSERC) through a Discovery Grant and a Discovery Launch Supplement (funding reference numbers RGPIN-2024-06222 and DGECR-2024-00144), as well as support from York University's Global Research Excellence Initiative.
GMM acknowledges financial support from Junta de Andalucia through the program Emergia (EMEC\_2023\_00533).

\section*{Data Availability}
Simulation data products (including stellar spectra) pertaining to this work will me made available upon reasonable request with the corresponding author.



\bibliographystyle{mnras}
\bibliography{refernces}




\appendix

\section{Thermo-chemical network}
\label{app:therm}
\begin{figure*}
  \begin{align}
    \dot{\mathcal{M}}_{\rm{HI}}= &\Gamma_{\rm{A}} n_{\rm{H}^{-}}+\Gamma_{\rm{B}} n_{\rm{H}_2^{+}}+2 \Gamma_{\rm{E}} n_{\rm{H}_2}+2 \Gamma_{\rm{LW}} n_{\rm{H}_2}-k_1 n_e n_{\rm{H}_{\rm{I}}}-k_2 n_{\rm{H}^{-}} n_{\rm{HI}}-k_3 n_{\rm{HII}} n_{\rm{HI}} -k_4 n_{\rm{H}_2^{+}} n_{\rm{H}_{\rm{I}}} -k_{26} n_{\rm{He}} n_{\rm{HI}}-2 k_{30} n_{\rm{HI}}^3\nonumber\\
    &-2 k_{31} n_{\rm{HI}}^2 n_{\rm{H}_2}-2 k_{32} n_{\rm{HI}}^2 n_{\rm{HeI}}+2 k_5 n_{\rm{HII}} n_{\rm{H}^{-}} +2 k_6 n_e n_{\rm{H}_2^{+}}+k_7 n_{\rm{H}_2} n_{\rm{HII}} +2 k_8 n_e n_{\rm{H}_2}+2 k_9 n_{\rm{HI}} n_{\rm{H}_2}+2 k_{10} n_{\rm{H}_2} n_{\rm{H}_2} \nonumber\\
    &+2 k_{11} n_{\rm{HeI}} n_{\rm{H}_2} +k_{14} n_e n_{\rm{H}^{-}}+k_{15} n_{\rm{HI}} n_{\rm{H}^{-}}+k_{21} n_{\rm{H}_2^{+}} n_{\rm{H}^{-}} +3 k_{22} n_{\rm{H}^{-}} n_{\rm{H}_2^{+}}+k_{23} n_e n_{\rm{H}_2} + k_{24} n_{\rm{HeII}} n_{\rm{H}_2} \nonumber\\
    &+k_{27} n_{\rm{He}} n_{\rm{HII}}+k_{28} n_{\rm{HeII}} n_{\rm{H}^{-}}+k_{29} n_{\rm{HeI}} n_{\rm{H}^{-}} + \alpha_{\rm HII}n_{\rm HII} n_e - \sigma_{\rm eHI}n_{\rm HI}n_e -\Gamma_{\rm HI}n_{\rm HI}~, \label{eq:HI}\\
    \dot{\mathcal{M}}_{\rm{HII}}= &~ \Gamma_{\rm{B}} n_{\rm{H}_2^{+}}+2 \Gamma_{\rm{C}} n_{\rm{H}_2^{+}}-k_3 n_{\rm{HI}} n_{\rm{HII}}-k_5 n_{\rm{H}^{-}} n_{\rm{HII}}-k_7 n_{\rm{H}_2} n_{\rm{HII}}-k_{16} n_{\rm{H}^{-}} n_{\rm{HII}} - k_{27}n_{\rm{HeI}} n_{\rm{HII}}+k_4 n_{\rm{H}_2^{+}} n_{\rm{HI}}  \nonumber\\
    &+k_{24} n_{\rm{He}} n_{\rm{H}_2}+k_{26} n_{\rm{H}_{\rm{I}}} n_{\rm{HeII}} - \alpha_{\rm HII}n_{\rm HII} n_e + \sigma_{\rm eHI}n_{\rm HI}n_e +\Gamma_{\rm HI}n_{\rm HI} ~,\label{eq:HII}\\
    \dot{\mathcal{M}}_{\rm{H}_2}= &-\Gamma_{\rm{D}} n_{\rm{H}_2}-\Gamma_{\rm{E}} n_{\rm{H}_2}-\Gamma_{\rm{LW}} n_{\rm{H}_2}-k_7 n_{\rm{H}_2} n_{\rm{HII}}-k_8 n_e n_{\rm{H}_2}-k_9 n_{\rm{H}_1} n_{\rm{H}_2}-k_{10} n_{\rm{H}_2} n_{\rm{H}_2} -k_{11} n_{\rm{He}_{\rm{I}}} n_{\rm{H}_2}-k_{23} n_e n_{\rm{H}_2}\nonumber\\
    &-k_{24} n_{\rm{He}_{\rm{II}}} n_{\rm{H}_2}-k_{25} n_{\rm{He}_{\rm{II}}} n_{\rm{H}_2}+k_2 n_{\rm{H}^{-}} n_{\rm{HI}_{\rm{I}}}+k_4 n_{\rm{H}_2^{+}} n_{\rm{H}_{\rm{I}}}+k_{21} n_{\rm{H}_2^{+}} n_{\rm{H}^{-}}\nonumber\\
    &+k_{30} n_{\rm{HI}}^3 +k_{31} n_{\rm{HI}}^2 n_{\rm{H}_2}+k_{32} n_{\rm{HI}}^2 n_{\rm{He}} + \alpha_{\rm H_2}^{\rm D} \left(\frac{D}{D_{\rm MW}}\right)n_{\rm H} n_{\rm HI}~, \label{eq:H2}\\
    \dot{\mathcal{M}}_{\rm{H}_2^{+}} = &-\Gamma_{\rm{B}} n_{\rm{H}_2^{+}}-\Gamma_{\rm{C}} n_{\rm{H}_2^{+}}+\Gamma_{\rm{D}} n_{\rm{H}_2}-k_4 n_{\rm{H}_{\rm{I}}} n_{\rm{H}_2^{+}}-k_6 n_e n_{\rm{H}_2^{+}}-k_{21} n_{\rm{H}^{-}} n_{\rm{H}_2^{+}}-k_{22} n_{\rm{H}^{-}} n_{\rm{H}_2^{+}} + k_3 n_{\rm{HI}} n_{\rm{HII}} +k_7 n_{\rm{H}_2} n_{\rm{HII}}\nonumber\\
    &+k_{16} n_{\rm{HII}} n_{\rm{H}^{-}}+k_{25} n_{\rm{H}_2} n_{\rm{HeII}} ~, \label{eq:H2plus } \\
    \dot{\mathcal{M}}_{\rm{H}^{-}} = &-\Gamma_{\rm{A}} n_{\rm{H}^{-}}-k_2 n_{\rm{HI}} n_{\rm{H}^{-}}-k_5 n_{\rm{H}} n_{\rm{H}^{-}}-k_{14} n_e n_{\rm{H}^{-}}-k_{15} n_{\rm{H}_{\rm{I}}} n_{\rm{H}^{-}}-k_{16} n_{\rm{H}_{\rm{II}}} n_{\rm{H}^{-}} -k_{21} n_{\rm{H}_2^{+}} n_{\rm{H}^{-}}-k_{22} n_{\rm{H}_2^{+}} n_{\rm{H}^{-}}  \nonumber\\
    & -k_{28} n_{\rm{He}} n_{\rm{H}^{-}}  -k_{29} n_{\rm{He}} n_{\rm{H}^{-}}+k_1 n_e n_{\rm{HI}_{\rm{I}}}+k_{23} n_e n_{\rm{H}_2}~, \label{eq:Hminus} \\
    \dot{\mathcal{M}}_{\rm{HeII} } =&~\alpha_{\rm HeIII} n_{\rm HeIII} n_e+\sigma_{e~ \rm HeI} n_e n_{\rm{HeI}} +n_{\rm{HeI}} \Gamma_{\rm{HeI}} - \alpha_{\rm{HeII}} n_{\rm{HeII}} n_e  -\sigma_{e~\rm HeII} n_e n_{\rm HeII}- n_{\rm{HeII}} \Gamma_{\rm{HeII}}, \label{eq:HeII}\\
    \dot{\mathcal{M}}_{\rm{HeIII} }=&~-\alpha_{\rm{He} \rm{III}} n_{\rm{He} \rm{III}} n_e+\sigma_{e~\rm HeII} n_e n_{\rm{He} \rm{II}} + n_{\rm{HeII}} \Gamma_{\rm{HeII}} \label{eq:HeIII} \\
    \dot{\mathcal{M}}_{\rm{U}} = &~\mathfrak{h}_{\rm{HI}} n_{\rm{HI} } + \mathfrak{h}_{\rm{HeI}} n_{\rm{HeI} } + \mathfrak{h}_{\rm{HeII}} n_{\rm{HeII} } + \mathfrak{h}_{\rm{H_2}} n_{\rm{H_2} } - \Lambda_M + \Lambda_{\rm{PE} }-\Lambda_D - \Lambda(n\to 0)_{\rm{H_2HI} }n_{\rm{H_2} }n_{\rm{HI} } - \Lambda(n \to 0)_{\rm{H_2H_2}}n_{\rm{H_2} }^2  \nonumber\\
    &-\Lambda_{\rm{H_2^+}e}n_{\rm{H_2^+}}n_e -\Lambda_{\rm{H_2^+ HI}}n_{\rm{H_2^+}}n_{\rm{HI}} - \Lambda_{\text{C}}n_e~.\label{eq:temperature-evolution}
  \end{align}
\end{figure*}

The seven-species thermochemical reaction network that computes the non-equilibrium abundances of $\rm{H_2}$, \ion{H}{I}, \ion{H}{II}, $\hmn$, $\hp$, \ion{He}{II}, and \ion{He}{III} is outlined in Eqs.~\ref{eq:HI} - \ref{eq:HeIII}. This is coupled to the equation governing the evolution of the internal energy (U), Eq.~\ref{eq:temperature-evolution}. The references we use to calculate the values of the various photoionization ($\Gamma$), photoheating ($\mathfrak{h}$), cooling ($\Lambda$) and other collisional ionization and recombination ($k_1 - k_{31}$) rates are outlined in \S~\ref{sec:thermochem}. We note that in this work we use a simplified version of this network by assuming kinetic equilibrium for $\hp$ and $\hmn$, i.e., $\dot{\mathcal{M}}_{\hp}, \dot{\mathcal{M}}_{\hmn}\thickapprox0$.

\section{Comparison of PopIII spectra for various IMFs}
\begin{figure*}
  \centering
  \includegraphics[width=1\linewidth]{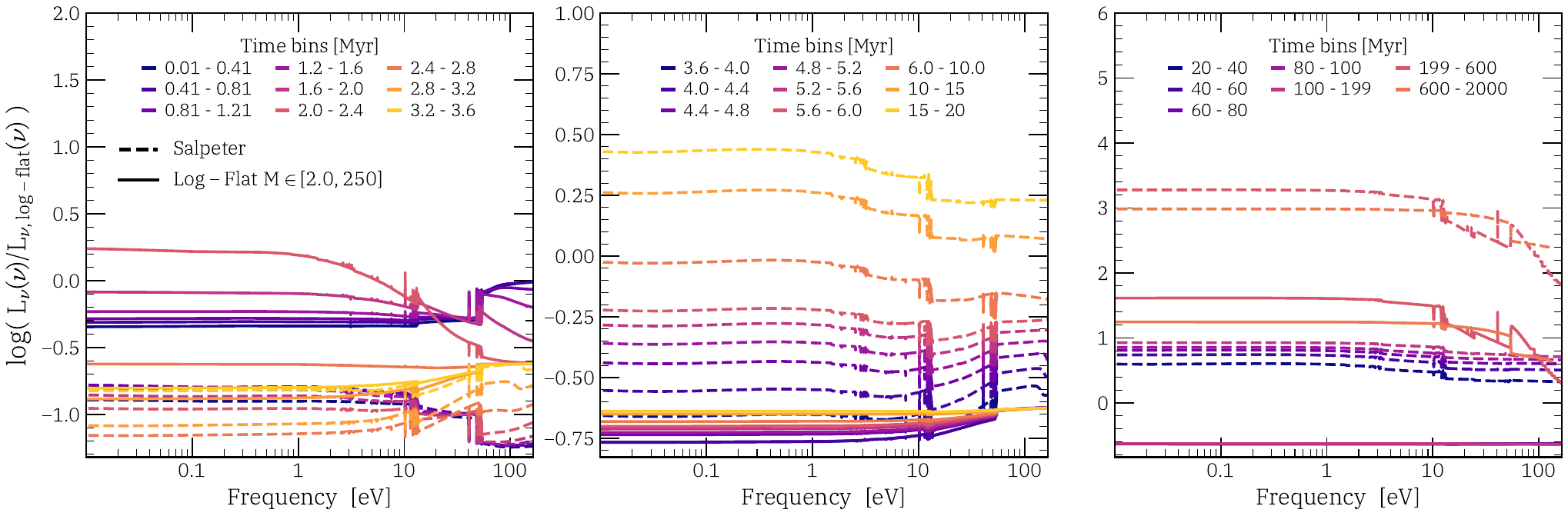}
  \caption{Comparison of IMF averaged spectra for different PopIII IMF choices at various stellar population ages used in the {\tt Pop3 fiducial}, {\tt Pop3 Salpeter}, and {\tt Pop3 M250} runs. Solid lines show the results for the {\tt Pop3 M250} run, while dashed lines represent the {\tt Pop3 Salpeter} run.}
  \label{fig:PopIII_IMF_comparison_spectra}
\end{figure*}

\label{sec:Spectra_Comparison}
Fig.~\ref{fig:PopIII_IMF_comparison_spectra} compares the IMF-averaged spectra (see Eq.~\ref{eq:IMF_averaged_spectra}) used in the {\tt Pop3 Salpeter} and {\tt Pop3 M250} runs with those of the {\tt Pop3 fiducial} run. The masses of stars in the {\tt fiducial} and {\tt Salpeter} IMFs range from $ 2.0 - 150 ~\M$, whereas in the {\tt Pop3 M250} model stars are allowed to form between $ 2.0 - 250 ~\M$. The solid lines show the spectra for the {\tt Pop M250} run, and the dashed lines show the spectra for the {\tt Pop3 Salpeter} run. The lines are colour-coded by the age of the stellar population (shown in the legends).
The {\tt Pop3 M250} model has a higher radiation output than the fiducial model ($\lesssim 4~\rm{Myrs}$) due to the inclusion of more massive stars at early times. However, this model also has a larger fraction of stars that die very young, and therefore, beyond very early times, the IMF-averaged spectra are generally softer than those of the fiducial model. Nevertheless, for $t < 2$ Myr, the {\tt Pop3 M250} model produces a harder spectrum in the helium-ionizing band $(\nu > 54.2~{\rm eV})$ compared to the {\tt Pop3 (fiducial)} model.
Conversely, in the {\tt Pop3 Salpeter} model, the stellar population is dominated by low-mass \pt stars. These low-mass \pt stars produce a softer spectrum but live longer. Thus, for $t < 10$ Myr, the {\tt Pop3 Salpeter} model has a softer spectrum than the {\tt Pop3 (fiducial)}, but for $t > 10$ Myr the spectrum is harder, as most of the massive stars in {\tt Pop3 (fiducial)} are past their MS lifetime.

\section{PopIII stellar yields}
\label{sec:yields}
\begin{figure}
  \centering
  \includegraphics[width=1\linewidth]{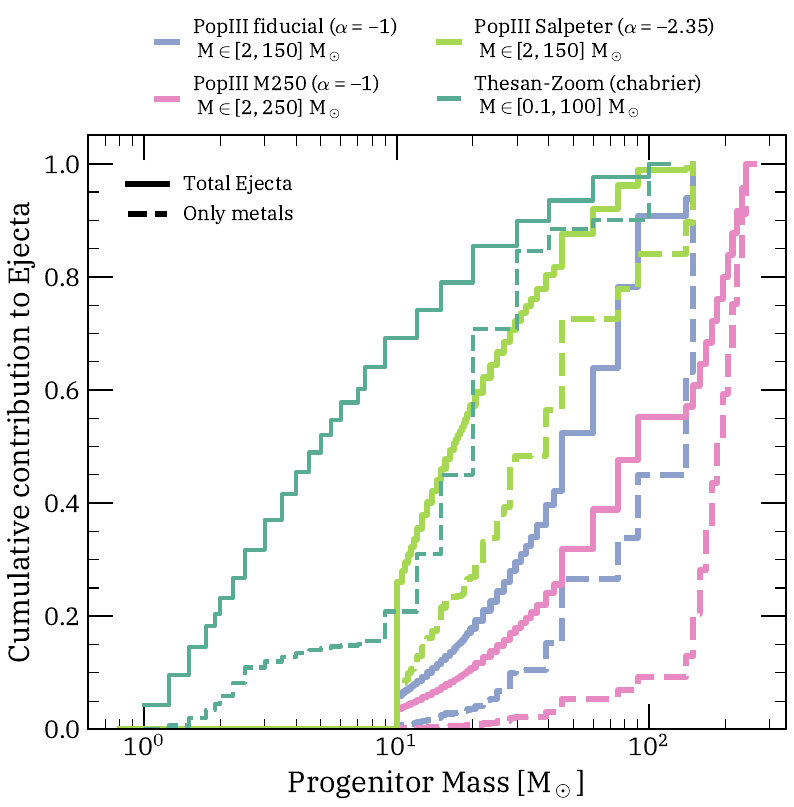}
  \caption{Cumulative contribution to the total ejected mass from SN as a function of stellar mass for different choices of IMF for both the \pt stars and non-\pt stars. For non-\pt stars we include contribution from both SNII and AGB. The solid lines is the cumulative contribution to entire ejecta from the SN, whereas the dashed lines is the contribution from only metals (i.e. everything other than Hydrogen and Helium)}
  \label{fig:cumulative_ejected_mass_in_SN_comparison}
\end{figure}

Fig.~\ref{fig:cumulative_ejected_mass_in_SN_comparison} shows the cumulative contribution to the total ejected mass (including metals) from SN as a function of stellar mass, for different choices of the \pt IMF. The cumulative mass is calculated as:
\begin{equation}
  \text{Cumulative yield}(< m) = \frac{\displaystyle\int_{M_{\min}}^{M_{\rm{max}}} Y(m) \phi(m) {{\rm d}m}}{\displaystyle\int_{M_{\min}}^{M_{\max}} Y(m) \phi(m) {{\rm d} m}}
\end{equation}
where $ Y(m) $ is the ejected mass from a star of mass $ m $, $ \phi(m) $ is the IMF, and $ M_{\min}, M_{\max} $ are the minimum and maximum masses of stars formed from the initial gas cloud. While the \pt stars contribute to the ejected mass only via SNII, the non-\pt stars include contributions from both SNII and AGB winds. The dashed lines shows the cumulative contribution to ejected metals (i.e. everything other than Hydrogen and Helium). We observe that for the log-flat IMF, the majority of the mass is contributed by high-mass stars (i.e., $ > 40 ~\M $), whereas for the Salpeter IMF, the mass contribution is more evenly distributed across the mass range. Additionally for the {\tt Pop3 M250} model, most of the ejected metals are produced by stars with $M_\star > 140~\M$

\begin{figure*}
  \centering
  \includegraphics[width=1\linewidth]{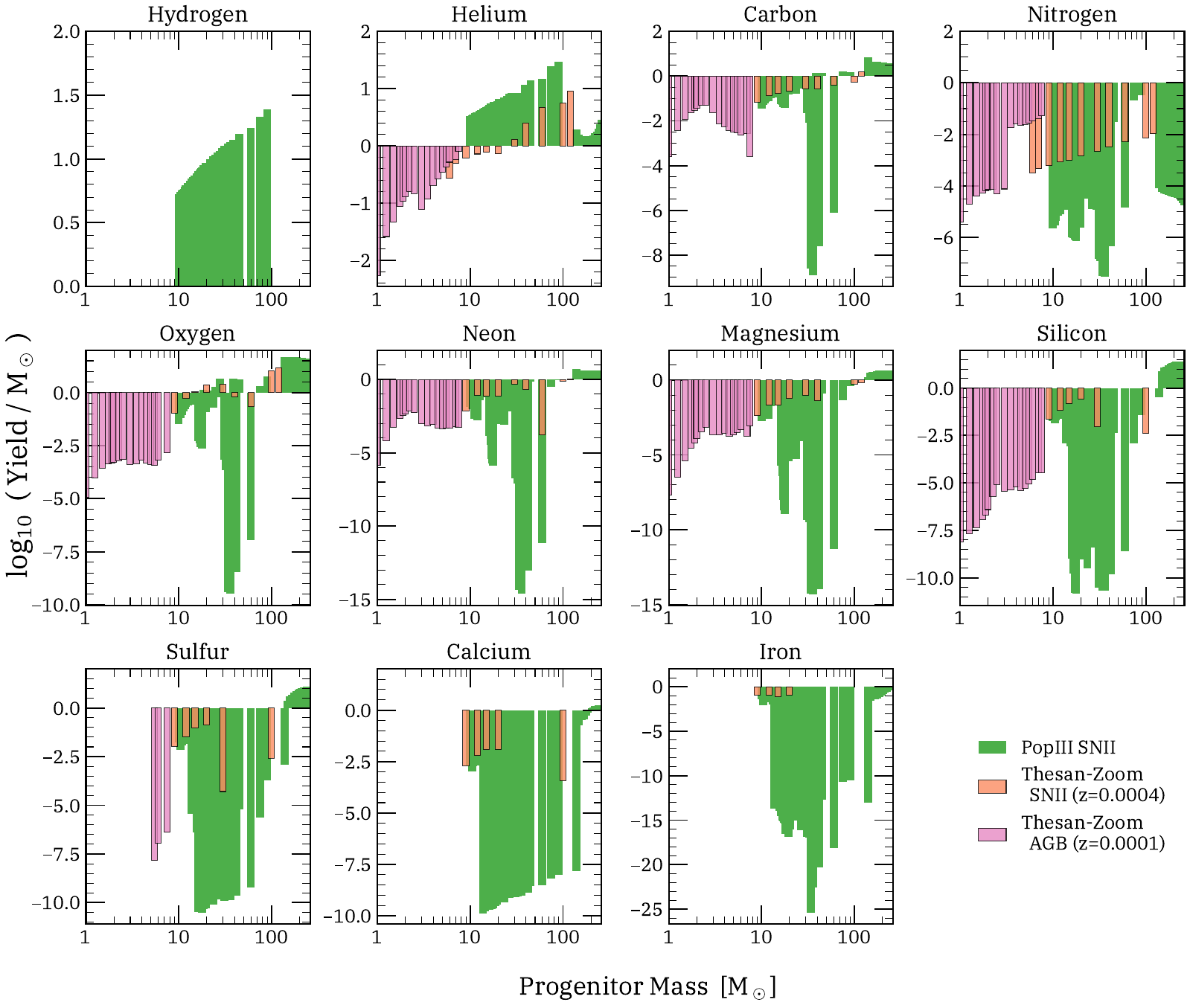}
  \caption{Comparison of absolute SN yields from \pt stars and non-\pt stars used in {\tt Thesan-Zoom} simulation for different species. The higher yields from the massive \pt stars (i.e $ > 140 ~\M $ ) which die as PISN, producing significantly more metals compared to both the lower mass \pt stars, and non-\pt stars is evident.}
  \label{fig:PopIII_SN_yields_comparison}
\end{figure*}

Finally for completeness, Fig.~\ref{fig:PopIII_SN_yields_comparison} plots the absolute yields from \pt stars and compares them with the yields from non-\pt stars used in the {\tt Thesan-Zoom} simulation. Note that the massive \pt stars (i.e., $ > 140 ~\M $) that die as PISN generally produce significantly more metals than both the lower-mass \pt stars and the non-\pt stars.

\section{Chemical and Radiative Feedback from PopIII stars}
\label{sec:chemical_feedback}
\begin{figure*}
    \centering
    \includegraphics[width=1\linewidth]{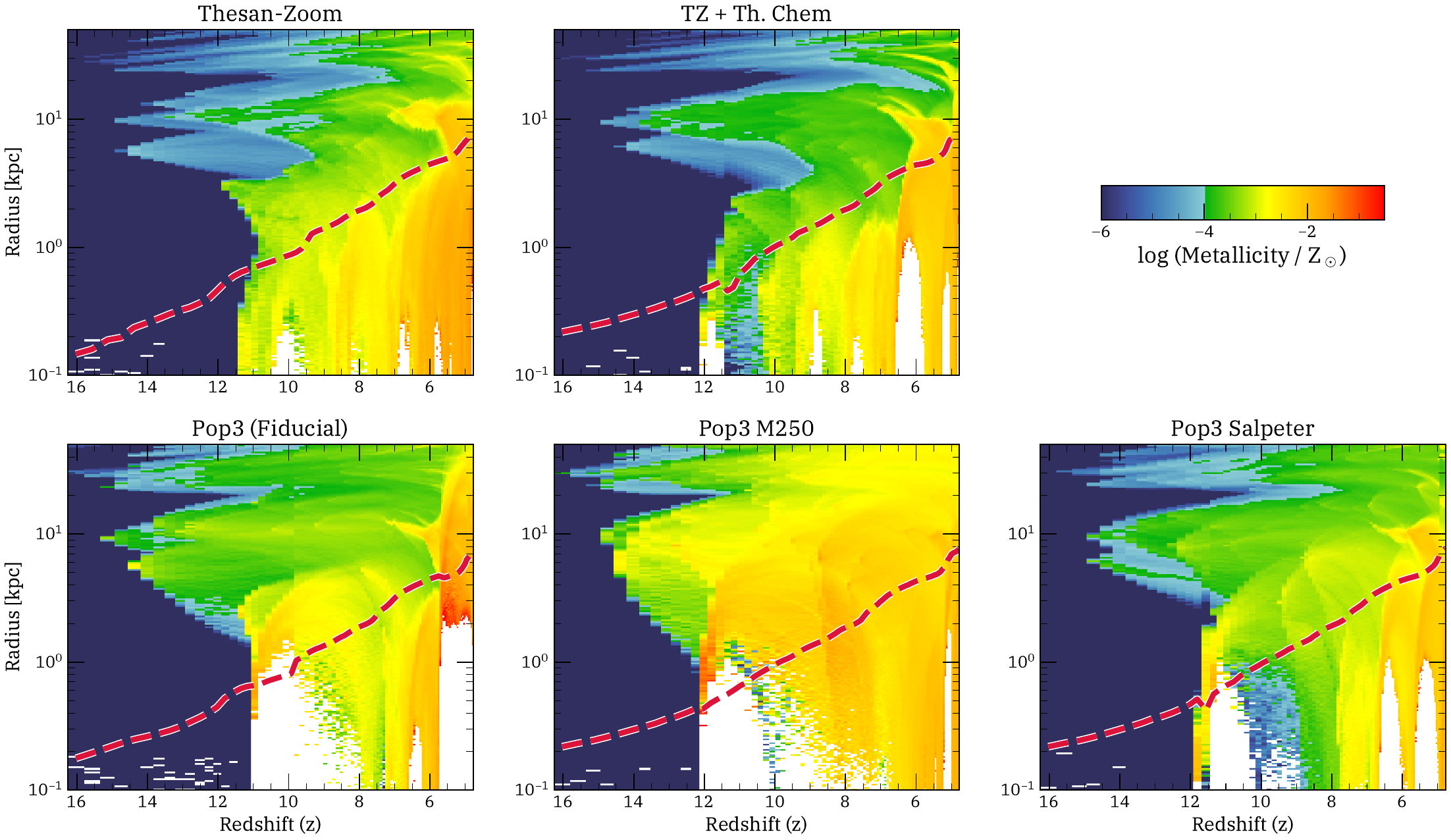}
    \caption{Mass-weighted metallicity in spherical shells centred on the primary halo across the simulation variations. The red dashed line indicates the evolving virial radius ($R_{\rm 200m}$). For visual clarity, the lower bound of the colour scale is capped at $Z=10^{-6}~Z_\odot$, capturing the initial star formation from pristine gas near the simulation floor of the metallicity; $Z=10^{-7}~Z_\odot$ (dark blue). The maps illustrate early, widespread intergalactic medium (IGM) enrichment by massive Pop III stars—most aggressively in the {\tt Pop3 M250} run—and the subsequent evacuation of inner-halo gas by intense supernova feedback (white gaps).}
    \label{fig:metallicity_shells}
\end{figure*}

\begin{figure*}
    \centering
    \includegraphics[width=1\linewidth]{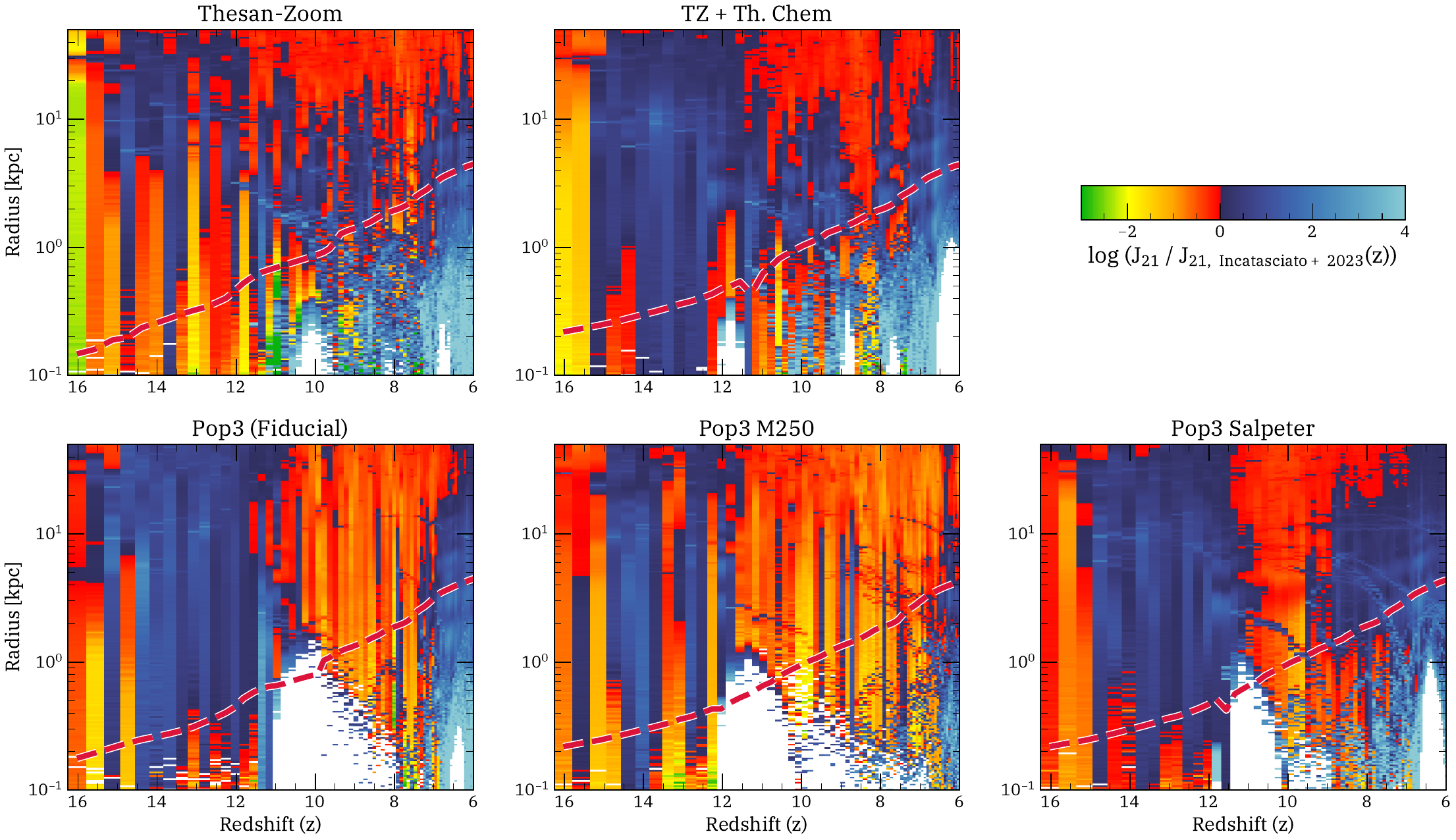}
    \caption{Radial and redshift evolution of the volume-weighted LW flux ($J_{21}$), normalized to the \citet{Incatasciato_2023} cosmological background, across the six simulation variations. The red dashed line tracks the virial radius ($R_{\rm 200m}$) of the central halo. The panels highlight how early, intense Pop III radiation strongly suppresses cooling in the IGM at high redshifts ($z > 10$), followed by a highly localized surge in flux at $z < 7$ driven by vigorous, metal-enriched Pop II star formation deep within the assembling halo.}
    \label{fig:J21_shells}
\end{figure*}

Fig.~\ref{fig:metallicity_shells} shows the mass-weighted metallicity in spherical shells centred on the target galaxy as a function of time across different model variations. The red dashed line denotes the virial radius, $R_{\rm 200m}$, of the target halo. In all runs, the first stars form around $z \approx 16-14$.
This occurs in pristine gas at the simulation's metallicity floor of $Z = 10^{-7}~Z_\odot$ (represented by the dark blue regions, as the colour-bar is strictly bounded at a minimum of $10^{-6}~Z_\odot$ for visual clarity). As soon as these first stars die, they begin to enrich the intergalactic medium (IGM), and by $z \sim 12$, the central halo is enriched above the \pt metallicity threshold. The simulations predict stronger early enrichment of the IGM in the {\tt Pop3}* runs. For instance, at $z > 12$, the IGM at $r > 10$ kpc is enriched to $Z > 10^{-4}~Z_\odot$ (indicated by the green/yellow contours) in the {\tt Pop3}* variations, whereas in the {\tt Thesan-Zoom} and {\tt TZ + Th. Chem} runs, the IGM remains largely below $Z < 10^{-4}~Z_\odot$. This effect is most pronounced in the {\tt Pop3 M250} run, where more massive \pt stars enrich the IGM more aggressively, driving metallicities to $Z \gtrsim 10^{-3}~Z_\odot$. Among the {\tt Pop3}* runs, metal enrichment in both the IGM and the central halo is weakest for the {\tt Pop3 Salpeter} variation. In this model, \pt star formation is dominated by lower-mass \pt stars, which yield fewer metals than their massive counterparts. Additionally, across all runs, there are prominent gaps in the data (visible as white regions, particularly at $r < 1$ kpc for $z < 10$). These arise when gas is evacuated by strong SN feedback. In the {\tt Pop3}* runs, these gaps are generally more extended due to the more energetic nature of \pt supernovae. Finally, we note that in the {\tt TZ + Th. Chem} and {\tt Pop3 Salpeter} runs, distinct dark blue streaks of gas remain below the \pt metallicity threshold and fall into the central halo between $10 < z < 12$, which will subsequently trigger late-stage \pt star formation.

Fig.~\ref{fig:J21_shells} shows the volume-weighted Lyman-Werner (LW) flux relative to the predicted LW background from \citet{Incatasciato_2023} in spherical shells centred on the central galaxy, as a function of time, across different model variations. The red dashed line denotes the $R_{\rm 200m}$ of the target halo. We limit the $x$-axis of the subplots to $z=6$, as the functional form for $J_{21}$ in \citet{Incatasciato_2023} is valid for $6 < z < 23$. The {\tt Pop3}* runs exhibit a stronger LW flux in the IGM than the non-\pt runs at early times. The first stars form outside the central galaxy across all variations; however, the \pt stars in the {\tt Pop3}* runs produce a stronger LW background for $11 < z < 16$ that suppresses cooling by destroying $\rm H_2$, thereby suppressing star formation in the surrounding medium. This sustained suppression is visually evident at $z < 10$, where the outer radii ($r > R_{\rm 200m}$) in the {\tt Pop3}* variations remain suppressed relative to the baseline, in stark contrast to the persistent elevated flux seen in the {\tt Thesan-Zoom} and {\tt TZ + Th. Chem} runs. Furthermore, the {\tt Pop3 M250} run predicts a rapid decline in the LW flux at $z < 11$, primarily due to stronger metal enrichment in the IGM, which produces metal-rich stars with lower LW radiation. Similar behaviour is predicted in the {\tt Pop3 (fiducial)} and {\tt Pop3 Salpeter} runs, but it is much weaker than in the {\tt Pop3 M250} run relative to the {\tt Thesan-Zoom} and {\tt TZ + Th. Chem} runs. Finally, at $z < 7$, all panels display an intense, highly localized surge in relative LW flux deep inside the halo ($r < R_{\rm 200m}$). This is driven by vigorous, sustained Population II star formation as metal-enriched gas funnels into the massive central potential well, producing an overwhelming local radiation field that completely outshines nearby sources.

\section{Affect of Gas-Phase Reactions on $\rm H_2$ formation}
\label{app:h2}
\begin{figure*}
  \centering
  \includegraphics[trim=0 35.5 0 0,clip,width=0.7\linewidth]{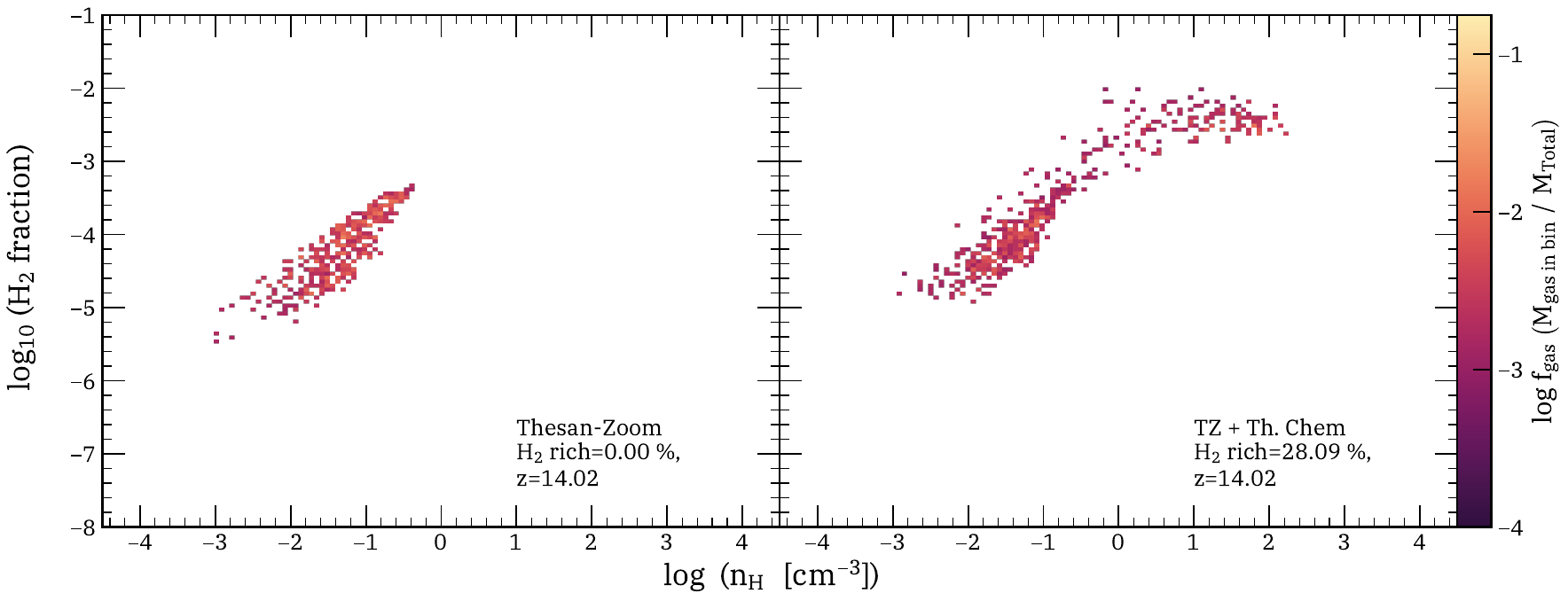}
  \includegraphics[trim=0 24 0 0,clip,width=0.7\linewidth]{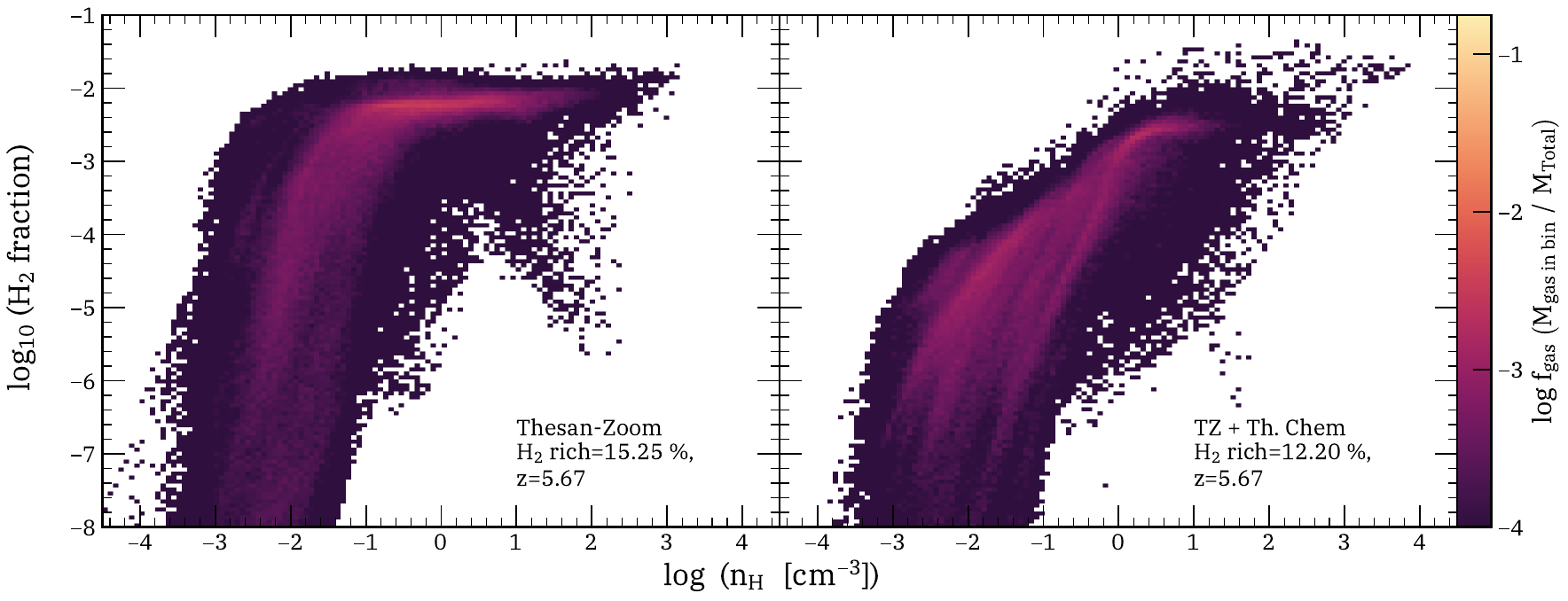}
  {\large $\log_{10}~(\text{Hydrogen number density}~[\pcc])$}
  \caption{$\rm H_2$ phase-space diagram of the central halo: $x,y$-axes shows hydrogen number density and $\rm H_2$ fraction respectively, colour-coded by the relative mass in each bin. Left and right column shows the {\tt Thesan-Zoom} and {\tt TZ+Th. Chem} variation at two different epochs (top at $z\thickapprox14$, bottom $z\thickapprox 5.6$). The annotations also shows the fraction of $\rm H_2$ rich gas $(n_{{\rm H}} >1~{\pcc}, x_{\rm{H_2}} > 10^{-4})$ At high redshift, the updated thermochemistry produces more $\rm H_2$ rich gas.}.
  \label{fig:H2_phaseSpace}
\end{figure*}

\begin{figure*}
  \centering
  \includegraphics[width=1\linewidth]{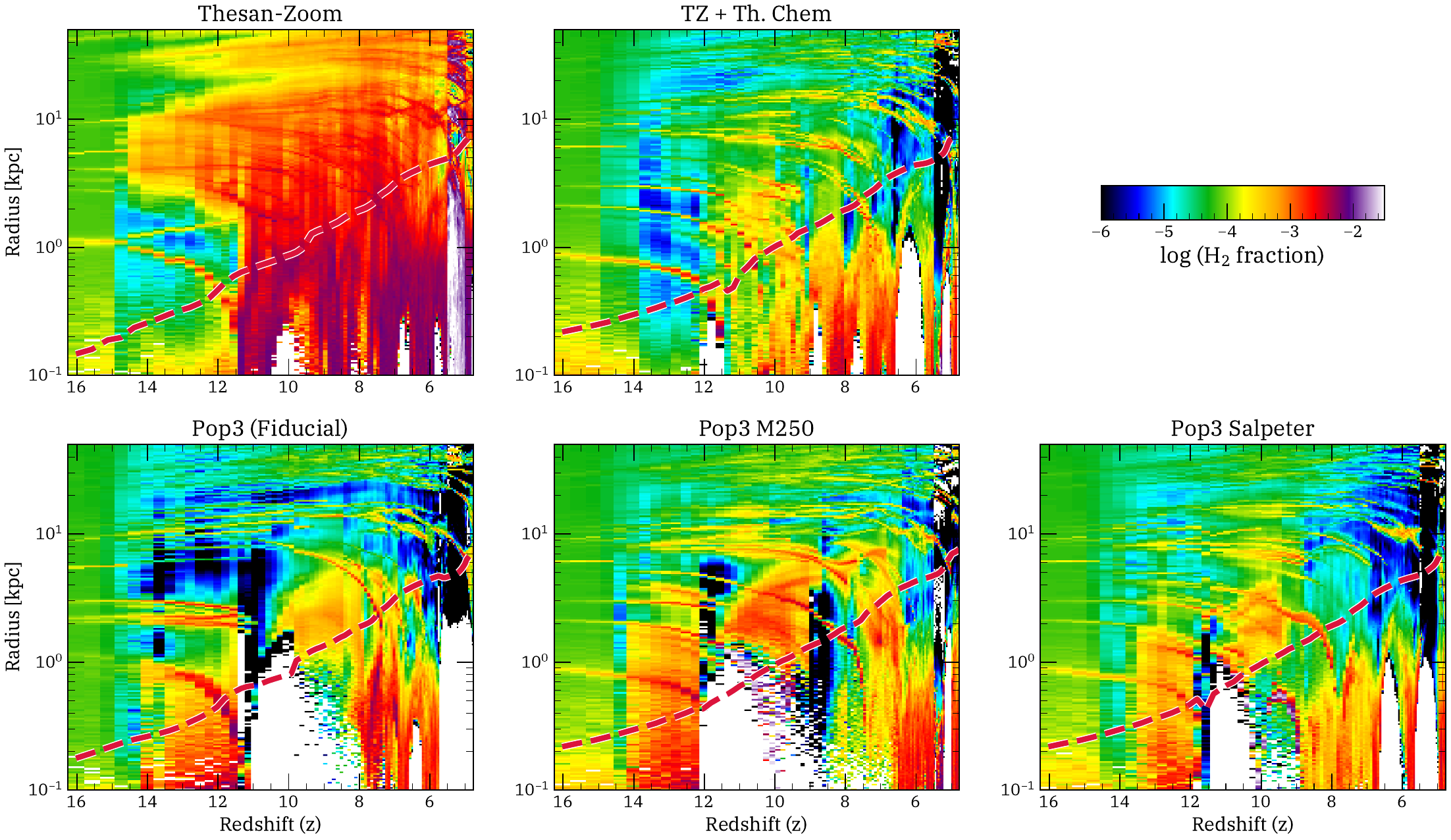}
  \caption{Mass-weighted gas $\rm H_2$ fraction in spherical shells centred on the primary halo across the simulation variations. The red dashed line indicates the evolving virial radius ($R_{\rm 200m}$). The maps illustrates that with the updated thermochemistry network, $\rm H_2$ in primarily confined within the dense substructures, unlike {\tt Thesan-Zoom}, where $\rm H_2$ is very diffused, and gas outside the halo exhibit high $\rm H_2$ fractions. Additionally, in the {\tt Pop3*} runs, the strong LW feedback from \pt further destroys $\rm H_2$ outside the halo creating pockets of $\rm H_2$ poor regions (black patches on the maps).}
  \label{fig:H2_shells}
\end{figure*}

To validate the enhanced thermochemical network, Fig.~\ref{fig:H2_phaseSpace} compares the distribution of $\rm H_2$-rich gas $(n_{\rm H} > 1~\pcc, x_{\rm H_2} > 10^{-4})$ in the central halo for the baseline {\tt Thesan-Zoom} and {\tt TZ + Th. Chem models}. At early epochs $(z\thickapprox14)$, prior to significant metal and dust enrichment, the baseline model forms negligible amounts of molecular hydrogen. In contrast, the enhanced network successfully catalyses $\rm H_2$ formation via the gas-phase $\hp, \hmn$ channels, yielding a significantly $\rm H_2$-rich gas. At later epochs $(z\thickapprox7)$, after supernova feedback has enriched the ISM, dust-driven $\rm H_2$ formation becomes dominant, and the two models predictably converge. While thermodynamic states at later times are subject to the stochastic divergence of sub-grid feedback, the $z\thickapprox14$ epoch provides verification of the primordial chemistry network producing higher $\rm H_2$

Fig.~\ref{fig:H2_shells} illustrates the radial evolution of the $\rm H_2$ mass fraction over time. The baseline {\tt Thesan-Zoom} model exhibits an unphysical excess of $\rm H_2$ in the diffuse IGM $(r > R_{\rm 200m})$ which is an effect of applying a dust-calibrated formation prescription to pristine, low-density gas. In contrast, the implementation of the explicit $\hp, \hmn$ primordial network i.e. the {\tt TZ + Th. Chem} correctly restricts $\rm H_2$ formation to the denser, collapsing regions of the halo, as the gas-phase catalytic channels are highly inefficient at IGM densities. Furthermore, the full {\tt Pop3*} runs vividly demonstrate the impact of Lyman-Werner feedback; the intense LW radiation from the first stars (as shown in Fig.~\ref{fig:J21_shells}) actively photodissociates molecular hydrogen in the halo outskirts at $z > 8$, creating distinct $\rm H_2$-poor voids that regulates star formation in neighbouring substructures.

While our network catalyses early $\rm H_2$ formation, later-epoch molecular fractions remain low. This underproduction is a known limitation in cosmological simulations. Recently \citet{Gurman_2025} with their sub-solar resolution simulations demonstrate that simulations systematically under-produce $\rm H_2$ by factors of 2 to 4 due to unresolved small-scale density peaks. They found that $\rm H_2$ formation is highly sensitive to unresolved density substructures, and that matching observed molecular fractions requires an explicit sub-grid clumping model to account for small-scale density enhancements. Given our mass resolution of $\thickapprox 1.14 \times 10^3~\M$, our models intrinsically miss these critical substructures. Implementing a sub-grid clumping factor to more accurately capture the transition from atomic to molecular gas remains a necessary and promising avenue for future iterations of this framework.

\section{Calculation of \ion{He}{ii} 1640 \AA\ Emission and Equivalent Widths}
\label{sec:methods_heII}
\begin{figure}
  \centering
  \includegraphics[width=1\linewidth]{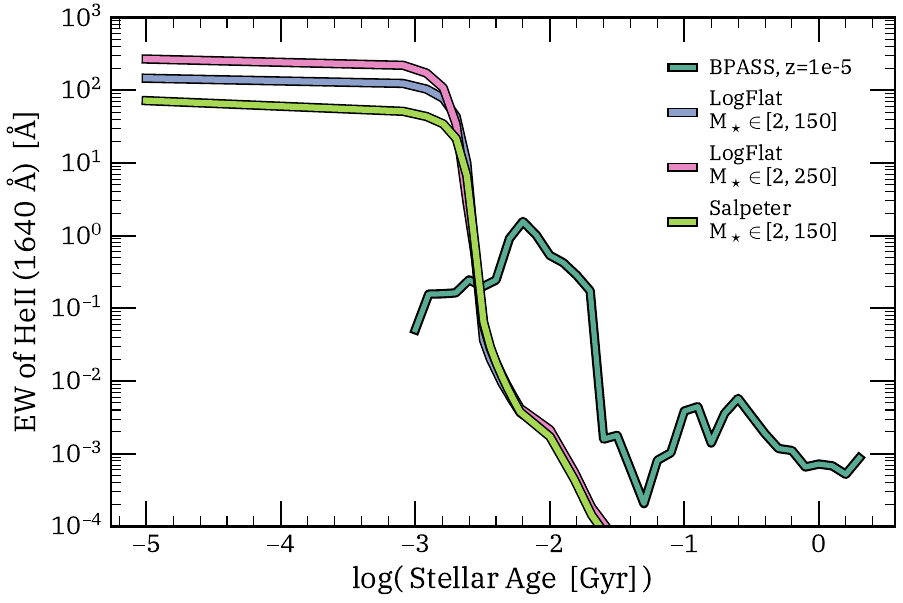}
  \caption{Expected equivalent width of the \ion{He}{II} $1640$ \AA~line as a function of the stellar age. The young \pt populations predict at-least 2 order of magnitude stronger emission line compared to the lowest metallicity non-\pt populations for $t\lesssim 2$ Myr}.
  \label{fig:Expected_HeII_EW}
\end{figure}

To model the \ion{He}{II} 1640 \AA\ emission from simulated galaxies, we construct composite synthetic spectra for each subhalo by combining the contributions of its constituent star particles. We select all star particles within a 30 physical kpc radius of the subhalo centre. The stellar populations are strictly partitioned into Metal-Free (Population III) and Metal-Enriched (Population II) stars, each evaluated with distinct baseline Spectral Energy Distribution (SED) models. For Population II stars, we use the BPASS v2.2.1 binary stellar population models \cite{Eldridge2017_BPASS}, assuming a Chabrier initial mass function (IMF) \cite{Chabrier_2003} with lower and upper mass limits of $0.1$ and $100\ \mathrm{M}_{\odot}$. For Population III stars, we employ three custom metal-free spectra, namely a log-flat IMF with a mass range of $2 - 150\ \mathrm{M}_{\odot}$, a log-flat IMF with a mass range of $2 - 250\ \mathrm{M}_{\odot}$, and a Salpeter IMF with a mass range of $2 - 150\ \mathrm{M}_{\odot}$ (see Tab.~\ref{tab:IMF_params} and Tab.~\ref{tab:sims}).

From these theoretical SEDs, we extract grids of specific continuum luminosities and ionizing photon rates as a function of stellar age and metallicity. For a given simulation snapshot, we calculate the age (based on the particle formation time) and metallicity of each star particle. We then determine the specific particle properties by performing 2D bilinear interpolation (over $\log(\mathrm{Age})$ and $\log(Z)$) across the BPASS grids for Pop II stars, and 1D interpolation (over $\log(\mathrm{Age})$) across the Pop III grids. The \ion{He}{II} 1640 \AA\ line emission is driven by the flux of photons capable of removing the last electron from \ion{He}{II} ($E > 54.42$~eV). For each star particle, we integrate the assigned SED to compute the total \ion{He}{II} ionizing photon rate per unit stellar mass, $Q_{\mathrm{HeII}}$ ($\mathrm{photons\ s}^{-1}$M$_\odot^{-1}$). Assuming ionization equilibrium and Case B recombination \cite[e.g.,][]{Osterbrock_2006}, roughly $45\%$ of \ion{He}{II} recombinations result in the emission of a $1640$ \AA\ photon. The predicted He II line luminosity ($L_{\mathrm{HeII}}$) for a given stellar mass is therefore calculated as:
\begin{equation}
  L_{\mathrm{HeII}} = M_* \times 0.45 \, Q_{\mathrm{HeII}} \left( \frac{hc}{\lambda} \right)
\end{equation}
where $M_*$ is the star particle mass, $\lambda = 1640.42\ \text{\AA}$, and $hc/\lambda \approx 1.21 \times 10^{-11}\ \mathrm{erg}$ is the energy of a single 1640 \AA\ photon.

Along with the ionizing photon rate, we interpolate the rest-frame UV continuum luminosity densities $L_\nu$ at 1640 \AA\ and 1500 \AA\ . To compute the expected He II Equivalent Width (EW), we convert the 1640 \AA\ continuum luminosity from frequency to wavelength space ($L_{\lambda, 1640} = L_{\nu, 1640} \cdot c / \lambda^2$). The total integrated observables for a given galaxy are then computed by summing the line and continuum contributions from all bound Pop II and Pop III star particles:
\begin{equation}
  \mathrm{EW}_{\mathrm{HeII}} = \frac{\sum L_{\mathrm{HeII}}}{\sum L_{\lambda, 1640}}
\end{equation}
The galaxy's absolute UV magnitude ($M_{1500}$) is derived simultaneously from the integrated $1500\ \text{\AA}$ continuum luminosity, defined in the standard AB magnitude system.

Fig. \ref{fig:Expected_HeII_EW} plots the expected equivalent width of the HeII $1640$\AA ~line for a given age of the stellar population at a fixed metallicity for the PopII stars and for different IMF choices for the PopIII stars. For the PopII population, we use the lowest metallicity bin available in the BPASS table. The simulations predict that for young stellar populations (age $< 2$ Myr), the expected EW from the \pt population is at least 2 orders of magnitude higher than that of the PopII population (primarily arising from Wolf-Rayet stars \citep[see e.g.][]{Leitherer_2019}). The choice of IMF also affects the HeII emission, with the Log-flat IMF with $M_{\rm max} = 250 \M$ predicting the highest \ion{He}{II} EW, followed by our fiducial \pt IMF and then the Salpeter IMF.


\bsp  
\label{lastpage}
\end{document}

%% file: Tab-thermochem.tex
\centering
\renewcommand{\arraystretch}{1.3}
\begin{tabular}{lcccccc}
  \hline
  \begin{tabular}[c]{@{}c@{}}Bins\\ (eV)
  \end{tabular} &
  \begin{tabular}[c]{@{}c@{}}Optical\\ $1-5.8$
  \end{tabular} &
  \begin{tabular}[c]{@{}c@{}}FUV\\ $5.8-11.2$
  \end{tabular} &
  \begin{tabular}[c]{@{}c@{}}LW\\ $11.2-13.6$
  \end{tabular} &
  \begin{tabular}[c]{@{}c@{}}EUV1\\ $13.6-24.6$
  \end{tabular} &
  \begin{tabular}[c]{@{}c@{}}EUV2\\ $24.6-54.4$
  \end{tabular} &
  \begin{tabular}[c]{@{}c@{}}EUV3\\ $54.4-\infty$
  \end{tabular} \\
  \hline\hline
  $\rm \sigma_{H_2}^A$ ($10^{-18}$~cm$^2$) & $23.17$ & $7.84$ & $4.43$ & $2.7$ & $1.21$ & $0.48$ \\
  $\rm \sigma_{H_2}^B$ ($10^{-18}$~cm$^2$) & $3.74\times10^{-4}$ & $2.81$ & $5.95$ & $0.65$ & $0$ & $0$ \\
  $\rm \sigma_{H_2}^C$ ($10^{-18}$~cm$^2$) & $0$ & $0$ & $0$ & $0$ & $0.28$ & $0.2$ \\
  $\rm \sigma_{H_2}^D$ ($10^{-18}$~cm$^2$) & $0$ & $0$ & $0$ & $5.23$ & $2.51$ & $0.44$ \\
  $\rm \sigma_{H_2}^E$ ($10^{-18}$~cm$^2$) & $0$ & $0$ & $0$ & $0.51$ & $0$ & $0$ \\
  $\rm \sigma_{H_2}^{LW}$ ($10^{-18}$~cm$^2$) & $0$ & $0$ & $0.21$ & $0$ & $0$ & $0$ \\
  $\rm \sigma_{HI}$ ($10^{-18}$~cm$^2$) & $0$ & $0$ & $0$ & $3.36$ & $0.71$ & $0.11$ \\
  $\rm \sigma_{HeI}$ ($10^{-18}$~cm$^2$) & $0$ & $0$ & $0$ & $0$ & $5.21$ & $1.53$ \\
  $\rm \sigma_{HeII}$ ($10^{-18}$~cm$^2$) & $0$ & $0$ & $0$ & $0$ & $0$ & $1.42$ \\
  \hline
  $\mathfrak{h}_{\rm D}$ (eV) & $0$ & $0$ & $0$ & $3.9$ & $13.98$ & $41.63$ \\
  $\mathfrak{h}_{\rm E}$ (eV) & $0$ & $0$ & $0$ & $0.49$ & $0$ & $0$ \\
  $\mathfrak{h}_{\rm LW}$ (eV) & $0$ & $0$ & $0$ & $0$ & $0$ & $0$ \\
  $\mathfrak{h}_{\rm HI}$ (eV) & $0$ & $0$ & $0$ & $3.19$ & $15.46$ & $43.18$ \\
  $\mathfrak{h}_{\rm HeI}$ (eV) & $0$ & $0$ & $0$ & $0$ & $5.32$ & $32.32$ \\
  $\mathfrak{h}_{\rm HeII}$ (eV) & $0$ & $0$ & $0$ & $0$ & $0$ & $2.41$ \\
  \hline
  $p_{\rm D}$ (eV) & $0$ & $0$ & $0$ & $19.1$ & $29.18$ & $56.83$ \\
  $p_{\rm E}$ (eV) & $0$ & $0$ & $0$ & $14.65$ & $0$ & $0$ \\
  $p_{\rm LW}$ (eV) & $0$ & $0$ & $12.35$ & $0$ & $0$ & $0$ \\
  $p_{\rm HI}$ (eV) & $0$ & $0$ & $0$ & $16.79$ & $29.06$ & $56.79$ \\
  $p_{\rm HeI}$ (eV) & $0$ & $0$ & $0$ & $0$ & $29.91$ & $56.91$ \\
  $p_{\rm HeII}$ (eV) & $0$ & $0$ & $0$ & $0$ & $0$ & $56.83$ \\
  \hline\hline
  & & & & & &
\end{tabular}